\documentclass[%pdftex,
smallcondensed,
epjc3
]{svjour3}          % twocolumn
\RequirePackage[T1]{fontenc}

\smartqed  % flush right qed marks, e.g. at end of proof

\RequirePackage{graphicx}
\RequirePackage{mathptmx}      % use Times fonts if available on your TeX system
\RequirePackage{flushend}
\RequirePackage[numbers,sort&compress]{natbib}
\RequirePackage[colorlinks,citecolor=blue,urlcolor=blue,linkcolor=blue]{hyperref}
\usepackage{amsmath}
\usepackage{txfonts}
\usepackage{amssymb}
\usepackage[figtopcap]{subfigure}

\graphicspath{{./Figs/}}

\journalname{Eur. Phys. J. C}

\begin{document}

\title{Non-trivial class of anisotropic compact stellar model in Rastall gravity}

\author{G.G.L. Nashed\thanksref{e1,addr1}
        \and
       W. El Hanafy\thanksref{e2,addr1} %etc.
}

%\thankstext[$\star$]{t1}{Thanks to the title}
\thankstext{e1}{e-mail: nashed@bue.edu.eg}
\thankstext{e2}{e-mail: waleed.elhanafy@bue.edu.eg}

\institute{Centre for Theoretical Physics, The British University in Egypt, P.O. Box 43, El Sherouk City, Cairo 11837, Egypt\label{addr1}
%          \and
%          Second Address, Street, City, Country\label{addr2}
%          \and
%          \emph{Present Address:} Street, City, Country\label{addr3}
}

\date{Received: date / Accepted: date}

\maketitle

\begin{abstract}
We investigated Rastall gravity, for an anisotropic star with a static spherical symmetry, whereas the matter-geometry coupling as assumed in Rastall Theory (RT) is expected to play a crucial role differentiating RT from General Relativity (GR). Indeed, all the obtained results confirm that RT is not equivalent to GR, however, it produces same amount of anisotropy as GR for static spherically symmetric stellar models. We used the observational constraints on the mass and the radius of the pulsar \textit{Her X-1} to determine the model parameters confirming the physical viability of the model. We found that the matter-geometry coupling in RT allows slightly less size than GR for a given mass. We confirmed the model viability via other twenty pulsars' observations. Utilizing the strong energy condition we determined an upper bound on compactness $U_\text{max}\sim 0.603$, in agreement with Buchdahl limit, whereas Rastall parameter $\epsilon=-0.1$. For a surface density compatible with a neutron core at nuclear saturation density the mass-radius curve allows masses up to $3.53 M_\odot$. We note that there is no equation of state is assumed, however the model fits well with linear behaviour. We split the twenty pulsars into four groups according to the boundary densities. Three groups are compatible with neutron cores while one group fits perfectly with higher boundary density $8\times 10^{14}$ g/cm$^3$ which suggests that those pulsars may have quark-gluon cores.
\end{abstract}

%%%%%%%%%%%%%%%%%%%%%%%%%%%%%%%%%%% Section 1 %%%%%%%%%%%%%%%%%%%%%%%%%%%%%%%%%%%%%%%%
\section{Introduction}\label{S1}
%%%%%%%%%%%%%%%%%%%%%%%%%%%%%%%%%%%%%%%%%%%%%%%%%%%%%%%%%%%%%%%%%%%%%%%%%%%%%%%%%%%%%%
The stellar structure topic has been developed over many years either in Newtonian gravity or later in General Relativity (GR) since the firstly obtained exterior/interior solution of a spherically symmetric object \cite{Schwarzschild:1916uq, Schwarzschild:1916ae}. Motivated by the claim that pressure at core of the compact star model could have anisotropic structure where the density exceeds the nuclear density $\sim 10^{15}$ g/cm$^{3}$, many models have been developed imposing the anisotropic pressure concept (assuming radial and tangential pressures are different) to derive realistic stellar models within the GR context \cite{ruderman1972pulsars} (see also \cite{PhysRevD.65.104011,Schunck:2003kk,Mak:2001eb,Usov:2004iz,Rahaman:2010mr,Rahaman:2010bt,Varela:2010mf,Rahaman:2011hd,Kalam:2012sh,
Deb:2015vda,Shee:2015kqa,Maurya:2016oml}) and in modified gravity as well \cite{Abbas:2014rja,Abbas:2015yma,Abbas:2015wea,Zubair:2015cpa,zubair2016some,zubair2016possible,Ilyas:2018tht,Yousaf:2017lto,Shamir:2017rjz,Shamir:2017yza,Das:2016mxq}. Possible sources of anisotropy of ultra-compact stars could be due to solidification \cite{Palmer:1974hb}, superfluidity \cite{Ramanan:2019kwf}, strong magnetic fields \cite{Weber:2006ep}, hyperons \cite{Rahmansyah:2020gar}, pion-condesation \cite{Sawyer:1972cq} and strong interactions \cite{bowers1974anisotropic}. Local anisotropy and its consequences on the stability and the gravitational collapse of self-gravitating systems has been investigated by Herrera and Santos \cite{herrera1997local} (see also \cite{herrera1992cracking,Herrera:2007kz,Herrera:2008bt,Herrera:2011cr}).

The GR theory has been proven to be a successful theory of gravity on solar system scales (spacetime curvature is weak) by many observational tests, c.f. \cite{misner1973gravitation}, and also on black hole scales (spacetime curvature is extremely strong) using black hole shadows observations by Event Horizon Telescope \cite{EventHorizonTelescope:2019dse} not to mention the perfect predictions of the gravitational waves due to binary black holes merging \cite{Abbott:2016blz,Abbott:2017oio,TheLIGOScientific:2017qsa}. On the cosmological scales, however, the GR does not provide answers for explaining the late accelerated expansion \cite{Riess:1998cb,Perlmutter:1998np,deBernardis:2000sbo,Knop:2003iy}. Even in presence of a cosmological constant $\Lambda$, the discrepancy of the current Hubble parameter $H_0$ value, between early universe observations by Planck satellite and late universe measurements by distance ladder or strong lensing, may point out the need to modify the GR theory \cite{Verde:2019ivm}, see also \cite{Hashim:2020sez,Hashim:2021pkq,DiValentino:2021izs}.

Many efforts have been done to generalize GR theory by using general function in Einstein-Hilbert action instead of the Ricci invariant, e.g. $f(R)$, $f(G)$, $f(T)$ and mimetic gravity \cite{Clifton:2011jh,Nojiri:2006ri,DeFelice:2010aj,Nashed:2019tuk,Nashed:2011fg,2014IJTP...53.3901W,Nashed:2018efg,Nashed:2020kdb,Cognola:2006eg,Li:2007jm,DeFelice:2008wz,Astashenok:2014nua,Linder:2010py,Cai:2015emx,Awad:2017tyz,Nashed:uja,Nashed:2013bfa,Nashed:2018qag,Nashed:2019yto,ElHanafy:2017sih,Hashim:2020sez,Hashim:2021pkq,ElHanafy:2020pek}. In fact, these modified theories kept the fundamental assumption that the covariant divergence of the energy-momentum vanishes, i.e. $\mathcal{T}^{\alpha}{_{\beta; \alpha}}=0$ where the semicolon denotes the Levi-Civita covariant derivative. On the contrary, Rastall attempted to modify GR by dropping this assumption replacing it by setting $\mathcal{T}^{\alpha}{_{\beta; \alpha}}=a_{\beta}$ where $a_{\beta}$ vanishes in flat spacetime (vacuum) and recovers GR, otherwise it does not \cite{Rastall:1972swe,Rastall:1976uh}. Rastall showed that $a_{\beta} \propto \partial_\beta \mathcal{R}$ is a reasonable choice which reflects the non-minimal coupling between matter and geometry. Interestingly, some cosmological models have been constructed using RT \cite{PhysRevD.85.084008,fabris2012rastall,moradpour2016thermodynamics} as well as black hole solutions \cite{heydarzade2017black,ma2017noncommutative,lobo2018thermodynamics,bamba2018thermodynamics,xu2018kerr}. Moreover, shadows of rotating black holes have been investigated within RT by Kumar et al. \cite{kumar2018rotating} and Heydarzade and Darabi \cite{heydarzade2017black}. Furthermore, the effects of the non-minimal coupling between matter and geometry on thermodynamics of black holes in RT has been investigated in comparison with the GR theory \cite{lobo2018thermodynamics,moradpour2016thermodynamic,Soroushfar2019,Cruz:2019jiq}.

Recently, Vissar claimed that RT is completely equivalent to GR \cite{Visser:2017gpz}. On the contrary, Darabi et al. investigated Visser's claim but they concluded that Visser misinterpreted the matter-geometry coupling term which led him to wrong conclusion \cite{Darabi:2017coc}. In addition, they showed that by applying Visser's approach to $f(R)$ theory one may conclude that it is equivalent to GR as well which is not true. Different studies have proven that RT is not equivalent to GR \cite{Hansraj:2018zwl,Hansraj:2020clg,abbas2018new,abbas2018isotropic}. Visser's conclusion is correct when Ricci scalar vanishes for black holes in general, otherwise the claim is incorrect and both theories are not equivalent. One of the good examples which may reveal the contribution of the matter-geometry coupling in RT in contrast to GR is the stellar models when the presence of matter plays a crucial role. It is the aim of the present study to derive a anisotropic static spherically symmetric interior solution using RT and confront it with pulsars observations.

The arrangement of this study is as follows: In Section \ref{S2} we briefly review the main assumptions and features of the non-conservative theory of Rastall's gravity. In Section \ref{S3}, we apply Rastall's field equations to a spherically symmetrical object in presence of anisotropic matter source. The system consists of three non-linear differential equations in five unknown functions (two metric potentials in addition to the energy-density, radial and tangential pressures). Therefore, two constraints have been imposed to close the system without using equation of state (EoS): We assumed a specific form of one of the metric potential, $g_{rr}$ as usually done in interior solutions, additionally we made the metric potential $g_{tt}$ contribution to the anisotropy to vanish. Consequently, we derive the analytic form of energy-density, radial, and tangential pressures which satisfy Rastall field equations. In Section \ref{S4}, we list the necessary and sufficient physical conditions that any stellar model must fulfill to be compatible with a realistic compact star. In Section \ref{S5}, we investigated the viability of the obtained solution with the conditions listed in Section \ref{S4}. In Section \ref{S6}, we matched our model with the exterior vacuum solution (Schwarzschild solution) fixing the model parameters and constants. The pulsar $\textit {Her X-1 }$, with mass $M= 0.85 \pm 0.15\, M_\odot,$ and  radius $R= 8.1 \pm 0.41$ km \cite{Abubekerov_2008} is used  to determine the constants of the model. In Section \ref{S7}, we investigated the stability of the model using a modified version of Tolman-Oppenheimer-Volkoff (TOV) equation of hydrostatic equilibrium and the adiabatic indices. In Section \ref{S8} we use other observational mass-radius constraints of twenty pulsars' data to test the viability of the model. Also we plot the mass-radius profile of the model for different surface densities compatible with neutron core pulsars to show the maximum masses allowed by the model. Finally, we summarize and conclude in Section \ref{S9}.
%%%%%%%%%%%%%%%%%%%%%%%%%%%%%%%%%%% Section 2 %%%%%%%%%%%%%%%%%%%%%%%%%%%%%%%%%%%%%%%%
\section{Rastall gravitational theory}\label{S2}
%%%%%%%%%%%%%%%%%%%%%%%%%%%%%%%%%%%%%%%%%%%%%%%%%%%%%%%%%%%%%%%%%%%%%%%%%%%%%%%%%%%%%%
In Riemann geometry, by making use of the contracted Bianchi Identity on one hand and the minimal coupling procedure on the other hand,
\begin{equation}\label{Bianchi}
    \mathcal{G}_{\alpha\beta;\alpha}= (\mathcal{R}_{\alpha\beta}-\frac{1}{2}g_{\alpha\beta}\mathcal{R})_{;\alpha}\equiv 0, \qquad \qquad  \mathcal{T}_{\alpha\beta;\alpha}=0,
\end{equation}
this led Einstein to formulate the consistent field equations of GR
\begin{equation}\label{GR}
    \mathcal{G}_{\alpha\beta}=\chi \mathcal{T}_{\alpha\beta},
\end{equation}
where $\chi=8\pi G_N/c^4$ where $G_N$ is the Newtonian gravitational constant and $c$ is the speed of light, $\mathcal{G}_{\alpha\beta}$ denotes Einstein tensor, $\mathcal{R}_{\alpha\beta}$ denotes Ricci tensor and $\mathcal{R}=g^{\alpha\beta}\mathcal{R}_{\alpha\beta}$ denotes Ricci invariant.
Rastall, however, dropped the minimal coupling procedure assuming non-divergence-free energy-momentum in curved spacetime \cite{Rastall:1972swe,Rastall:1976uh}
\begin{equation}\label{Rastall}
    \mathcal{T}^\alpha{}_{\beta;\alpha}\neq 0, \qquad \mathcal{T}^\alpha{}_{\beta;\alpha}=a_{\beta}=\tilde{\epsilon}\, \partial_{\beta} \mathcal{R},
\end{equation}
where the constant of proportionality $\tilde{\epsilon}$ measures how much the conservation law is locally violated. According to this assumption Rastall obtained a consistent set of field equations
\begin{equation}\label{RT}
    \mathcal{G}_{\alpha\beta}=\mathcal{R}_{\alpha\beta}-\frac{1}{2}g_{\alpha\beta}\mathcal{R}=\chi(\mathcal{T}_{\alpha\beta}-\tilde{\epsilon} g_{\alpha\beta}\mathcal{R}).
\end{equation}
Obviously Rastall gravity reflects non-minimal coupling features of matter and geometry whereas Einstein's field equations are recovered when the spacetime is flat. Alternatively, Eq. \eqref{RT} can be rewritten as
\begin{equation}\label{RT2}
    \mathcal{R}_{\alpha\beta}-\left(\frac{1}{2}-\chi \tilde{\epsilon}\right) g_{\alpha\beta}\mathcal{R}=\chi \mathcal{T}_{\alpha\beta}.
\end{equation}
Contracting the above equation gives
\begin{equation}\label{R-contraction}
    (1-4\chi\tilde{\epsilon})\mathcal{R}=-\chi \mathcal{T},
\end{equation}
where $\mathcal{T}=g^{\alpha\beta}\mathcal{T}_{\alpha\beta}$ is the trace of the energy-momentum tensor. Thus the field equations of RT read
\begin{equation}\label{RT3}
     \mathcal{G}_{\alpha\beta}=\chi \widetilde{\mathcal{T}}_{\alpha\beta}.
\end{equation}
where
\begin{equation}\label{RTmn}
    \widetilde{\mathcal{T}}_{\alpha\beta}=\mathcal{T}_{\alpha\beta}+\frac{\chi \tilde{\epsilon}}{1-4\chi\tilde{\epsilon}}g_{\alpha\beta}\mathcal{T}, \qquad 1-4\chi\tilde{\epsilon}\neq 0.
\end{equation}
It proves convenient to use a dimensionless Rastall's parameter $\epsilon= \tilde{\epsilon} \chi$, c.f. \cite{Oliveira:2015lka}. Then, Eq. \eqref{RT2} becomes
\begin{equation} \label{e2}
\mathcal{R}_{\alpha \beta}-\left(\frac{1}{2}-\epsilon\right){ g}_{\alpha \beta}\mathcal{R}=\chi \mathcal{T}_{\alpha \beta},
\end{equation}
and the tensor $\widetilde{\mathcal{T}}_{\alpha\beta}$ in Eq. \eqref{RT3} reads
\begin{equation} \label{e3}
\widetilde{\mathcal{T}}_{\alpha \beta}= \mathcal{T}_{\alpha \beta}+\frac{\epsilon}{1-4\epsilon} g_{\alpha \beta} \mathcal{T}, \qquad \epsilon \neq \frac{1}{4}.
\end{equation}
For $\epsilon=0$ case, the conservation law is restored and the GR version of gravity is recovered. In this sense, RT generalizes Einstein's one by assuming a local violation of conservation law in curved spacetime due to non-minimal coupling between matter and geometry. Otherwise, flat spacetime, both theories are equivalent. Therefore, one of the important applications, which differentiate both theories, is stellar structure models when presence of the matter sector plays a crucial role in interior solutions.
%%%%%%%%%%%%%%%%%%%%%%%%%%%%%%%%%%% Section 3 %%%%%%%%%%%%%%%%%%%%%%%%%%%%%%%%%%%%%%%%
\section{spherically symmetric interior solution}\label{S3}
%%%%%%%%%%%%%%%%%%%%%%%%%%%%%%%%%%%%%%%%%%%%%%%%%%%%%%%%%%%%%%%%%%%%%%%%%%%%%%%%%%%%%%
Providing that the static spherically symmetrical spacetime is given by the following metric\footnote{We take the geometric units which set $\chi=c=1$.}
\begin{equation}\label{met12}
ds^2=-F(r)dt^2+G(r)\,dr^2+r^2(d\theta^2+\sin^2{\theta}\, d\phi^2)\,,
\end{equation}
where $F(r)$ and $G(r)$ are unknown functions. For the spacetime metric (\ref{met12}), we obtain the non-vanishing components of  Levi-Civita  connection
\begin{eqnarray} \label{cons}
&&{\Gamma_{tt}}^r=\frac{F'}{2G}\,, \qquad \qquad {\Gamma_{tr}}^t=\frac{F'}{2F} \,, \qquad \qquad {\Gamma_{rr}}^r=\frac{G'}{2G} \,, \qquad \qquad  {\Gamma_{r\theta}}^\theta={\Gamma_{r\phi}}^\phi=\frac{1}{r} \,,\nonumber\\
&& {\Gamma_{\theta \theta}}^r=-\frac{r}{G} \,, \qquad \qquad {\Gamma_{\theta \phi}}^\phi=\cot\theta \,, \qquad \qquad {\Gamma_{\phi \phi}}^r=-\frac{r \sin^2\theta}{G}\,, \qquad  \qquad {\Gamma_{\phi \phi}}^\theta=-\sin\theta\, \cos\theta\,,
\end{eqnarray}
where prime (double prime) denotes first (second) derivative with respect to the radial coordinate. We write the Ricci invariant
\begin{eqnarray} \label{Ricci}
  \mathcal{R}(r)={-2F''G F r^2+F'^2 G r^2+r F F' (r G'-4G)+4 F^2 [r G' + G(G-1)] \over 2F^2 G^2 r^2}\,.
\end{eqnarray}
We assume the energy-momentum tensor for a anisotropic fluid with spherical symmetry, i.e.
\begin{equation}\label{Tmn-anisotropy}
    \mathcal{T}{^\alpha}{_\beta}=(p_{t}+\rho)u{^\alpha}u{_\beta}+p_{t} \delta ^\alpha _\beta + (p_{r}-p_{t})\zeta{^\alpha}\zeta{_\beta},
\end{equation}
where $\rho=\rho(r)$ is the fluid energy density, $p_{r}=p_{r}(r)$ its radial pressure (in the direction the time-like four-velocity $u_\alpha$), $p_{t}=p_{t}(r)$ its tangential pressure (perpendicular to $u_\alpha$) and $\zeta{^\alpha}$ is the unit space-like vector in the radial direction. Then, the energy-momentum tensor takes the diagonal form $\mathcal{T}{^\alpha}{_\beta}=diag(-\rho,\,p_{r},\,p_{t},\,p_{t})$.

Applying Rastall's field equations \eqref{RT3} to the spacetime \eqref{met12} where the matter sector is as given by \eqref{Tmn-anisotropy} we obtain, respectively, the components $t\,t$, $r\,r$ and $\theta\,\theta$ ($= \phi\,\phi$) as follows:
\begin{eqnarray}
\rho&=&{r G' +G(G-1) \over G^2 r^2}-{\epsilon \over 1-4\epsilon}(\rho-p_r-2p_t)\,,\nonumber\\[5pt]
p_r&=&{F' r- F(G-1) \over F G r^2}+{\epsilon \over 1-4\epsilon}(\rho-p_r-2p_t)\,,\nonumber\\[5pt]
p_t&=&{F[2G(F''r+F')-G' F' r]-2 G' F^2-F'^2 G r \over 4 F^2 G^2 r}+{\epsilon \over 1-4\epsilon}(\rho-p_r-2p_t)\,.
\label{feq}
\end{eqnarray}
Additionally, we define the anisotropy of the system (\ref{feq}) using the parameter
\begin{equation}\label{anio}
\Delta(r)=p_t-p_r={2F'' G F r^2-F'^2 G r^2 - r F F' (r G'+2 G)-2 F^2 [r G' -2 G (G-1)] \over 4 F^2 G^2 r^2}\,.
\end{equation}
We note that the Rastall parameter quantitatively contributes in the radial and the tangential pressures by equal amount, i.e. \textit{Rastall parameter has no contribution in the anisotropy parameter, and this conclusion is valid for any spherically symmetric interior solution}. For the $\epsilon=0$ case, the differential equations \eqref{feq} coincide with Einstein field equations of an interior spherically symmetrical spacetime \cite{Nashed:2020buf,Nashed:2020kjh}. The system (\ref{feq}) consists of three independent non-linear differential equations in five unknowns $G$, $F$, $\rho$, $p_r$ and $p_t$. Therefore, two conditions need to be imposed to close the above system. We follow the ansatz given in \cite{Das:2019dkn} (also used in \cite{Nashed:2020kjh}). First, we assume the metric potential $G$ to have the form
\begin{equation}\label{metg}
G(r)=\frac{1}{\left(1-\frac{a_2^2 r^2}{R^2}\right)^4}\,,
\end{equation}
with $a_2$ is a dimensionless constant to be determined by boundary condition and $R$ is the radius at the star boundary. We note that the above ansatz is regular everywhere inside the star, i.e. $0\leq r\leq R$, where $|a_2|<1$. Substituting (\ref{metg}) in the anisotropy parameter (\ref{anio}), we get
\begin{equation}\label{anis1}
\Delta(r)={a_2^4 r^2 (6R^4-8R^2 a_2^2 r^2+3 a_2^4 r^4)\over R^8}+{{(R^2-a_2^2 r^2)^3} [r (2F F'' - F'^2) (R^2-a_2^2 r^2)-2F F' (R^2+3a_2^2 r^2)]\over 4 r F^2 R^8}\,.
\end{equation}
 Now, we impose the second condition by assuming that the component $g_{tt}$ has no contribution on the anisotropy parameter, i.e.
\begin{equation}\label{anis2}
\Delta(r)={a_2^4 r^2 (6R^4-8R^2 a_2^2 r^2+3 a_2^4 r^4)\over R^8}\,.
\end{equation}
This choice clearly gives no anisotropy at the center, $r=0$, which is physically a reasonable feature \cite{Das:2019dkn}, see the physical conditions in Sec. \ref{S4}. Using Eqs. (\ref{anis1}) and (\ref{anis2}) and by solving for the metric potential $F$, we obtain:
\begin{equation}\label{metf}
F(r)={[a_0 R^2+2 a_1 a_2^2 (R^2-a_2^2 r^2)]^2 \over 8 a_2^4 (R^2-a_2^2 r^2)^2}\,,
\end{equation}
where the constants of integration $a_0$ and $a_1$ are dimensionless to be fixed by matching conditions. Up to this step the obtained results are the same as given by \cite{Das:2019dkn}. However, the matching conditions with the exterior solution give the set of constants \{$a_0$, $a_1$, $a_2$\} in terms of Rastall parameter $\epsilon$. Therefore, any deviation from the GR in the following results should be related to the assumed matter-geometry coupling in Rastall field equations \eqref{feq}, i.e. Rastall parameter $\epsilon$.

Substituting the metric potentials (\ref{metg}) and (\ref{metf}) into the system (\ref{feq}), we get the energy-density, radial and tangential pressures in the form
\begin{eqnarray}\label{sol}
\rho&=& \frac{12 a_2^2}{R^8[2a_1 a_2^2 (a_2^2 r^2-R^2) -a_0 R^2]}\Bigg\{
{\frac{3a_1 a_2^{10}}{2} (2\epsilon-1)}r^8
- \frac{a_2^6}{2}{\left[\frac{a_1 a_2^2}{3}(74\epsilon-37)+\frac{a_0}{2}(2\epsilon-3)\right]}R^2 r^6
 \nonumber\\
&&+\frac{a_2^4}{3}{\left[a_1 a_2^2(58\epsilon-29)+a_0 (5\epsilon-7)\right]}R^4 r^4-2 a_2^2{\left[\frac{7a_1 a_2^2}{2} (2\epsilon-1)+\frac{a_0}{4}(4\epsilon-5)\right]}R^6 r^2
+{\left[(4a_1 a_2^2+a_0)\epsilon-(2a_1 a_2^2+a_0)\right]}R^8 \Bigg\}\,, \nonumber\\
p_r&=&\frac{12 a_2^2}{R^8[2a_1 a_2^2 (R^2-a_2^2 r^2)+ a_0 R^2]}\Bigg\{
{\frac{a_1 a_2^{10}}{6} (18\epsilon-1)}r^8
- \frac{a_2^6}{2}{\left[\frac{a_1 a_2^2}{3}(74\epsilon-5)+\frac{a_0}{2}(2\epsilon+1)\right]}R^2 r^6
 \nonumber\\
&&+\frac{a_2^4}{3}{\left[a_1 a_2^2(58\epsilon-5)+a_0 (5\epsilon+2)\right]}R^4 r^4-2 a_2^2{\left[\frac{a_1 a_2^2}{6} (42\epsilon-5)+\frac{a_0}{4}(4\epsilon+1)\right]}R^6 r^2
+{\left[(4a_1 a_2^2+a_0)\epsilon-\frac{2}{3}a_1 a_2^2\right]}R^8 \Bigg\}\,, \nonumber\\
p_t&=&\frac{12 a_2^2}{R^8[2a_1 a_2^2 (R^2-a_2^2 r^2)+ a_0 R^2]}\Bigg\{
{\frac{a_1 a_2^{10}}{3} (9\epsilon-2)}r^8
- \frac{a_2^6}{2}{\left[\frac{2a_1 a_2^2}{3}(74\epsilon-8)+a_0\epsilon\right]}R^2 r^6
+\frac{a_2^4}{3}{\left[a_1 a_2^2(58\epsilon-12)+5 a_0 \epsilon\right]}R^4 r^4 \nonumber\\
&&-2 a_2^2{\left[\frac{a_1 a_2^2}{3} (42\epsilon-4)+a_0\epsilon\right]}R^6 r^2
+{\left[(4a_1 a_2^2+a_0)\epsilon-\frac{2}{3}a_1 a_2^2\right]}R^8 \Bigg\}\,,
\end{eqnarray}
Equations (\ref{sol}) coincide with the GR version when Rastall parameter $\epsilon$ vanishes \cite{Nashed:2020buf}. It is to be mentioned that the anisotropic force, $F_{a}=\frac{2\Delta}{r}$, becomes attractive if $p_t-p_r<0$ and repulsive if $p_t-p_r>0$. The mass contained within a radius $r$ of the sphere is defined as
\begin{equation}\label{mas}
M(r)=4\pi \int_0^r \rho(\zeta) \zeta^2 d\zeta\,.
\end{equation}
Using the energy-density as defined in Eqs. (\ref{sol}) and the above equation (\ref{mas}), we get
\begin{eqnarray}\label{mas1}
&&M(r)=\frac{-3\pi}{a_1^4 a_2^{10} R^8\aleph}\Bigg\{\frac{\sqrt{2}a_0 R^9 \epsilon}{2}(a_0+2a_1 a_2^2)\tanh^{-1}\left(\frac{\sqrt{2}a_1 a_2^2 r}{R \aleph}\right)+a_2^2 \aleph r \left[\frac{256 a_1^4 a_2^{10} r^2}{3}  (2\epsilon-1) \left(2R^2-a_2^2 r^2\right) \left(2R^4+ \right.\right.\nonumber\\
&&\left.\left.a_2^4 r^4-2a_2^2 R^2 r^2\right)+a_0 R^2 \epsilon \left(a_0^3 R^6+\frac{2 r^2 a_1 a_2^4}{8}  (4 a_1^2 a_2^4-2 a_1 a_2^2 a_0+a_0^2) R^4+\frac{4  a_1^2 a_2^8 r^4}{5} (a_0 -4 a_1 a_2^2) R^2+\frac{8a_1^3 a_2^{12} r^6 }{7} \right) \right]\Bigg\},\qquad
\end{eqnarray}
where $\aleph=\sqrt{(a_0+2a_1a_2^2)a_1}$. It proves convenient to use the compactness parameter \cite{Singh:2019ykp,Roupas:2020mvs} of a spherically symmetric source with radius $r$,
\begin{eqnarray}\label{gm1}
&&u(r)=\frac{2M(r)}{r},
\end{eqnarray}
to study the stability of compact objects. Similarly we use the gravitational red-shift parameter $Z$ which is related to the metric potential as
\begin{equation}\label{gravitational-red-shift}
    1+Z=\frac{1}{\sqrt{-g_{tt}}}.
\end{equation}
These parameters will be used later in Sec \ref{S6}. In the next section, however, we are going to discuss the physical requirements to derive viable stellar structures in order to test the viability of the obtained model.
%%%%%%%%%%%%%%%%%%%%%%%%%%%%%%%%%%%%%% Section 4 %%%%%%%%%%%%%%%%%%%%%%%%%%%%%%%%%%%%%%%%%%
\section{Physical conditions for a stellar model}\label{S4}
%%%%%%%%%%%%%%%%%%%%%%%%%%%%%%%%%%%%%%%%%%%%%%%%%%%%%%%%%%%%%%%%%%%%%%%%%%%%%%%%%%%%%%%%%%%
For a stellar model to be physically well behaved, it needs to satisfy the following conditions:\\
\noindent(\textbf{i}) For the geometric sector, the metric potentials $F$ and $G$ should be free from coordinate and physical singularities within the interior region of the star $0\leq r\leq R$, where the center (boundary) is at $r=0$ ($r=R$) respectively.

\noindent(\textbf{ii})  The metric potentials of the interior solution and the exterior\footnote{In our case the exterior solution is nothing rather Schwarzschild's one, since vacuum solutions of both GR and RT are equivalent.} should match smoothly at the boundary.

\noindent(\textbf{iii})  For the matter sector, the fluid density, radial and the tangential pressures should be free from coordinate or physical singularities within the interior region of the star. In addition, they should be maximum at the center of the star and monotonically decrease towards the boundary of the star. i.e.
\begin{itemize}
  \item[a.] $\rho(r=0)>0$, $\rho'(r=0)=0$, $\rho''(r=0)<0$ and $\rho'(0< r \leq R)< 0$,
  \item[b.] $p_r(r=0)>0$, $p_r'(r=0)=0$, $p_r''(r=0)<0$ and $p_r'(0< r \leq R)< 0$,
  \item[c.] $p_t(r=0)>0$, $p_t'(r=0)=0$, $p_t''(r=0)<0$ and $p_t'(0< r \leq R)< 0$.\\
\end{itemize}

\noindent(\textbf{iv})  At the center of the star ($r=0$), the anisotropy parameter $\Delta$ should vanish, i.e. $p_r(r=0)=p_t(r=0)$, and increasing toward the boundary, i.e. $\Delta'(0 \leq r\leq R)>0$.

\noindent(\textbf{v})  At the boundary of the star ($r=R$), the radial pressure should vanish, i.e. $p_r(r=R)=0$. However, the tangential pressure at the boundary should not necessarily vanish.

\noindent(\textbf{vi})  Within the star ($0 < r < R$), the density, radial and tangential pressures should be positive, i.e. $\rho(0 < r < R)>0$, $p_r(0 < r < R)>0$ and $p_t(0 < r < R)>0$.

\noindent(\textbf{vii})  The fluid density, radial and tangential pressures should fulfill the following energy conditions:
\begin{itemize}
  \item[a.] Null energy condition (NEC): $\rho c^2+ p_t > 0$, $\rho> 0$,
  \item[b.] Weak energy condition (WEC): $\rho c^2+p_r > 0$, $\rho> 0$,
  \item[c.] Dominant energy conditions (DEC): $\rho c^2 \geq \lvert p_r\lvert$ and $\rho c^2\geq \lvert p_t\lvert$,
  \item[d.] Strong energy condition (SEC): $\rho c^2+p_r > 0$, $\rho c^2+p_t > 0$, $\rho c^2-p_r-2p_t > 0$.\\
\end{itemize}

\noindent(\textbf{viii}) The causality condition should be satisfied, that is the speed of sound should be smaller than unity
everywhere inside the star and monotonically decrease toward the boundary, i.e. for the radial velocity $0\leq v_r/c=\frac{1}{c}\sqrt{\frac{dp_r}{d\rho}}\leq 1$ and $v'_r{^2}<0$, and for the tangential velocity $0\leq v_t/c=\frac{1}{c}\sqrt{\frac{dp_t}{d\rho}}\leq 1$ and $v'_t{^2}<0$.

\noindent(\textbf{ix}) The stability condition should be satisfied, i.e. $-1< (v_t^2-v_r^2)/c^2 < 0$ within the star \cite{HERRERa1992206}.

\noindent(\textbf{x}) The gravitational red-shift should be finite and positive everywhere inside the star and decreases monotonically toward the boundary, i.e. $Z>0$ and $Z'<0$.

\noindent(\textbf{xi}) The adiabatic index stability condition for anisotropic star should be fulfilled, i.e. the adiabatic index $\Gamma> \gamma$ where $\gamma=4/3$ is the adiabatic index corresponds to the isotropic case.\\

We note that the stellar model which fulfills the above mentioned conditions is physically viable and well behaved. In the following sections we are going to examine the model at hand with these conditions investigating possible roles of Rastall parameter.
%%%%%%%%%%%%%%%%%%%%%%%%%% Section 5 %%%%%%%%%%%%%%%%%%%%%%%%%%%%%%%%%
\section{Physical properties of the model}\label{S5}
%%%%%%%%%%%%%%%%%%%%%%%%%%%%%%%%%%%%%%%%%%%%%%%%%%%%%%%%%%%%%%%%%%%%%%
We test the model showing that the metric potentials \eqref{metg} and \eqref{metf}, and the solution (\ref{sol}) are free from physical or geometric singularities. Then we derive some physical quantities which will be used later in Sec. \ref{S6} to examine the viability of the present model. In addition, we use the matching conditions at the boundary surface of the star to set the constraining equations on the model parameters \{$\epsilon, a_0, a_1, a_2$\}.
\subsection{Non singular model}
From Eqs. \eqref{metg} and \eqref{metf} one finds that the metric potentials at the center read
\begin{equation}\label{sing}
F(r=0)={(a_0+2a_1 a_2^2)^2 \over 16 a_2^4}\,\qquad  \qquad \textrm{and} \qquad \qquad G(r=0)=1.
\end{equation}
This ensures that the gravitational potentials are finite at the center of the star. Moreover, the derivatives of these potentials are finite at the center, i.e. $F'(r=0)=G'(r=0)=0$. Equation \eqref{sing} ensures that the metric is regular at the center. This satisfies condition (i) in Sec. \ref{S4}.

From Eqs. \eqref{sol} one finds that the density, radial and tangential pressures at the center are
\begin{align}\label{reg}
\rho(r=0)={-12 a_2^2 [a_0(\epsilon-1)+2a_1 a_2^2(2\epsilon-1)] \over R^2(a_0+2a_1 a_2^2)}\,,
\qquad p_{r}(r=0)=p_{t}(r=0)={12 a_2^2 [a_0 \epsilon+\frac{2}{3}a_1 a_2^2 (6\epsilon-1)] \over R^2(a_0+2a_1 a_2^2)}.
\end{align}
These ensure that the anisotropy parameter has a vanishing value at the center as stated in condition (iv) on Sec. \ref{S4}. The density and the pressures should be positive at the center which sets two constraints on the model parameters/constants. Additionally, the Zeldovich condition \cite{1971reas.book.....Z} states that the radial pressure must be less than or equal to the density at the center, i.e. $\frac{p_r(0)}{\rho(0)}\leq 1$, i.e.
\begin{align}\label{reg1}
{-3(a_0+4a_1 a_2^2)\epsilon+2a_1 a_2^2 \over 3(a_0+4a_1 a_2^2)\epsilon -3(a_0+2a_1 a2^2)}\leq 1.
\end{align}

Using Eqs. \eqref{sol} we give the derivative of energy density, radial and tangential pressures, respectively, as follows
\begin{eqnarray}\label{dsol1}
\rho'&=&\frac {2r{a_2}^{4}  }{{R}^{8} \left( a_0{R}^{2}+2\,a_1{a_2}^{2}{R}^{2}-2a_1{a_2}^{4}{r}^{2} \right)^{2}}\biggl\{ 216{r}^{8}{a_1}^{2}{a_2}^{12}\epsilon-108\,{r}^{8}{a_1}^{2}{a_2 }^{12}+440{r}^{6}{ R}^{2}{a_1}^{2}{a_2}^{10}-880{r}^{6}{R}^{2}{a_1}^{2}{ a_2}^{10}\epsilon\nonumber\\
&&-676\,{r}^{4}{R}^{4}{a_1}^{2}{a_2}^{8} +1352\,{r}^{4}{R}^{4}{a_1}^{2}{a_2}^{8}\epsilon-168\,{r}^{6} {R}^{2}a_1\,{a_2}^{8}a_0\,\epsilon+108\,{r}^{6}{R}^{2}a_1\,{a_2}^{8}a_0+464\,{r}^{2}{R}^{6}{a_1}^{2}{a_2}^{6}-928\,{r}^{2}{R}^{6}{a_1}^{2}{a_2}^{6}
\epsilon\nonumber\\
&&+ 520\,{r}^{4}{R}^{4}a_1\,{a_2}^{6}a_0\,\epsilon-332\,{r} ^{4}{R}^{4}a_1\,{a_2}^{6}a_0-120\,{R}^{8}{a_2}^{4} {a_1}^{2}+240\,{R}^{8}{a_2}^{4}{a_1}^{2}\epsilon-544\,x
{ R}^{6}{a_2}^{4}{r}^{2}a_0\,a_1\,\epsilon+344\,{R}^{6}{a_2}^{4}{r}^{2}a_0\,a_1\nonumber\\
&&-27\,{R}^{4}{a_2}^{4}{r}^{4} {a_0}^{2}+18\,{R}^{4}{a_2}^{4}{r}^{4}{a_0}^{2}\epsilon- 120\,a_0\,{R}^{8}{a_2}^{2}a_1+192\,a_0\,{R}^{8}{a_2}^{2}a_1\,\epsilon+56\,{a_0}^{2}{R}^{6}{a_2}^{2} {r}^{2}-40\,{a_0}^{2}{R}^{6}{a_2}^{2}{r}^{2}\epsilon-30\,{a_0}^{2}{R}^{8}\nonumber\\
&&+24\,{a_0}^{2}{R}^{8}\epsilon \biggr\}\,,
\end{eqnarray}
\begin{eqnarray}\label{dsol2}
p'_r&=&\frac {-2r{a_2}^{4}}{{R}^{8} \left( a_0\,{R}^{2}+2\,a_1\,{a_2}^{2}{R}^{2}-2\,a_1\,{a_2}^{4}{r}^{2} \right) ^{2}}\biggl\{ 216\,{r}^{8}{a_1}^{2}{a_2}^{12}\epsilon-12\,{r}^{8}{a_1}^{2}a_2^{12}+56\,{r}^{6}{R }^{2}{a_1}^{2}{a_2}^{10}-880\,{r}^{6}{R}^{2}{a_1}^{2}{a_2}^{10}\epsilon\nonumber\\
&&+1352\,{r}^{4}{R}^{4}{a_1}^{2}{a_2}^{8} \epsilon-100\,{r}^{4}{R}^{4}{a_1}^{2}{a_2}^{8}-168\,{r}^{6}{ R}^{2}a_1\,{a_2}^{8}a_0\,\epsilon-4\,{r}^{6}{R}^{2}a_1\,{a_2}^{8}a_0+80\,{r}^{2}{R}^{6}{a_1}^{2}{a_2}^{6}-928\,{r}^{2}{R}^{6}{a_1}^{2}{a_2}^{6}\epsilon\nonumber\\
&&+520\, {r}^{4}{R}^{4}a_1\,{a_2}^{6}a_0\,\epsilon+4\,{r}^{4}{R} ^{4}a_1\,{a_2}^{6}a_0+240\,{R}^{8}{a_2}^{4}{a_1}^{2}\epsilon-24\,{R}^{8}{a_2}^{4}{a_1}^{2}-544\,{R}^{6}{ a_2}^{4}{r}^{2}a_0\,a_1\,\epsilon+8\,{R}^{6}{a_2}^ {4}{r}^{2}a_0\,a_1\nonumber\\
&&+9\,{R}^{4}{a_2}^{4}{r}^{4}{a_0} ^{2}+18\,{R}^{4}{a_2}^{4}{r}^{4}{a_0}^{2}\epsilon-8\,a_0\,{R}^{8}{a_2}^{2}a_1+192\,a_0\,{R}^{8}{a_2}^{2}a_1\,\epsilon-16\,{a_0}^{2}{R}^{6}{a_2}^{2}{r}^{2}-40\,{ a_0}^{2}{R}^{6}{a_2}^{2}{r}^{2}\epsilon+6\,{a_0}^{2}{R} ^{8}\nonumber\\
&&+24\,{a_0}^{2}{R}^{8}\epsilon \biggr\}\,,
\end{eqnarray}
\begin{eqnarray}\label{dsol3}
p'_t&=&\frac {-4r{a_2}^{4}}{{R}^{8} \left( a_0{R}^{2}+2a_1{a_2}^{2}{R}^{2}-2a_1{a_2}^{4}{r}^{2} \right) ^{2}} \biggl\{ 108{r }^{8}{a_1}^{2}{a_2}^{12}\epsilon-24{r}^{8}{a_1}^{2}{a_2}^{12}+96{r}^{6}{R}^{2}{a_1 }^{2}{a_2}^{10}-440{r}^{6}{R}^{2}{a_1}^{2}{a_2}^{10} \epsilon-144 {r}^{4}{R}^{4}{a_1}^{2}{a_2}^{8}\nonumber\\
&&+676{r}^{4}{R}^{4}{a_1}^{2}{a_2}^{8}\epsilon-84{r}^{6}{R}^{2}a_1 {a_2}^{8}a_0\epsilon+16{r}^{6}{R}^{2}a_1{a_2}^{8}a_0-464{r}^{2}{R}^{6}{a_1}^{2}{a_2}^{6} \epsilon+96{r}^{2}{R}^{6}{a_1}^{2}{a_2}^{6}+260{r}^{4}{R }^{4}a_1{a_2}^{6}a_0\epsilon+12{a_0}^{2}{R}^{8}\epsilon\nonumber\\
&&-48\,{r}^{4}{R}^{4}a_1\,{a_2}^{6}a_0-24\,{R}^{8}{a_2}^{4}{a_1}^{2 }+120\,{R}^{8}{a_2}^{4}{a_1}^{2}\epsilon-272\,{R}^{6}{a_2}^{4}{r}^{2}a_0\,a_1\,\epsilon+48\,{R}^{6}{a_2}^{4}{ r}^{2}a_0\,a_1+9\,{R}^{4}{a_2}^{4}{r}^{4}{a_0}^{2} \epsilon-16\,a_0\,{R}^{8}{a_2}^{2}a_1\nonumber\\
&&+96\,a_0\,{R} ^{8}{a_2}^{2}a_1\,\epsilon-20\,{a_0}^{2}{R}^{6}{a_2}^{2}{r}^{2}\epsilon \biggr\}\,.
\end{eqnarray}
We use Eqs. \eqref{dsol1}--\eqref{dsol3} to show that the gradients of the energy-density, radial and tangential pressures are negative later in Sec. \ref{S6}.\\

The radial and tangential sound velocities are given
\begin{eqnarray}\label{rad_v}
v_r^2&=&\frac{dp_r}{d\rho}=- \biggl\{216\,{r}^{8 }{a_1}^{2}{a_2}^{12}\epsilon -12\,{r}^{8}{a_1}^{2}{a_2}^{12}+56\,{r}^{6}{R}^{2}{a_1}^{2 }{a_2}^{10}-880\,{r}^{6}{R}^{2}{a_1}^{2}{a_2}^{10} \epsilon+1352\,{r}^{4}{R}^{4}{a_1}^{2}{a_2}^{8}\epsilon-100 \,{r}^{4}{R}^{4}{a_1}^{2}{a_2}^{8}\nonumber\\
&&-168\,{r}^{6}{R}^{2}a_1\,{a_2}^{8}a_0\,\epsilon-4\,{r}^{6}{R}^{2}a_1\,{a_2}^{8}a_0+80\,{r}^{2}{R}^{6}{a_1}^{2}{a_2}^{6}- 928\,{r}^{2}{R}^{6}{a_1}^{2}{a_2}^{6}\epsilon+520\,{r}^{4}{R }^{4}a_1\,{a_2}^{6}a_0\,\epsilon+4\,{r}^{4}{R}^{4}a_1\,{a_2}^{6}a_0\nonumber\\
&&+240\,{R}^{8}{a_2}^{4}{a_1}^{2} \epsilon-24\,{R}^{8}{a_2}^{4}{a_1}^{2}-544\,{R}^{6}{a_2 }^{4}{r}^{2}a_0\,a_1\,\epsilon+8\,{R}^{6}{a_2}^{4}{r}^{ 2}a_0\,a_1+9\,{R}^{4}{a_2}^{4}{r}^{4}{a_0}^{2}+18 \,{R}^{4}{a_2}^{4}{r}^{4}{a_0}^{2}\epsilon\nonumber\\
&&-8\,a_0\,{R}^ {8}{a_2}^{2}a_1+192\,a_0\,{R}^{8}{a_2}^{2}a_1 \,\epsilon-16\,{a_0}^{2}{R}^{6}{a_2}^{2}{r}^{2}-40\,{a_0}^{2}{R}^{6}{a_2}^{2}{r}^{2}\epsilon+6\,{a_0}^{2}{R}^{8}+24 \,{a_0}^{2}{R}^{8}\epsilon \biggr\} \biggl\{216\,{r}^{8}{a_1}^{2}{a_2}^{12} \epsilon\nonumber\\
&&-108\,{r}^{8}{a_1}^{2}{a_2}^{12}+440\,{r}^{6}{R}^{2}{a_1}^{2}{a_2}^{10}-880\,{r}^{6} {R}^{2}{a_1}^{2}{a_2}^{10}\epsilon-676\,{r}^{4}{R}^{4}{a_1}^{2}{a_2}^{8}+1352\,{r}^{4}{R}^{4}{a_1}^{2}{a_2}^{8 }\epsilon-168\,{r}^{6}{R}^{2}a_1\,{a_2}^{8}a_0\, \epsilon\nonumber\\
&&+108\,{r}^{6}{R}^{2}a_1\,{a_2}^{8}a_0+464\,{r}^ {2}{R}^{6}{a_1}^{2}{a_2}^{6}-928\,{r}^{2}{R}^{6}{a_1}^{ 2}{a_2}^{6}\epsilon+520\,{r}^{4}{R}^{4}a_1\,{a_2}^{6}a_0\,\epsilon-332\,{r}^{4}{R}^{4}a_1\,{a_2}^{6}a_0- 120\,{R}^{8}{a_2}^{4}{a_1}^{2}\nonumber\\
&&+240\,{R}^{8}{a_2}^{4}{a_1}^{2}\epsilon-544\,{R}^{6}{a_2}^{4}{r}^{2}a_0\,a_1\,\epsilon+344\,{R}^{6}{a_2}^{4}{r}^{2}a_0\,a_1-27\, {R}^{4}{a_2}^{4}{r}^{4}{a_0}^{2}+18\,{R}^{4}{a_2}^{4}{r }^{4}{a_0}^{2}\epsilon-120\,a_0\,{R}^{8}{a_2}^{2}a_1\nonumber\\
&&+192\,a_0\,{R}^{8}{a_2}^{2}a_1\,\epsilon+56\,{a_0}^{2}{R}^{6}{a_2}^{2}{r}^{2}-40\,
{a_0}^{2}{R}^{6}{a_2}^{2}{r}^{2}\epsilon-30\,{a_0}^{2}{R}^{8}+24\,{a_0}^{2}{R}^ {8}\epsilon\biggr\}^{-1}\,,
\end{eqnarray}
\begin{eqnarray}\label{tan_v}
v_t^2&=&\frac{dp_t}{d\rho}=-2 \,\biggl\{108\,{r} ^{8}{a_1}^{2}{a_2}^{12}\epsilon-24\,{r}^{8}{a_1}^{2}{a_2}^{12}+96\,{r}^{6}R^{2}{a_1} ^{2}{a_2}^{10}-440\,{r}^{6}R^{2}{a_1}^{2}{a_2}^{10} \epsilon+676\,{r}^{4}R^{4}{a_1}^{2}{a_2}^{8}\epsilon-144\, {r}^{4}R^{4}{a_1}^{2}{a_2}^{8}\nonumber\\
&&-84\,{r}^{6}R^{2}a_1 \,{a_2}^{8}a_0\,\epsilon+16\,{r}^{6}R^{2}a_1\,{a_2}^{8}a_0-464\,{r}^{2}R^{6}{a_1}^{2}{a_2}^{6} \epsilon+96\,{r}^{2}R^{6}{a_1}^{2}{a_2}^{6}+260\,{r}^{4}{R }^{4}a_1\,{a_2}^{6}a_0\,\epsilon-48\,{r}^{4}R^{4}a_1\,{a_2}^{6}a_0\nonumber\\
&&-24\,R^{8}{a_2}^{4}{a_1}^{2 }+120\,R^{8}{a_2}^{4}{a_1}^{2}\epsilon-272\,R^{6}{a_2}^{4}{r}^{2}a_0\,a_1\,\epsilon+48\,R^{6}{a_2}^{4}{ r}^{2}a_0\,a_1+9\,R^{4}{a_2}^{4}{r}^{4}{a_0}^{2} \epsilon-16\,a_0\,R^{8}{a_2}^{2}a_1+96\,a_0\,R ^{8}{a_2}^{2}a_1\,\epsilon\nonumber\\
&&-20\,{a_0}^{2}R^{6}{a_2}^{2}{r}^{2}\epsilon+12\,{a_0}^{2}R^{8}\epsilon \biggr\}\biggl\{-108\,{r}^{8}{a_1}^{2}{a_2}^{12}+216\,{r}^{8}{a_1}^{ 2}{a_2}^{12}\epsilon+440\,{r}^{6}R^{2}{a_1}^{2}{a_2}^ {10}-880\,{r}^{6}R^{2}{a_1}^{2}{a_2}^{10}\epsilon-676\,{r} ^{4}R^{4}{a_1}^{2}{a_2}^{8}\nonumber\\
&&+1352\,{r}^{4}R^{4}{a_1} ^{2}{a_2}^{8}\epsilon-168\,{r}^{6}R^{2}a_1\,{a_2}^{8} a_0\,\epsilon+108\,{r}^{6}R^{2}a_1\,{a_2}^{8}a_0+464\,{r}^{2}R^{6}{a_1}^{2}{a_2}^{6}-928\,{r}^{2}R^{6}{a_1}^{2}{a_2}^{6}\epsilon
+520\,{r}^{4}R^{4}a_1\,{a_2}^{6}a_0\,\epsilon\nonumber\\
&&-332\,{r}^{4}R^{4}a_1\,{a_2}^{6} a_0-120\,R^{8}{a_2}^{4}{a_1}^{2}+240\,R^{8}{a_2}^{4}{a_1}^{2}\epsilon-544\,R^{6}{a_2}^{4}{r}^{2}a_0 \,a_1\,\epsilon+344\,R^{6}{a_2}^{4}{r}^{2}a_0\,a_1-27\,R^{4}{a_2}^{4}{r}^{4}{a_0}^{2}\nonumber\\
&&+18\,R^{4}{a_2}^{4}{r}^{4}{a_0}^{2}\epsilon-120\,a_0\,R^{8}{a_2}^{ 2}a_1+192\,a_0\,R^{8}{a_2}^{2}a_1\,\epsilon+56\, {a_0}^{2}R^{6}{a_2}^{2}{r}^{2}-40\,{a_0}^{2}R^{6}{a_2}^{2}{r}^{2}\epsilon-30\,{a_0}^{2}R^{8}+24\,{a_0}^{ 2}R^{8}\epsilon\biggr\}^{-1}\,.
\end{eqnarray}
We use Eqs. \eqref{rad_v} and \eqref{tan_v} to show that the sound speeds satisfy the causality and the stability conditions later in Sec. \ref{S6}.
%%%%%%%%%%%%%%%%%%%%%%%%%%%%%%%%%%%%%%%%%%%
\subsection{Matching conditions}
%%%%%%%%%%%%%%%%%%%%%%%%%%%%%%%%%%%%%%%%%%%
We note that the exterior spacetime of a static spherically symmetric star is the same for both GR and RT, since the exterior region is vacuum. Thus no reason to expect any solution rather the exterior Schwarzschild one for Rastall's theory, that is
\begin{eqnarray}\label{Eq1}
ds^2= -\Big(1-\frac{2M}{r}\Big) dt^2+\Big(1-\frac{2M}{r}\Big)^{-1}dr^2+r^2(d\theta^2+d\phi^2),
 \end{eqnarray}
where $M$ is the total mass $r>2M$. We are going to match the interior spacetime metrics (\ref{metg}) and (\ref{metf}) and
the exterior Schwarzschild spacetime metric (\ref{Eq1}) at the boundary of the star $r =R$. Therefore, the continuity of the metric
functions, as stated by condition (ii) in Sec. \ref{S4}, across the boundary gives the conditions
\begin{eqnarray}\label{Eq2}
F(r=R)={[a_0-2a_1 a_2^2 (a^2-1)]^2 \over 12 a_2^4 (a_2^2-1)^2}=\left(1-\frac{2M}{R}\right), \qquad \qquad G(r=R)=(a_2^2-1)^4=\left(1-\frac{2M}{R}\right).
\end{eqnarray}
In addition, the radial pressure \eqref{sol} approaches zero at the star boundary, $p_{r|r=R}= 0$, which reads
\begin{eqnarray}\label{Eq3}
&&2a_1a_0^{10}-10a_1a_0^8+(20a_1+3a_0)a_0^6-4(5a_1+2a_0)a_0^4+2(4a_1+3a_0)a_0^2 \nonumber\\
&-&\left[36a_1a_2^{10}-148a_1a_2^8+2(116 a_1-3a_0)a_2^6-4(42a_1-5a_0)a_2^4+24(2a_1-a_0)a_2^2-12a_0\right]\epsilon=0\,
\end{eqnarray}
The above constraint ensures that condition (v) in Sec. \ref{S4} is fulfilled. From the above conditions, namely \eqref{Eq2} and \eqref{Eq3}, we get the constraints on the set os constants $\{a_0,~ a_1,~ a_2\}$ in terms of the start mass $M$, radius $R$ in addition to the Rastall parameter $\epsilon$. Using observational pulsars data, knowing the observed values of $M$ and $R$, we obtain the corresponding numerical values for a particular choice of $\epsilon$.
%%%%%%%%%%%%%%%%%%%%%%%%%%%%%%%%%% Section 6 %%%%%%%%%%%%%%%%%%%%%%%%%%%%%%%%%%%%%%%%%%%%
\section{Astrophysical observational constraints}\label{S6}
%%%%%%%%%%%%%%%%%%%%%%%%%%%%%%%%%%%%%%%%%%%%%%%%%%%%%%%%%%%%%%%%%%%%%%%%%%%%%%%%%%%%%%%%%
We convert back to physical standard units in order to correctly determine the numerical values of the model parameters. Using masses and radii of observed pulsars along with the physical conditions mentioned in the previous section, one can determine the constant parameters of the obtained model (\ref{sol}) and test its viability. We use the observational constraints of the particular pulsar \textit{Her X-1}, whose mass $M = 0.85\pm 0.15 M_\odot$ and radius $R = 8.1\pm0.41$~km\\ \cite{Gangopadhyay:2013gha}, where $M_\odot$ ($=1.989\times 10^{30}$ kg) denotes the solar mass. Then, the boundary conditions \eqref{Eq2} and \eqref{Eq3} are adopted  to determine the dimensionless constants in terms of the Rastall parameter $\epsilon$
\[ a_0={2.564\, \epsilon- 0.4694 \over 4.542\, \epsilon-1.514},\, a_1=-6.192\, a_0 +1.661\, \textrm{and}\, a_2=0.298.\]
Noting that we select $a_2<1$ which is required by the regularity condition of ansatz \eqref{metg}. Substituting the above expressions into Zeldovich condition \eqref{reg1}, keeping in mind that the RT predictions are not expected to be far from GR ones (i.e. $\epsilon$ should be small), we obtain the following constraints on Rastall parameter $-1.880 \lesssim \epsilon \lesssim 0.259$.\\
\begin{figure}
\centering
\subfigure[~The energy-density]{\label{fig:density}\includegraphics[scale=0.28]{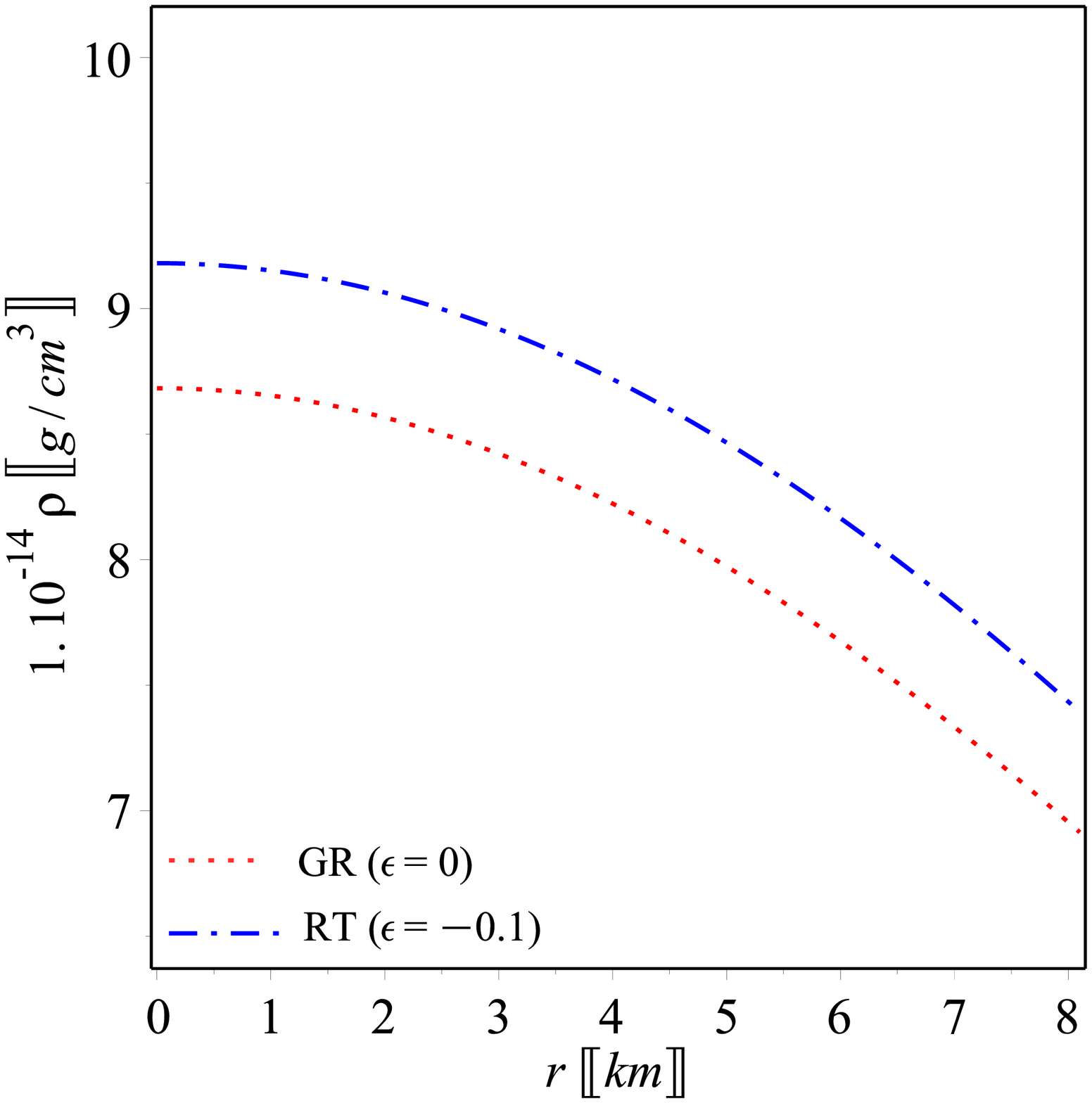}}
\subfigure[~The radial pressure]{\label{fig:radpressure}\includegraphics[scale=.28]{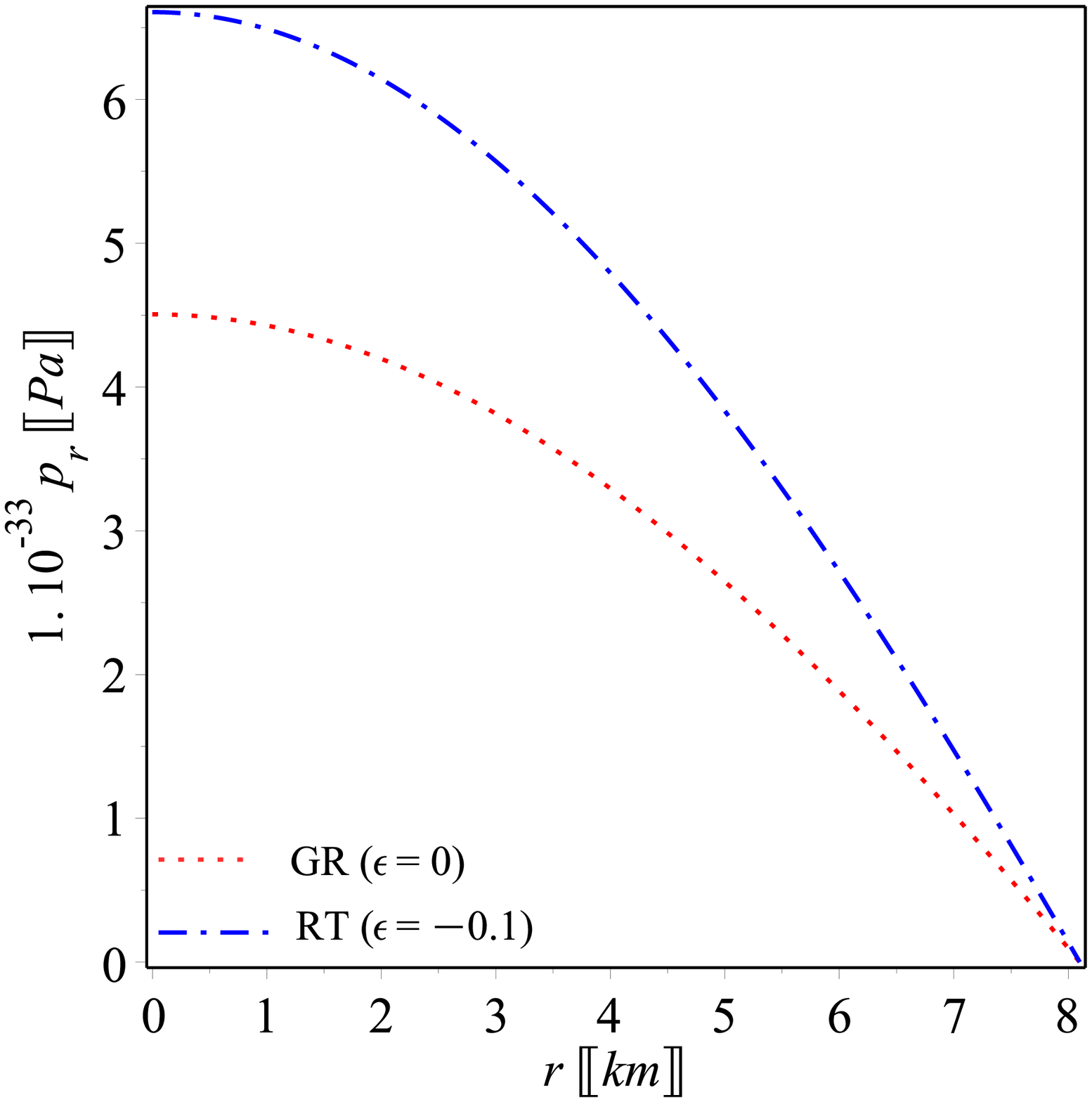}}
\subfigure[~The tangential pressure]{\label{fig:tangpressure}\includegraphics[scale=.28]{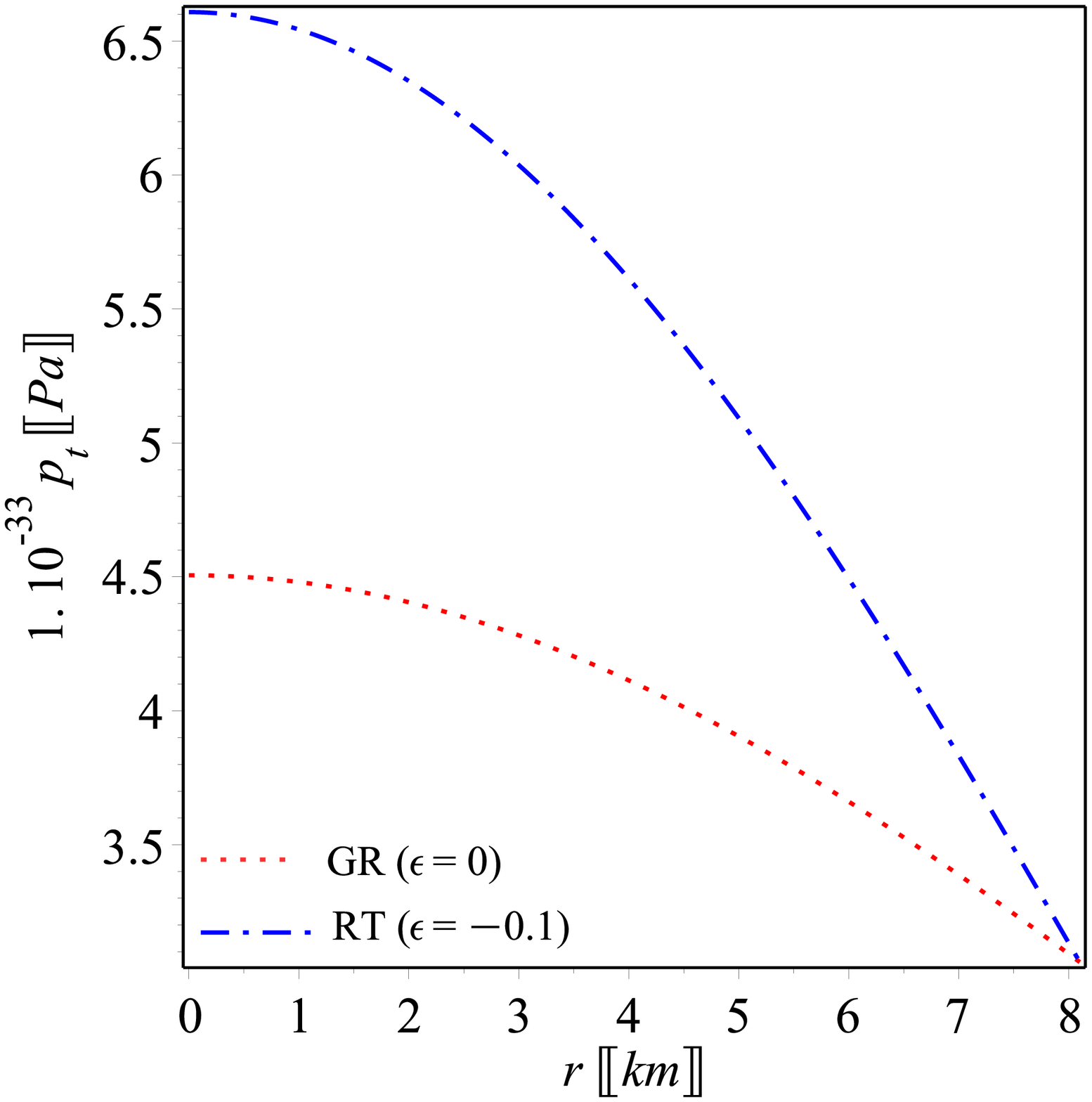}}
\caption[figtopcap]{\small{Plots of the density, radial and tangential pressures given by (\ref{sol}) versus the radial coordinate $r$ in km of the pulsar \textit{Her X-1} ($M = 0.85\pm 0.15 M_\odot$, $R = 8.1\pm0.41$ km). We set $\epsilon=-0.1$, $a_0 \approx 0.369$, $a_1\approx -0.622$ and $a_2\approx 0.298$.}}
\label{Fig:1}
\end{figure}
\begin{figure}
\centering
\subfigure[~Anisotropy and anisotropic force]{\label{fig:An}\includegraphics[scale=0.28]{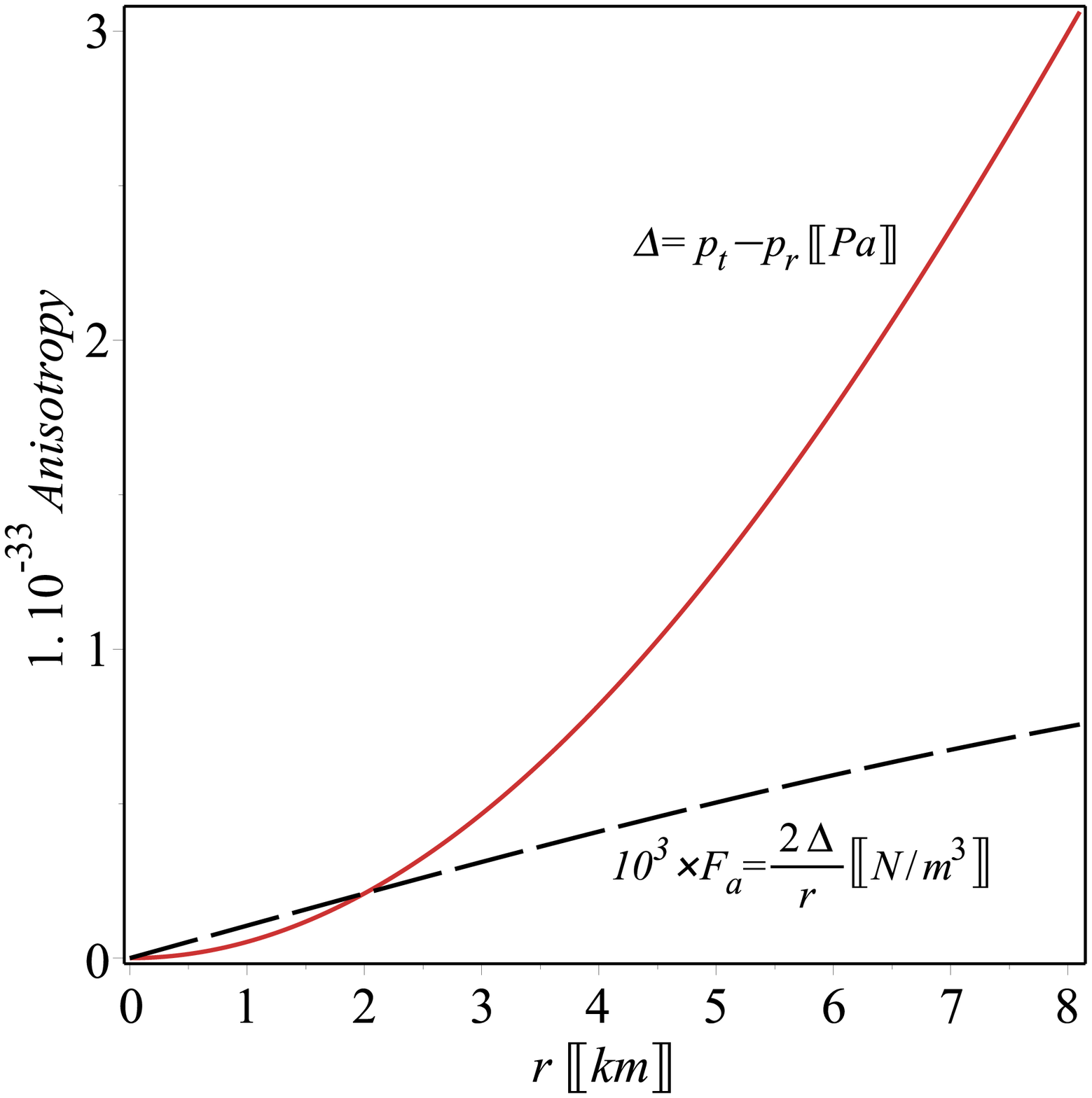}}
\subfigure[~Gradients ($\epsilon=0$)]{\label{fig:grdgr}\includegraphics[scale=.28]{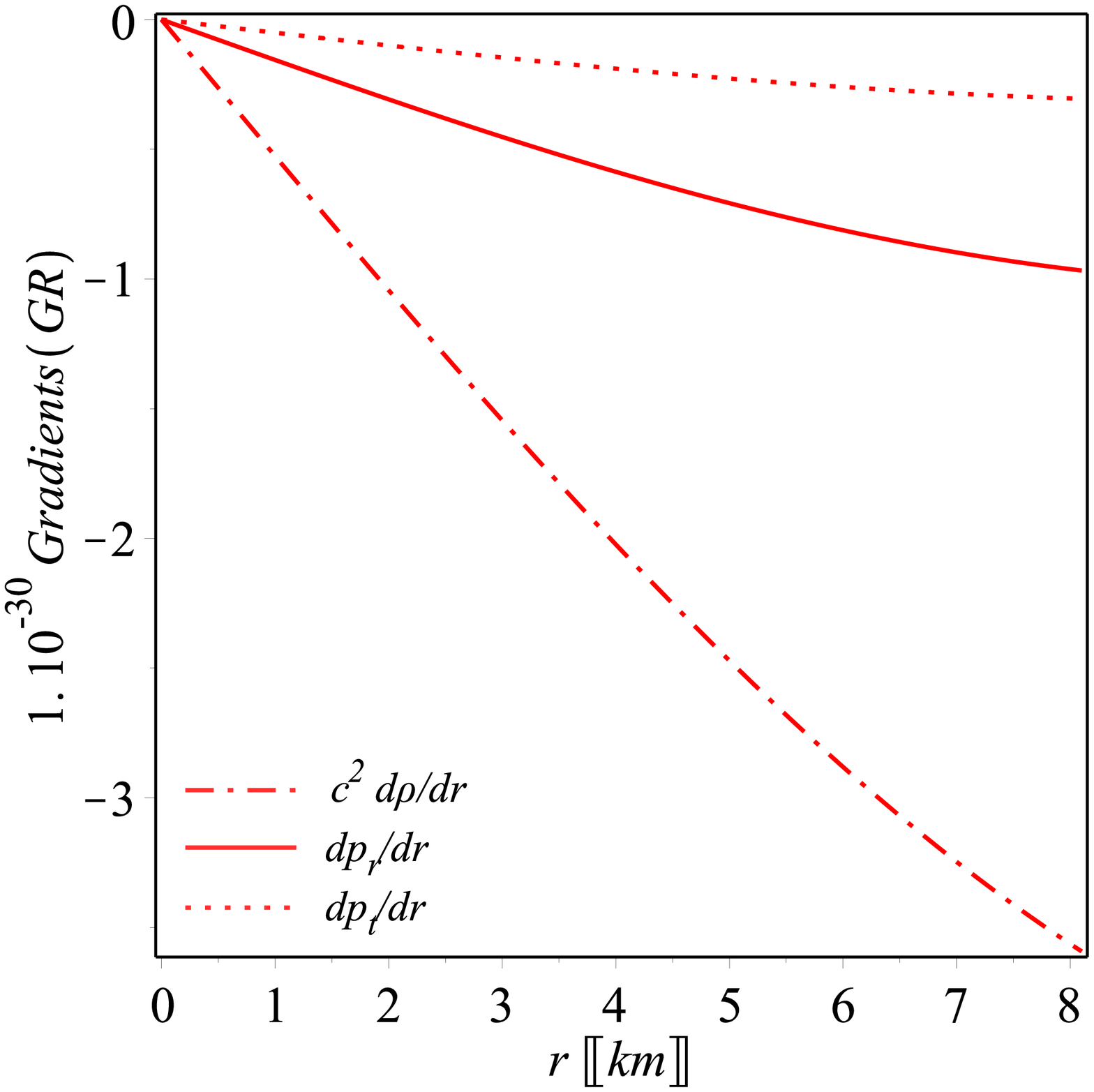}}
\subfigure[~Gradients ($\epsilon=-0.14$)]{\label{fig:grdrast}\includegraphics[scale=.28]{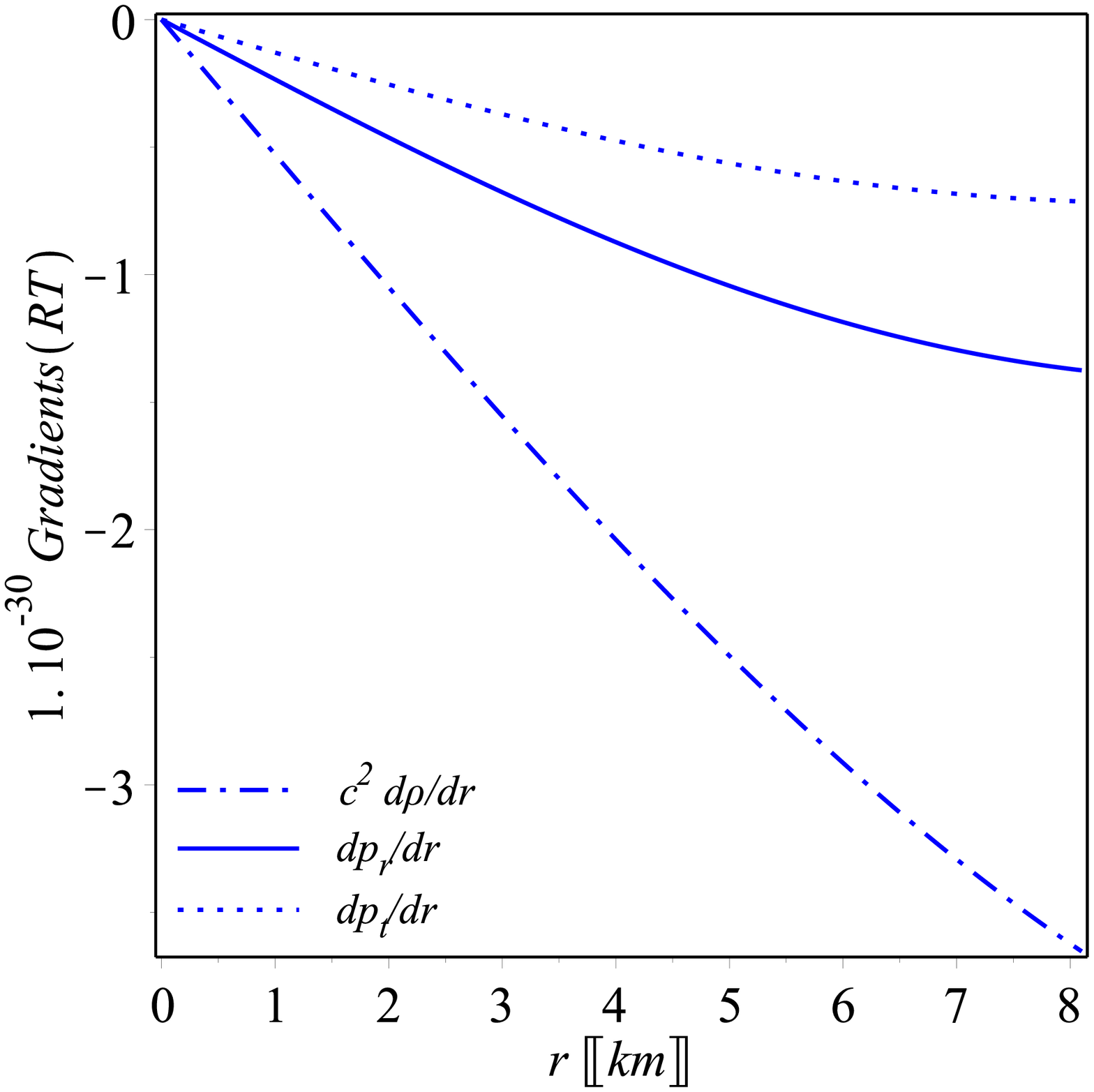}}
\caption[figtopcap]{\small{Plot of the anisotropy parameter \eqref{anis2}, anisotropic force $F_a=2\Delta/r$. We note that the Rastall parameter has no contribution in the anisotropy, therefore GR and RT predicts same anisotropy in the case of spherical symmetry as discussed after \eqref{anio}. For $\epsilon=0$ and $\epsilon\neq 0$, the gradients of the density, tangential and radial pressures given by Eqs. \eqref{dsol1}--\eqref{dsol3} versus the radial coordinate $r$ in km of the pulsar \textit{Her X-1}.}}
\label{Fig:2}
\end{figure}

Next, we use the above expressions to plot some physical quantities for $\epsilon=0$ (GR case) and $\epsilon\neq 0$ (RT case) which enable us to test the model viability and investigate the role of Rastall parameter. We note that one should be careful for the choice of Rastall parameter in order to satisfy the physical conditions in Sec. \ref{S4} with a particular compact object. In agreement with Oliveira et al. \cite{Oliveira:2015lka} we confirm that the present model is physically stable whereas the Rastall parameter $\epsilon$ is negative\footnote{We note that Rastall parameter in the present paper is related to Oliveira et al. notations as $\epsilon=-\eta$. In the latter negative $\eta$ values have been found to be problematic since they lead to violation of the energy conditions and non-stable stellar configurations as well.}. For the particular pulsar \textit{Her X-1}, setting $\epsilon=-0.1$, we determine the set of constants $a_0 \approx 0.369$, $a_1\approx -0.622$ and $a_2\approx 0.298$, which clearly satisfies Zeldovich condition \eqref{reg1}.\\
\begin{figure}
\centering
\subfigure[~Radial speed of sound]{\label{fig:spr}\includegraphics[scale=0.28]{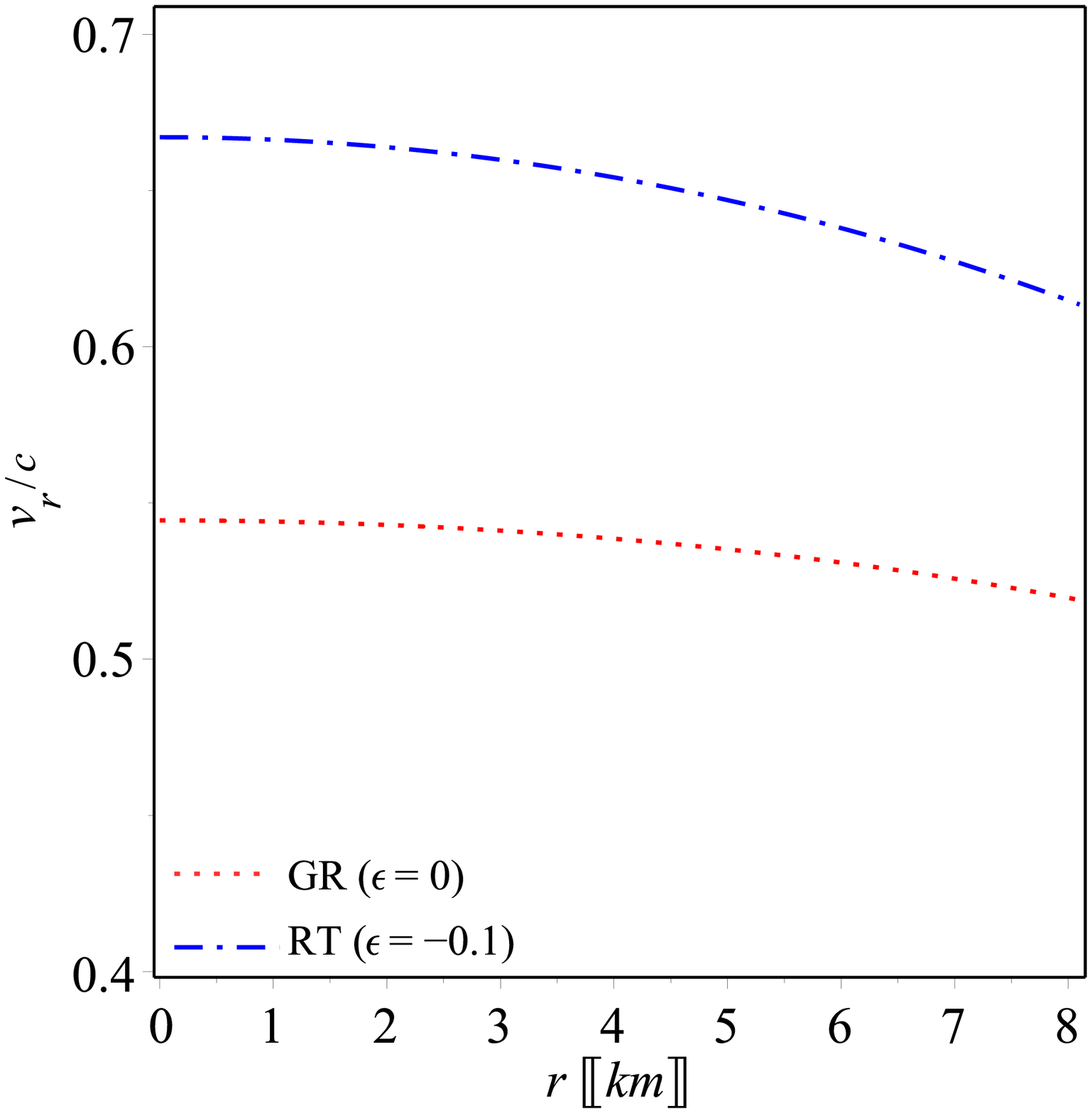}}
\subfigure[~Tangential speed of sound]{\label{fig:pressure}\includegraphics[scale=.28]{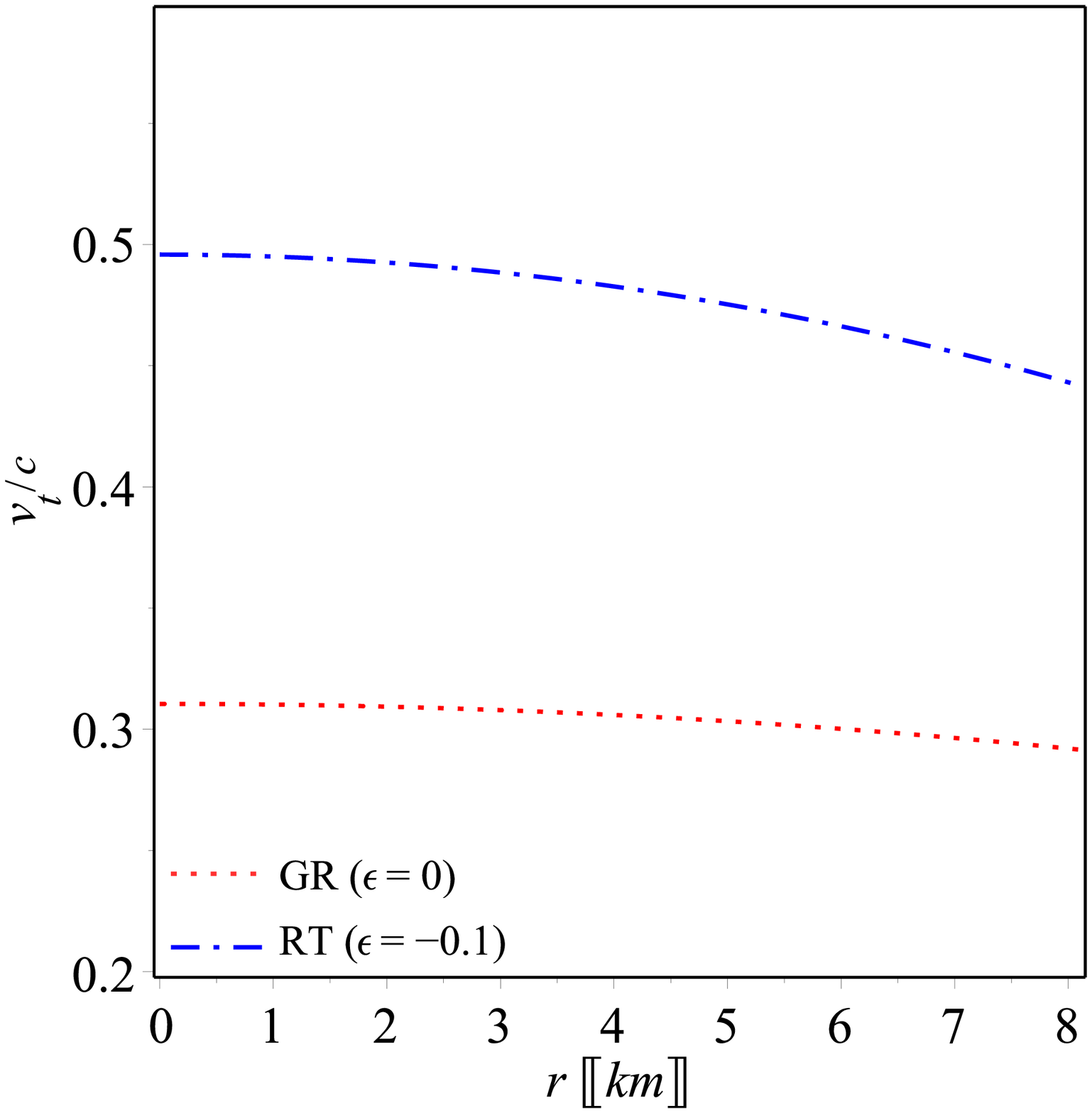}}
\subfigure[~difference between radial and tangential  speed of sounds]{\label{fig:pressure}\includegraphics[scale=.28]{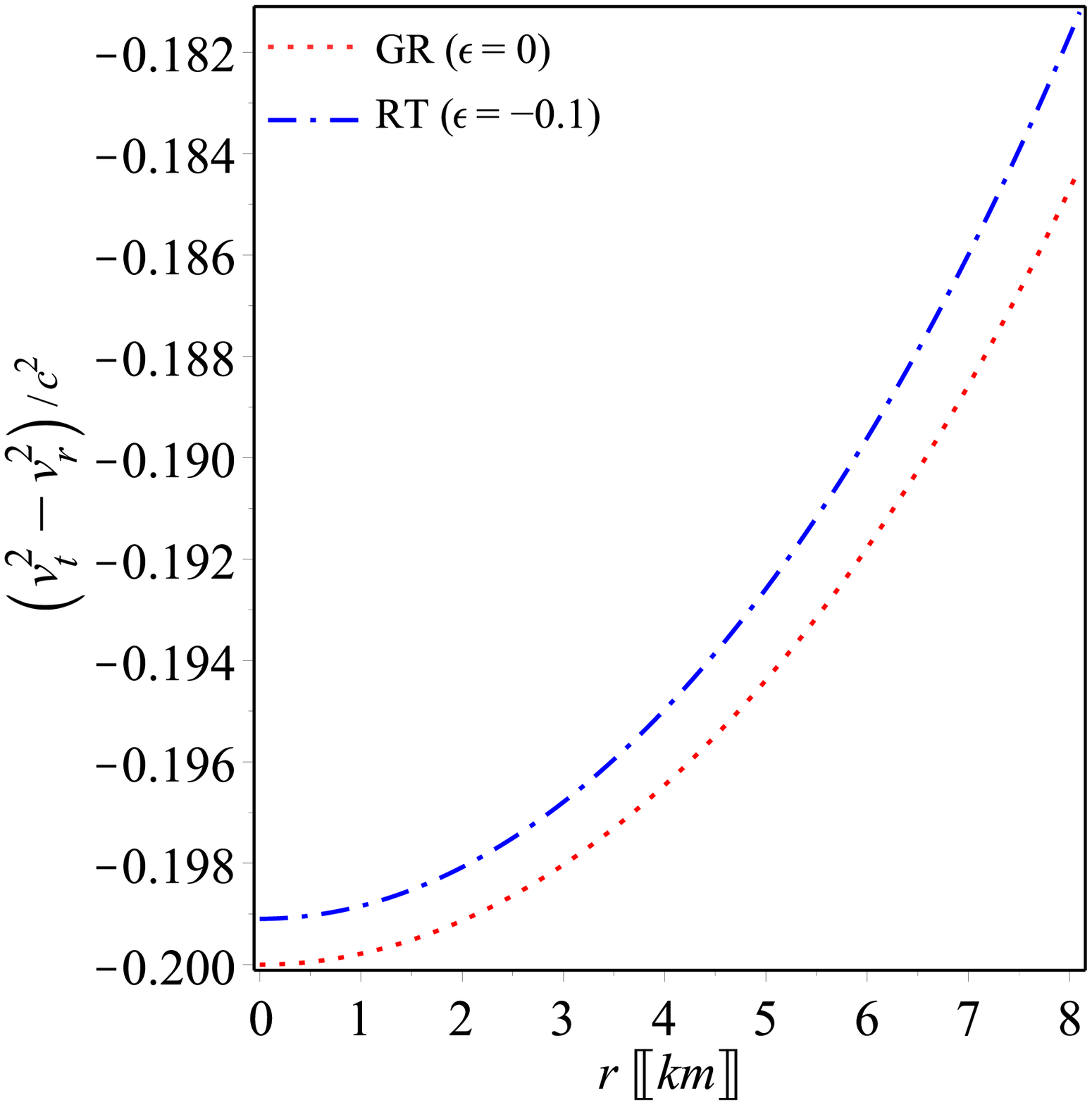}}
\caption[figtopcap]{\small{The radial and tangential sound speeds \eqref{dsol2} versus the radial coordinate $r$ in km for the pulsar \textit{Her X-1}. The plots confirm that the model fulfill the causality and the stability conditions (viii) and (ix) in Sec. \ref{S4}.}}
\label{Fig:3}
\end{figure}
\begin{figure}
\centering
\subfigure[~Weak energy conditions]{\label{fig:WEC}\includegraphics[scale=0.25]{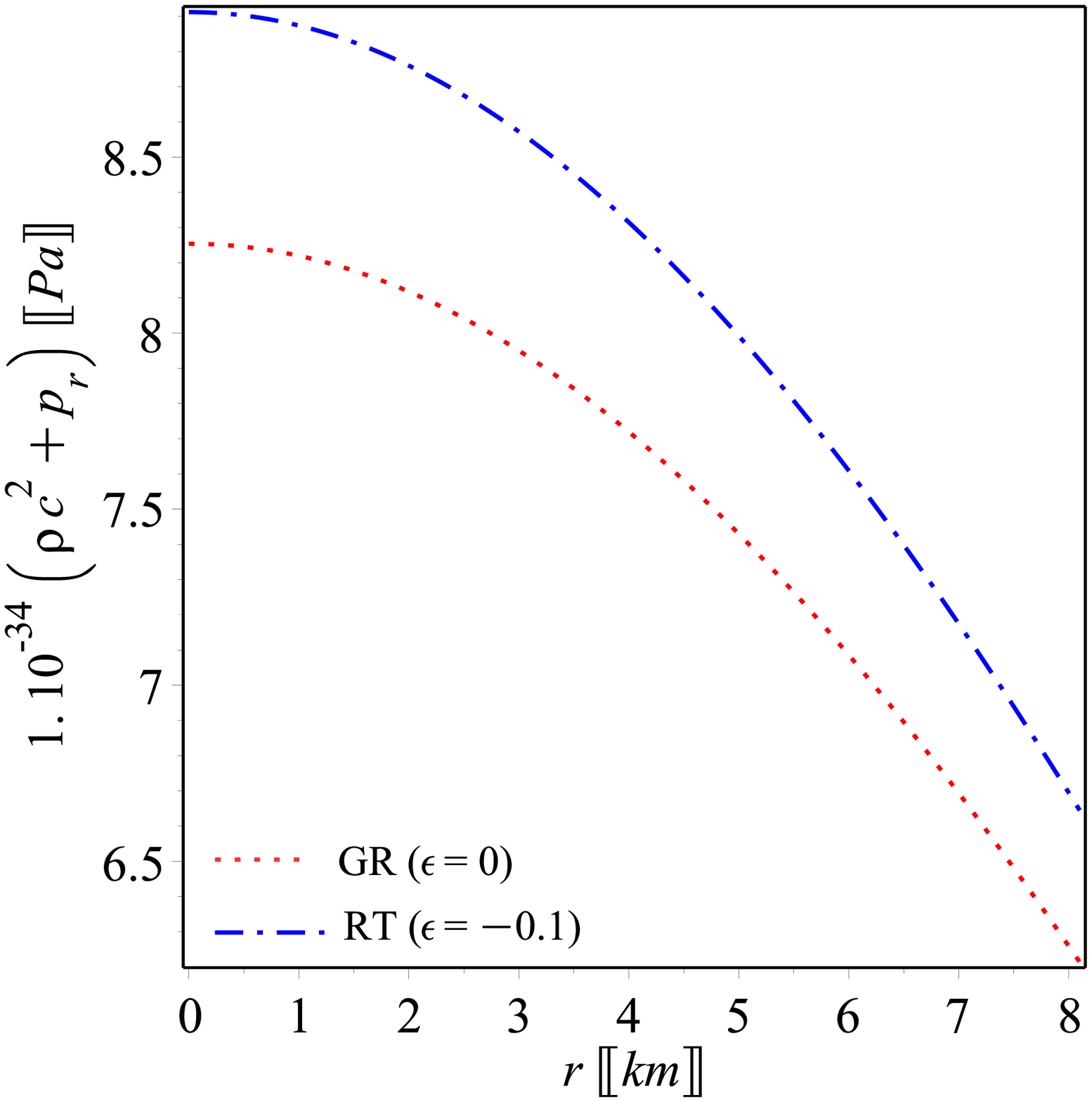}}\hspace{0.5cm}
\subfigure[~Null energy conditions]{\label{fig:NEC}\includegraphics[scale=0.25]{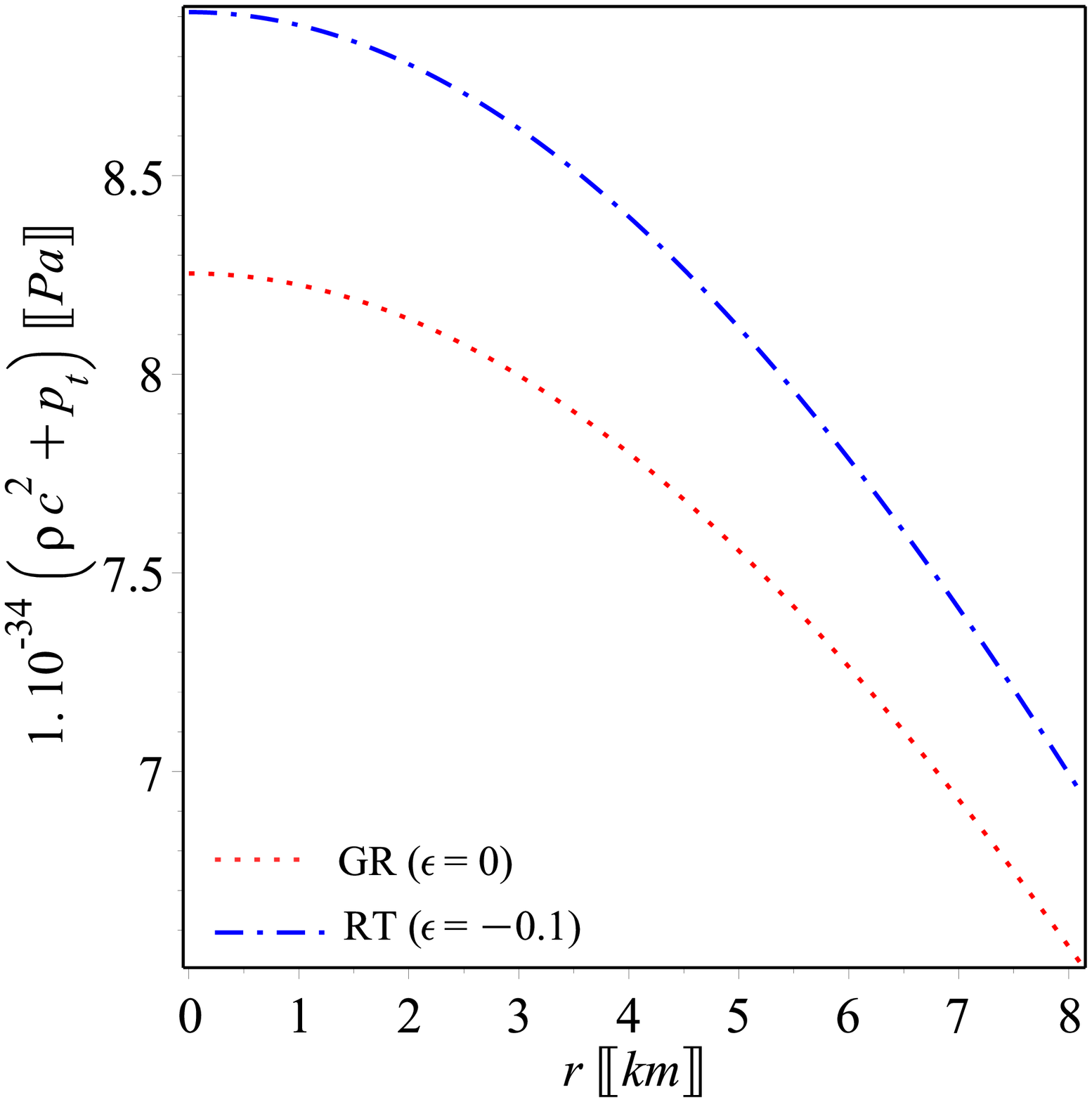}}\hspace{0.5cm}
\subfigure[~Strong energy condition]{\label{fig:SEC}\includegraphics[scale=.25]{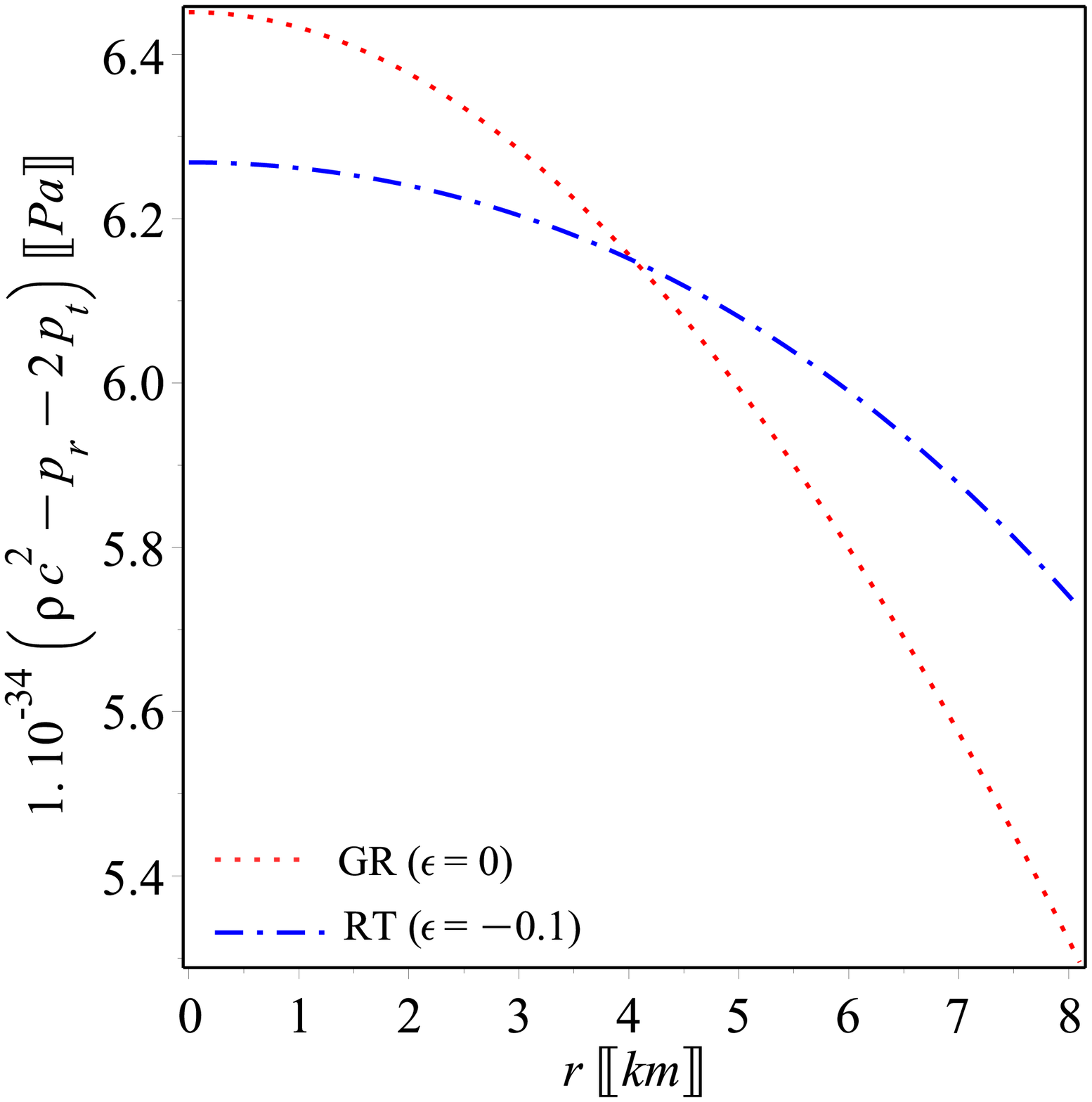}}\\
\subfigure[~Dominant energy condition (radial)]{\label{fig:DEC}\includegraphics[scale=.25]{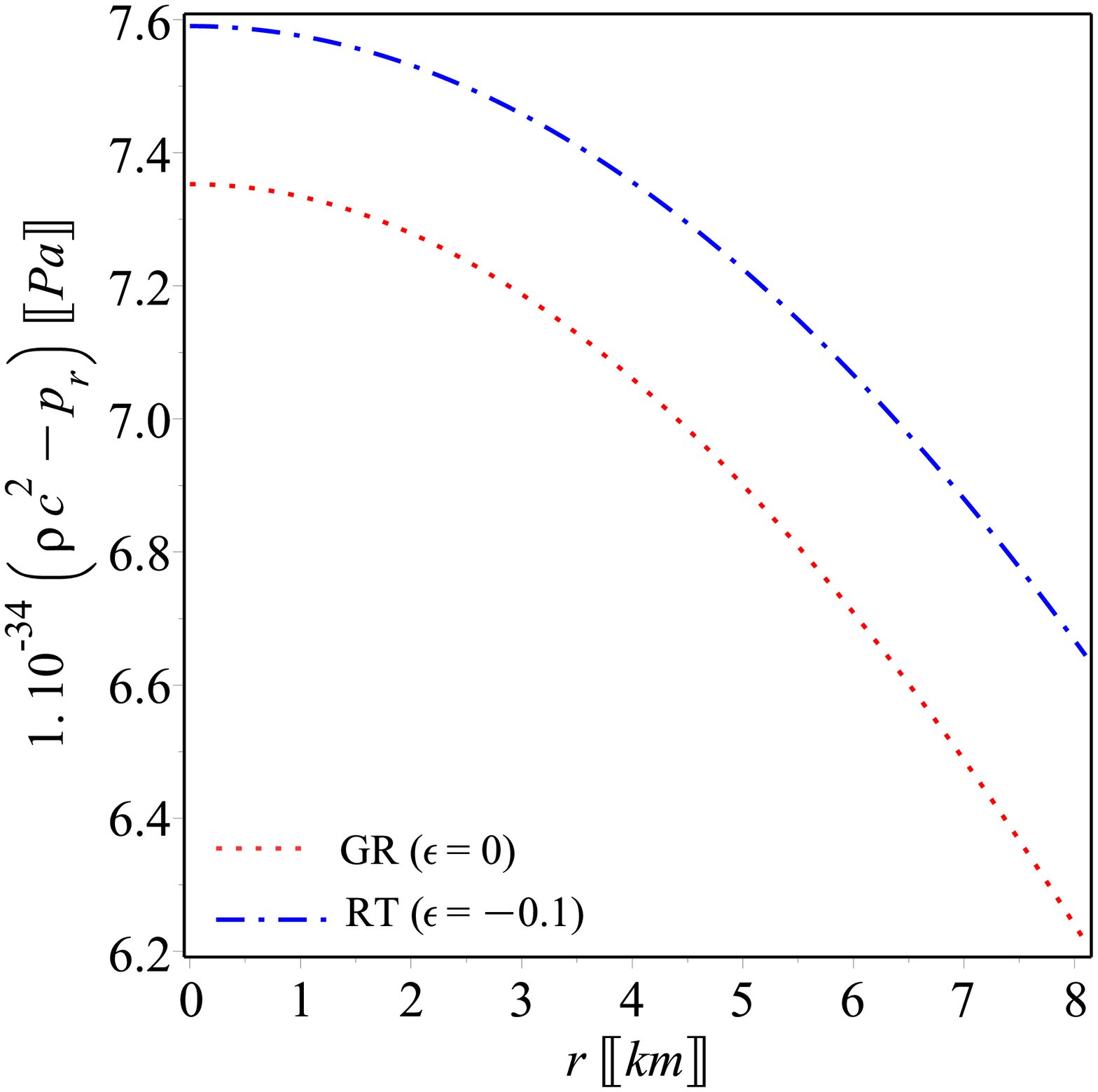}}\hspace{1cm}
\subfigure[~Dominant energy condition (tangential)]{\label{fig:DEC2}\includegraphics[scale=.25]{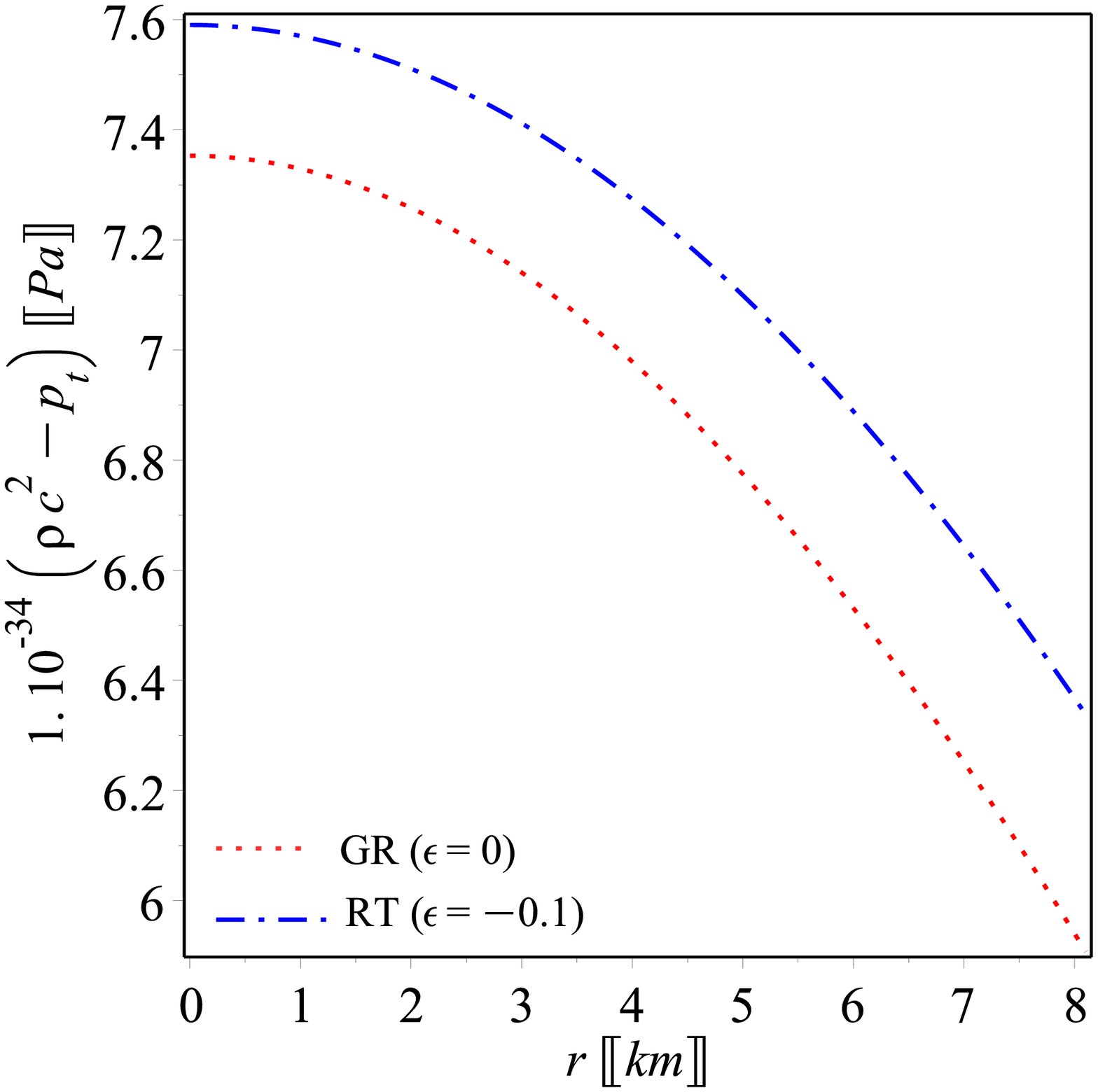}}
\caption[figtopcap]{\small{The weak, null, strong and dominant energy conditions, using Eqs. (\ref{sol}), versus the radial coordinate $r$ in km for the pulsar \textit{Her X-1}. The plots show that the model fulfill the energy conditions (vii) in Sec. \ref{S4}.}}
\label{Fig:4}
\end{figure}

In Figs. \ref{Fig:1}\subref{fig:density}--\subref{fig:tangpressure}, recalling condition (iii) in Sec. \ref{S4}, we represent the behavior of energy-density, radial and tangential pressures for the pulsar \textit{Her X-1} from its center to the boundary surface. This shows that the density, radial and tangential pressures are positive as required for realistic stellar configuration with maximal finite values at the center decrease monotonically towards the surface. Also, as Fig. \ref{Fig:1} \subref{fig:density} shows, the surface density is above $7\times 10^{14}$ g/cm$^3$ while the central density is comparable to the nuclear density $\sim 10^{15}$ g/cm$^3$ which indicate that the pulsar \textit{Her X-1} may has no neutron core but quark-gluon one. This will be revisited in Sec. \ref{S8}. In addition, Fig. \ref{Fig:1}\subref{fig:radpressure} shows that the radial pressure at the surface is null in agreement with condition (v) in Sec. \ref{S4}.\\

In Fig. \ref{Fig:2}\subref{fig:An}, recalling condition (iv) in Sec. \ref{S4}, we show that the anisotropy parameter \eqref{anis2} is null at the center and increases monotonically toward the surface of the star. Also, it shows that the anisotropic force $F_a=\frac{2\Delta}{r}$ is positive reflecting the repulsive behaviour of the force as $p_t-p_r>0$. Recalling Eq. \eqref{anio} we verify that the Rastall parameter has no contribution to the anisotropy parameter so both GR ($\epsilon=0$) and RT ($\epsilon\neq 0$) cases coincide. However, this is not the case with other ingredients. For the GR ($\epsilon=0$) and RT ($\epsilon\neq 0$) cases, Figs. \ref{Fig:2} \subref{fig:grdgr} and \subref{fig:grdrast} respectively confirm that the gradients of energy-density, radial and tangential pressures, given by Eqs. \eqref{dsol1}--\eqref{dsol3}, are negative everywhere inside the star as required by condition (iii) in Sec. \ref{S4}.\\

Recalling the causality and stability conditions (viii) and (ix) in Sec. \ref{S4}, in  Fig. \ref{Fig:3}, for GR ($\epsilon=0$) and RT ($\epsilon\neq 0$) cases, we plot the radial and tangential sound speeds \eqref{dsol2} where both have positive values less than one. This result confirms that the obtained stellar model is fulfilling the causality condition. In addition, we investigated the energy conditions as visualized in Fig. \ref{Fig:4}. The plots of Figs. \ref{Fig:4}\subref{fig:WEC}--\subref{fig:DEC}, for $\epsilon\neq 0$, prove that the model satisfies the WEC, NEC, SEC and DEC since the graphs are always obtained in positive energy regions. This confirms the fulfillment of the energy conditions (vii) in Sec. \ref{S4}.\\

\begin{figure}
\centering
\subfigure[~radial EoS parameter]{\label{fig:wr}\includegraphics[scale=0.22]{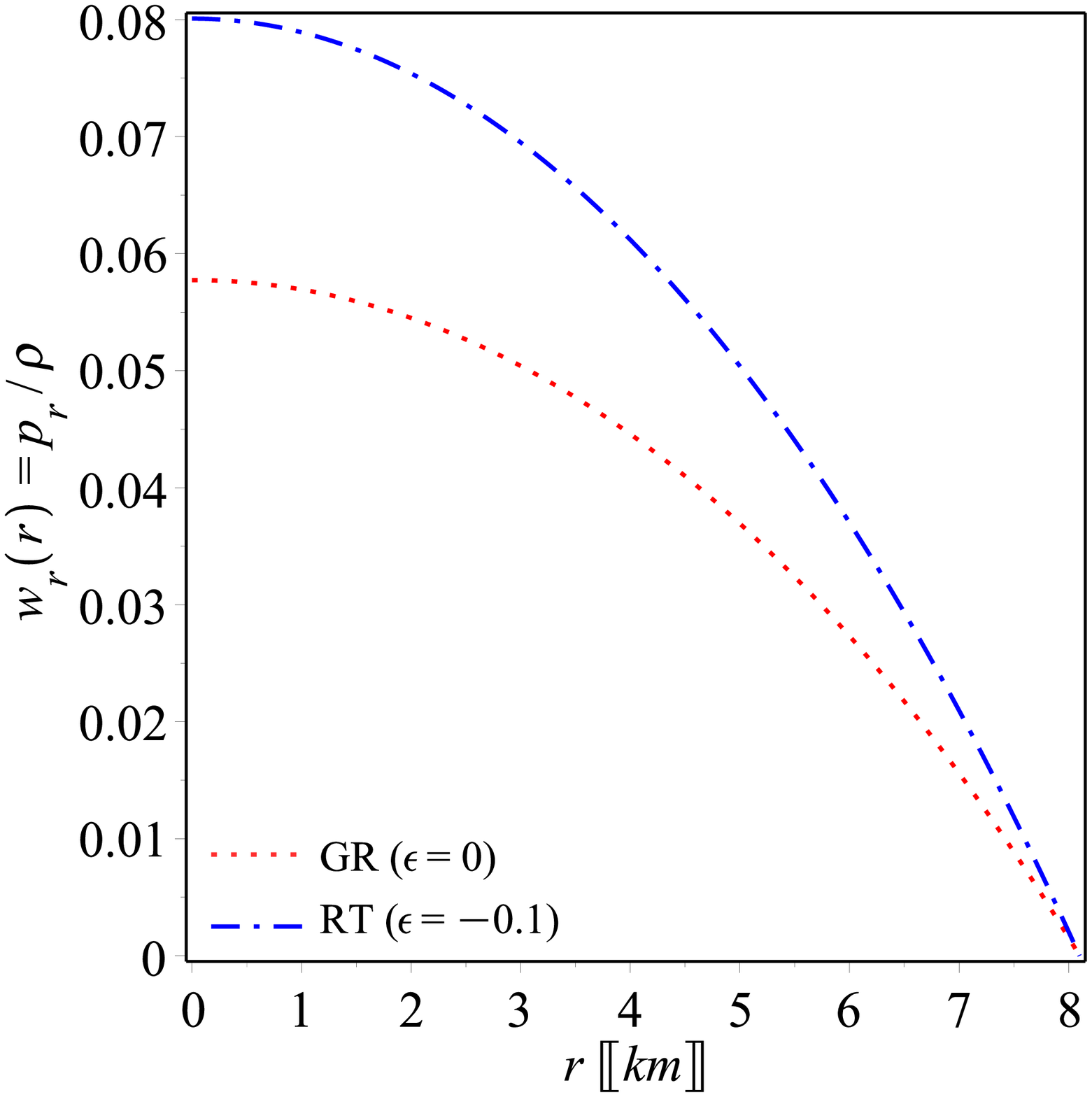}}
\subfigure[~linear radial EoS]{\label{fig:EoSr}\includegraphics[scale=0.22]{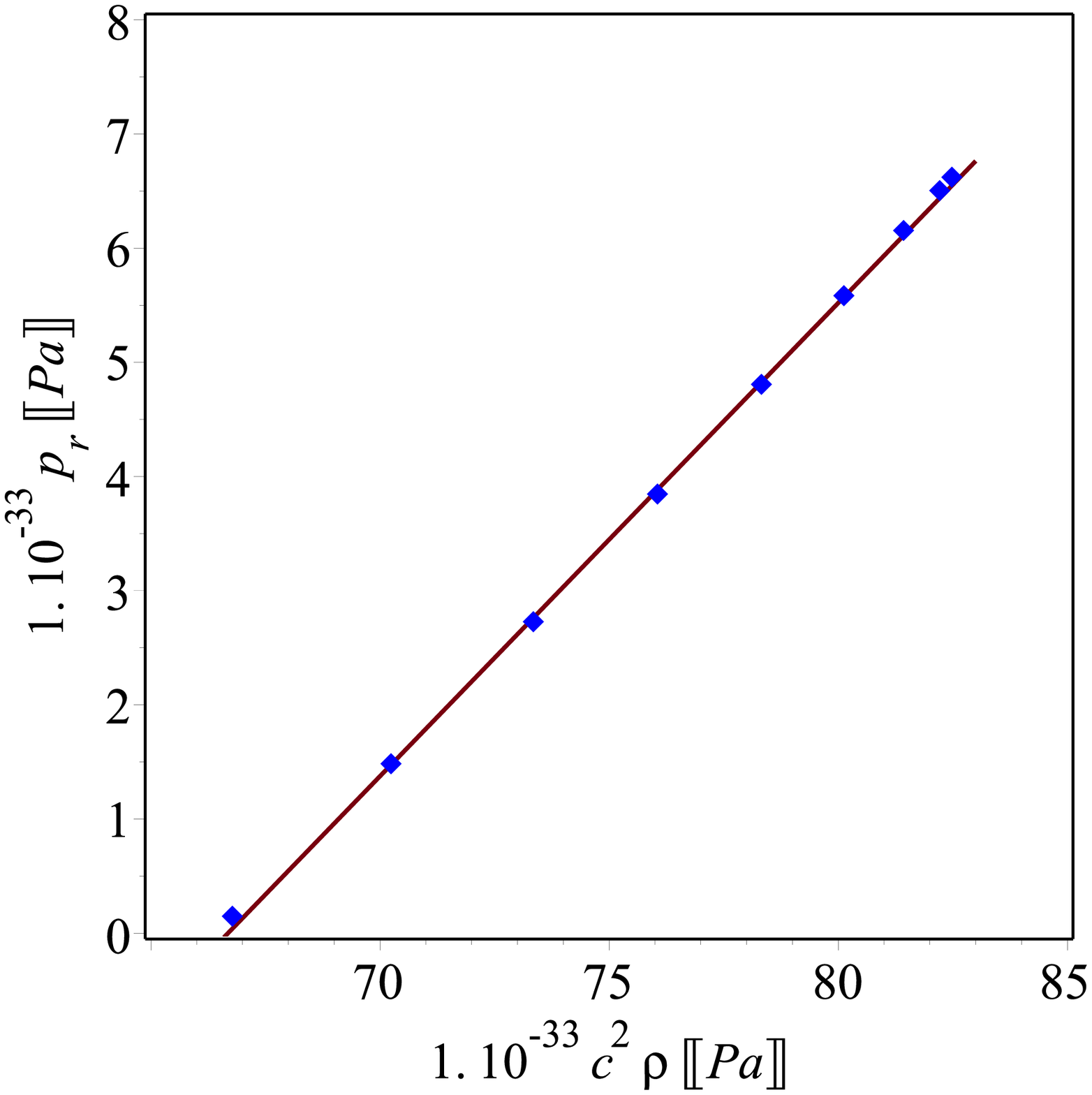}}
\subfigure[~tangential EoS parameter]{\label{fig:wt}\includegraphics[scale=0.22]{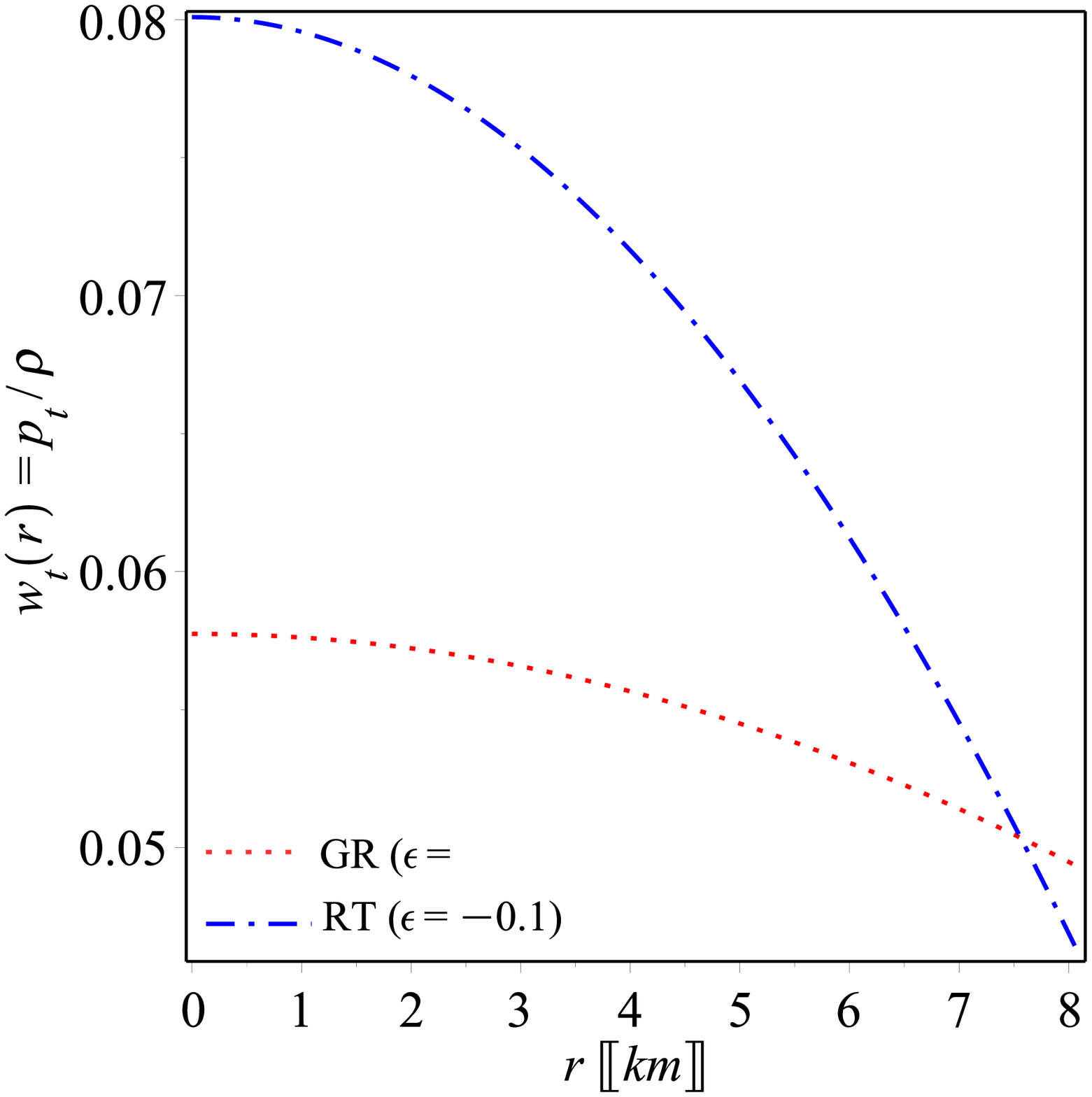}}
\subfigure[~linear tangential EoS]{\label{fig:EoSt}\includegraphics[scale=0.22]{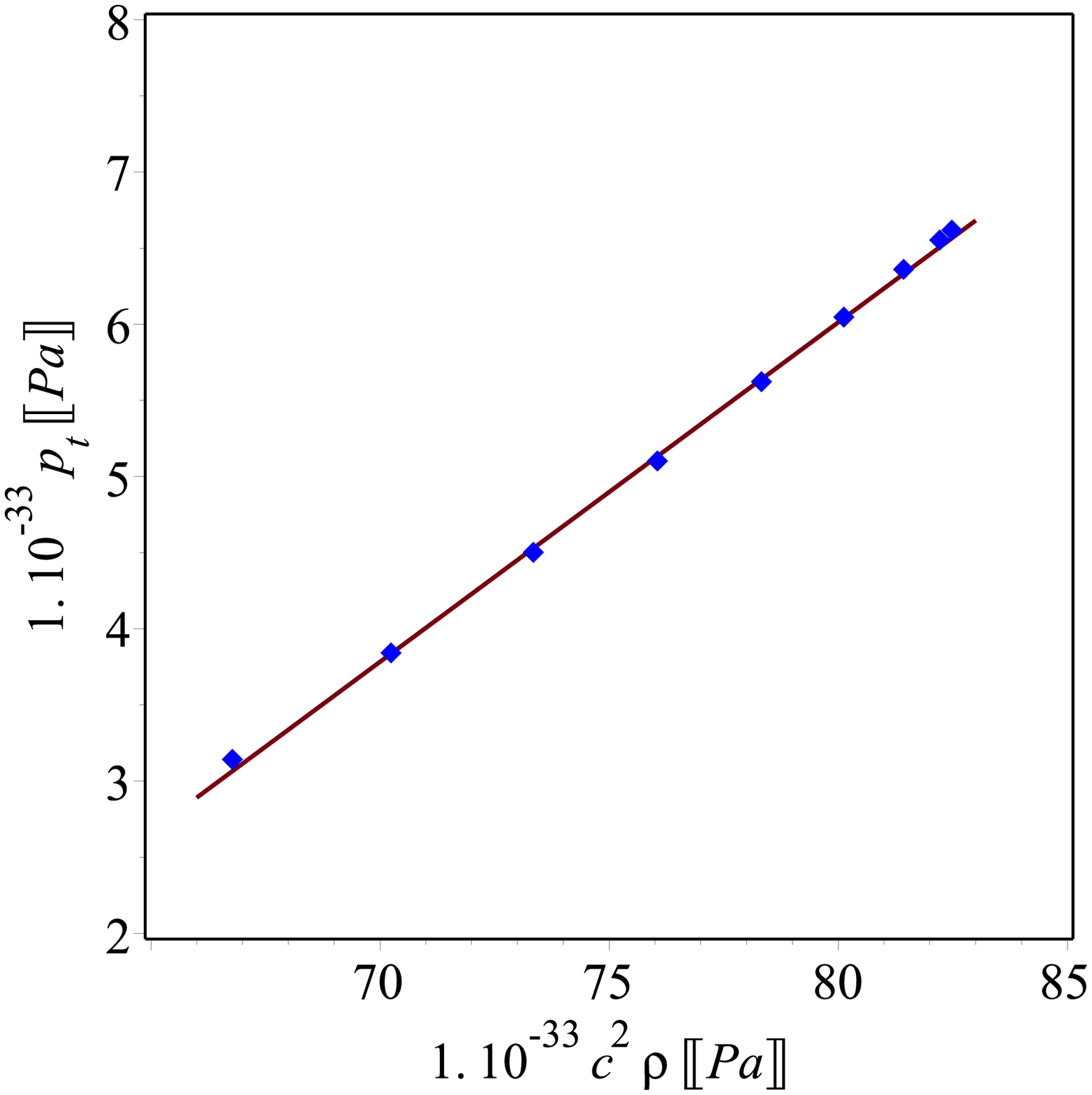}}
\caption[figtopcap]{\small{Figs. \subref{fig:wr} and \subref{fig:wr} show the behaviours of the EoS parameters, defined as $w_r(r)=p_r/\rho$ and $w_t(r)=p_t/\rho$, at different radial distances within the pulsar \textit{Her X-1} as predicted by RT and GR. We note that no EoS are imposed at any stage of the present work, while it is evidently that the result fit well with the linear behaviour whereas the best fit lines in Fig. \subref{fig:EoSr} and \subref{fig:EoSt} are given by $p_r=0.414\,\rho-27.6$ and $p_t=0.223\, \rho-11.8$ in RT case.}}
\label{Fig:5}
\end{figure}

In Figs. \ref{Fig:5}\subref{fig:wr} and \subref{fig:wt}, for $\epsilon=0$ and $\epsilon\neq 0$ cases, we plot the evolution of the radial and the tangential EoS parameters, i.e. $w_r(r)=p_r/\rho$ and $w_t(r)=p_t/\rho$, respectively, verses the radial distance from the center. The plots show that the EoS parameters slightly vary from maximum at the center with a monotonically decreasing behavior toward the surface whereas $w_r$ drops to zero at the surface as expected. We note that the variation of the EoS parameters is more severe in RT than GR, which should be understood as an effect of the matter-geometry coupling. On the other hand, we remind that there are no EoS are preassumed in the model setup, Figs. \ref{Fig:5}\subref{fig:EoSr} and \subref{fig:EoSt}, however, prove that the EoS $p(\rho)$ in both radial and tangential directions is in well fit with linear patterns, whereas the best fit are given by $p_r=0.414\,\rho-27.6$ and $p_t=0.223\, \rho-11.8$ for the pulsar \textit{Her X-1}. Notably the slopes of the fitted lines $dp_r/d\rho=0.414=v_r^2$ and $dp_t/d\rho=0.223=v_t^2$ are consistent with the corresponding values of Table \ref{Table2} for the pulsar under investigation.\\

\begin{figure}
\centering
\subfigure[~Mass function]{\label{fig:mass}\includegraphics[scale=0.28]{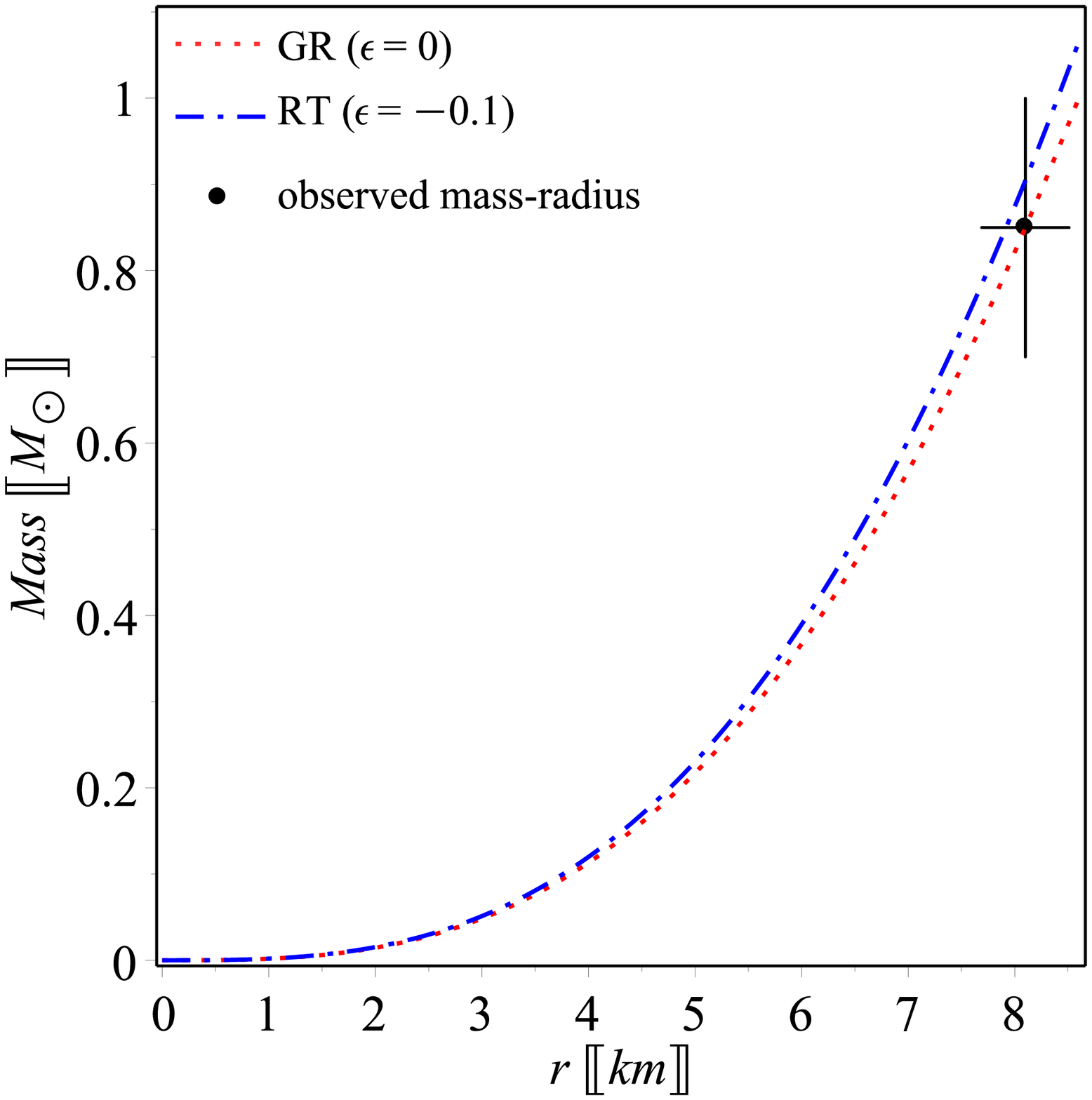}}
\subfigure[~Compactness function]{\label{fig:comp}\includegraphics[scale=.28]{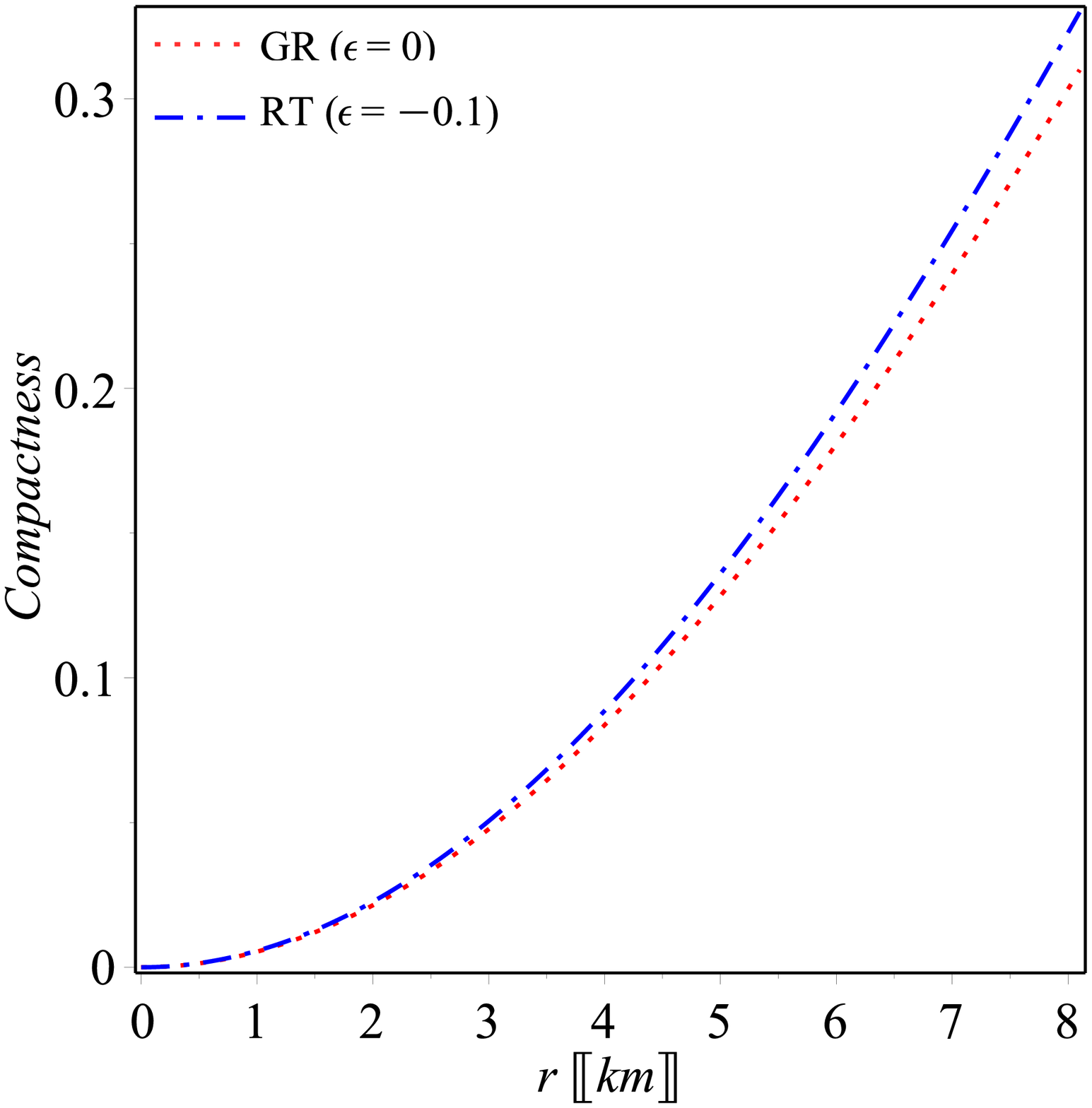}}
\subfigure[~Red-shift function]{\label{fig:red}\includegraphics[scale=.28]{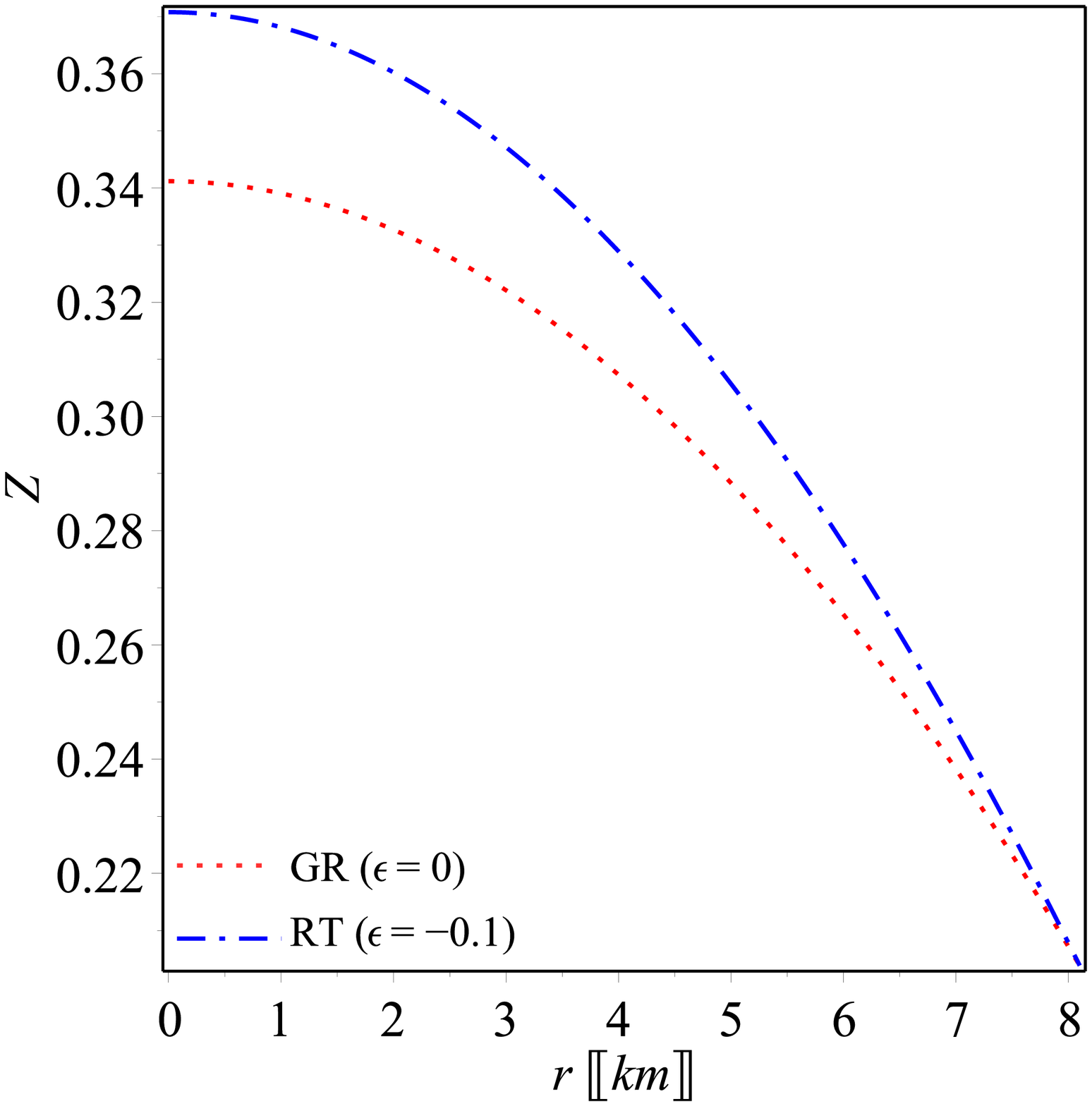}}
\caption[figtopcap]{\small{The mass function plot confirm that the model can predict the a of the pulsar \textit{Her X-1} in agreement with observational data. The plot shows that RT predicts compactness values higher than GR which reflects the role of the matter-geometry coupling to allow more compactness values. The redshift is finite everywhere within the pulsar and decreases toward the surface as stated by condition (x) in Sec. \ref{S4} and also predict a surface redshift consistent with the upper limit constraints as given by \cite{Bohmer2006}.}}
\label{Fig:6}
\end{figure}

The mass function given by Eq. (\ref{mas}) is plotted in Fig. \ref{Fig:6} \subref{fig:mass} which shows that it is a monotonically increasing function of the radial coordinate whereas the mass at the center $M(r=0) = 0$. The plot shows that theoretical $M(r)$ curve of the pulsar \textit{Her X-1} according to RT is compatible with the observed values of the mass and the radius. In Fig. \ref{Fig:6} \subref{fig:comp} we show the behavior of the compactness parameter of the pulsar. Interestingly the RT curve allows less size than the GR for same mass which may evidently reflects the role of the matter-geometry coupling to obtain slightly higher compactness. Also Fig. \ref{Fig:6} \subref{fig:red} shows the redshift of the pulsar is maximum at the center and decreases monotonically toward the surface. Notably the redshift at the surface is found to be $Z_{R}\sim 0.204$ which is in an agreement with the upper bound constraints $Z_{R}\leq 2$ as given by Buchdahl \cite{PhysRev.116.1027}, for anisotropic spheres see \cite{Ivanov:2002xf,PhysRevD.67.064003} and for the upper bound on the surface redshift in presence of a cosmological constant see \cite{Bohmer2006}.
%%%%%%%%%%%%%%%%%%%%%%%% Section 7 %%%%%%%%%%%%%%%%%%%%%%
\section{Stability of the model}\label{S7}
%%%%%%%%%%%%%%%%%%%%%%%%%%%%%%%%%%%%%%%%%%%%%%%%%%%%%%%%%
In addition to the stability condition (ix) in Sec. \ref{S4} which has been shown to be verified we are going to discuss the stability of the obtained stellar model using two different techniques in the present section; that are the modified TOV equations and the adiabatic index.
%%%%%%%%%%%%%%%%%%%%%%%%%%%%%%%%%%%%%%%%%%%%%%
\subsection{Equilibrium analysis via Tolman-Oppenheimer-Volkoff equation}
\begin{figure}
\centering
\subfigure[~TOV equation ($\epsilon=0$)]{\label{fig:TOVgr}\includegraphics[scale=0.3]{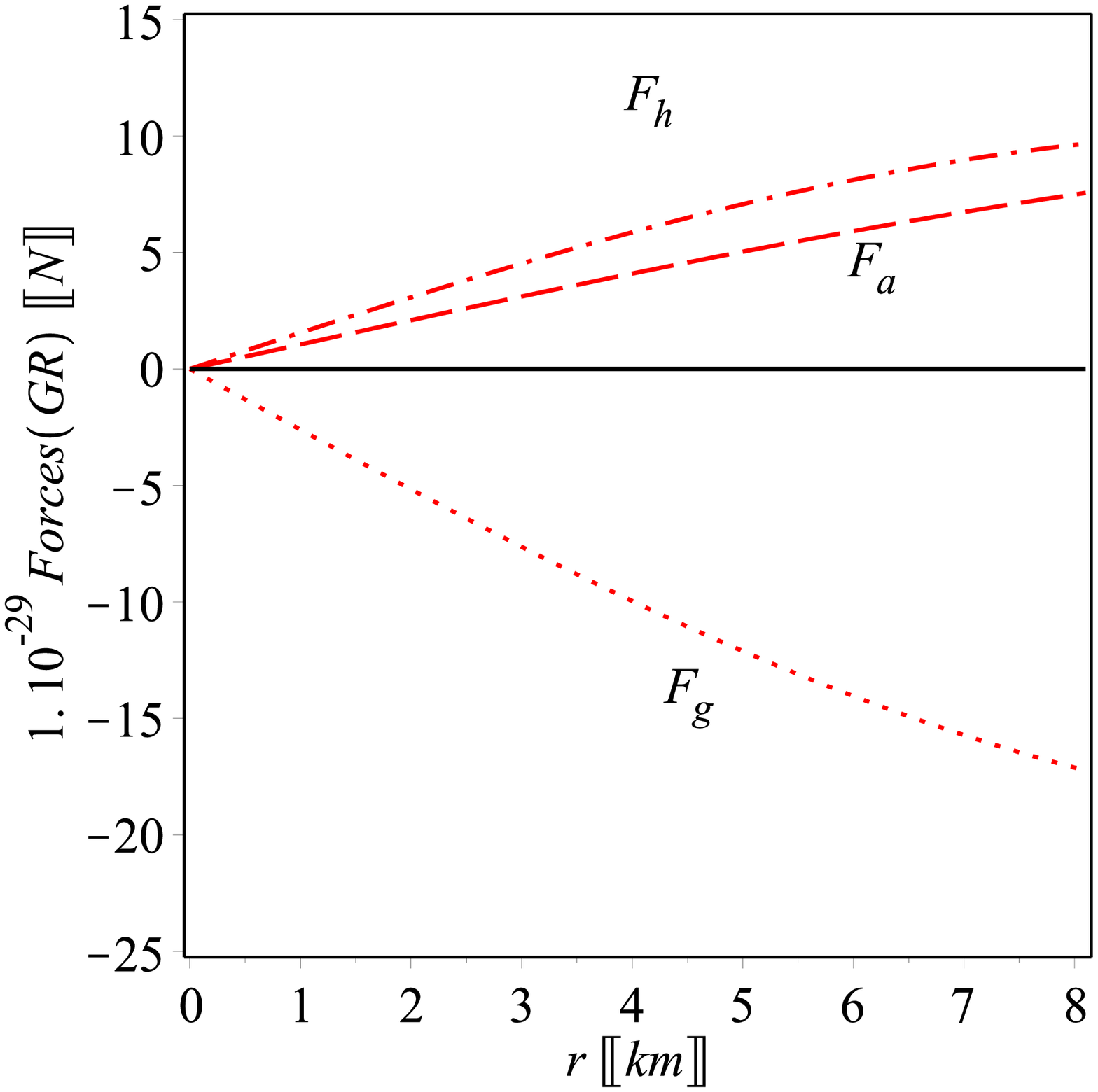}}\hspace{1cm}
\subfigure[~TOV equation ($\epsilon=-0.1$)]{\label{fig:TOVras}\includegraphics[scale=.3]{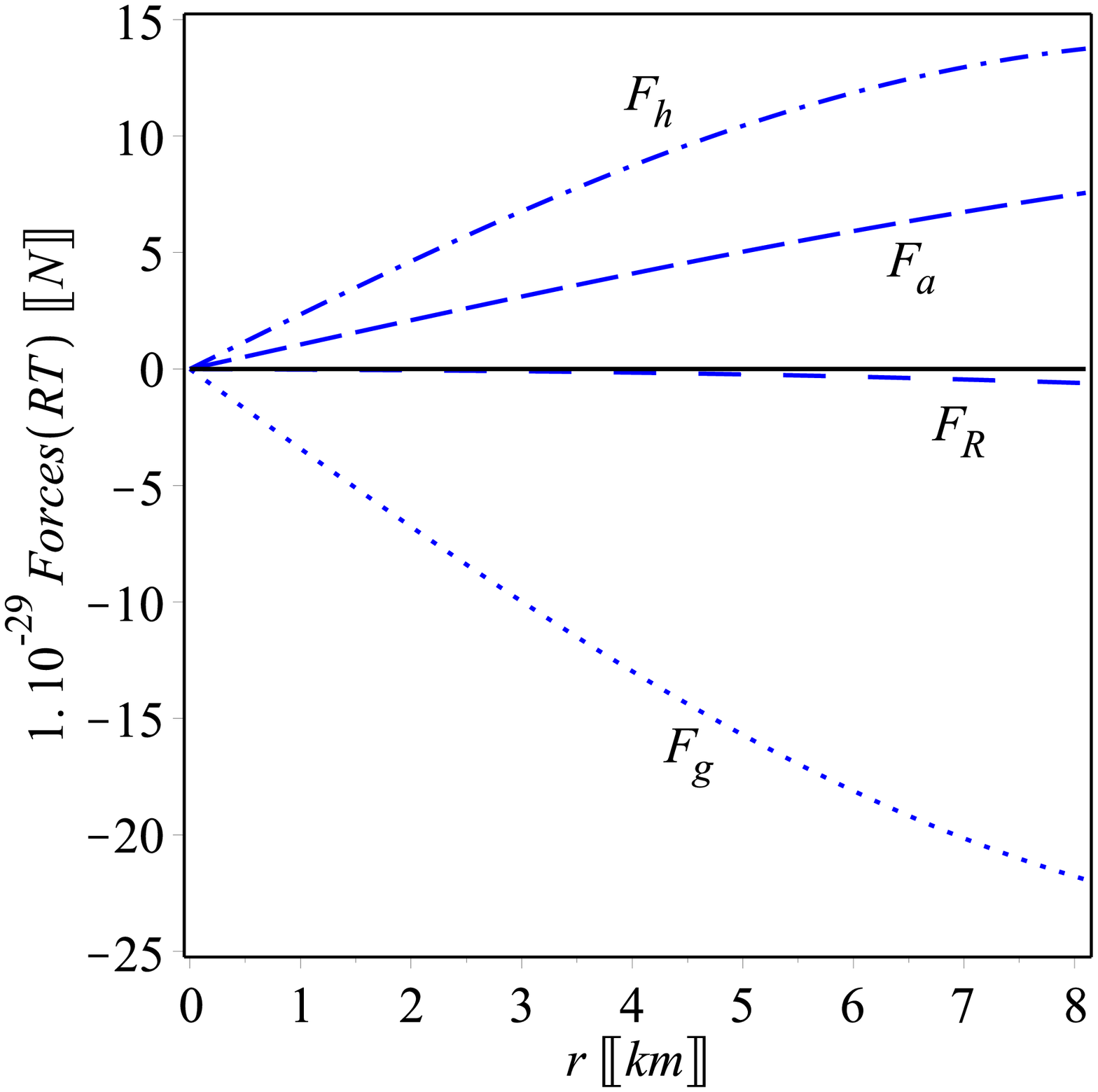}}
\caption[figtopcap]{\small{Plots of the forces of TOV equation \eqref{ma_1} in cases $\epsilon=0$ and $\epsilon=-0.1$ versus the radius $r$ using the constants constrained from $\textit{Her X-1}$. In the RT case the negative gravitational force is the dominant one over the hydrostatic and the anisotropic forces. This guarantees stable equilibrium configuration for the pulsar.}}
\label{Fig:7}
\end{figure}

%%%%%%%%%%%%%%%%%%%%%%%%%%%%%%%%%%%%%
We assume hydrostatic equilibrium to be everywhere within the stable compact star. This configuration, then, can be described by the GR based TOV equation \cite{PhysRev.55.364,PhysRev.55.374,PoncedeLeon1993} which gives the following stability constraint
\begin{equation}\label{TOV}
\frac{2(p_t-p_r)}{r}-\frac{M_g (\rho+p_r)\sqrt{F}}{r \sqrt{G}}-\frac{dp_r}{dr}=0,
\end{equation}
where $M=M_g(r)$ is the gravitational mass within a radius $r$, which is defined by the Tolman-Whittaker mass formula
\begin{equation}\label{ma}
M_g(r)=4\pi{\int_0}^r\Big({T_t}^t-{T_r}^r-{T_\theta}^\theta-{T_\phi}^\phi\Big)r^2\sqrt{FG}dr=\frac{r F' \sqrt{G}}{2 F \sqrt{F}}\,.
\end{equation}
Inserting Eq. (\ref{ma}) into (\ref{TOV}),  we get
\begin{equation}\label{ma_1}
\frac{2}{r}(p_t-p_r)-\frac{F'}{2 F}(\rho+p_r)-\frac{dp_r}{dr}=F_a+F_g+F_h=0\,,
\end{equation}
where $F_g=-\frac{F'}{2 {F}}(\rho+p_r)$ and $F_h=-\frac{dp_r}{dr}$ are the gravitational and the hydrostatic forces respectively, in addition to the anisotropic force $F_a$. We note that the TOV equation should be modified in RT due to the non-minimal coupling constraint, $\mathcal{T}{^\alpha}{_{\beta;\alpha}}=\epsilon\, \partial_{\beta} \mathcal{R}$, to include one more force $F_R$ as following
\begin{equation}\label{RS_TOV}
F_a+F_g+F_h+F_R=0\,,
\end{equation}
where $F_R=-\frac{\epsilon}{1-4\epsilon}\frac{d}{dr}(\rho-p_r-2p_t)$. These different forces, for GR ($\epsilon=0$) and RT ($\epsilon\neq0$), are plotted in Fig. \ref{Fig:7} using the pulsar $\textit {Her X-1}$ data. For the RT case, the figure clearly shows that the pulsar is dominated by the negative gravitational force over the positive hydrostatics and anisotropic forces. Therefore, it holds the pulsar in a static equilibrium. In conclusion, we verify the stability of the model via TOV equation using the pulsar \textit{Her X-1} data.
%
%%%%%%%%%%%%%%%%%%%%%%%%%%%%%%%%%%%
\subsection{Relativistic adiabatic indices}
%%%%%%%%%%%%%%%%%%%%%%%%%%%%%%%%%%%%%
Another verification of the stable equilibrium configuration of a spherically symmetric object can be done via the adiabatic index, that is defined as the ratio of two specific heats and can be given as follows \cite{1964ApJ...140..417C,1989A&A...221....4M,10.1093/mnras/265.3.533}
\begin{equation}\label{a_11}
\Gamma=\frac{\rho+p_r}{p_r} v_r^2\,.
\end{equation}
For the general case of anisotropic spheroid fluid, it has been shown that the object is in a neutral equilibrium if its adiabatic index $\Gamma =\gamma$ and in a stable equilibrium if $\Gamma>\gamma$ \cite{10.1093/mnras/265.3.533}, whereas
\begin{equation}\label{ai}
\gamma=\frac{4}{3}\left(1+\frac{F_a}{2 |p'_r|}\right)_{max}\,.
\end{equation}
Clearly, for an isotropic fluid, the object is in a neutral equilibrium if the adiabatic index $\Gamma=\frac{4}{3}$, while for $\Gamma>\frac{4}{3}$ the object is in a stable equilibrium \cite{1975A&A....38...51H}. Using Eq. (\ref{ai}),  we get
\begin{eqnarray}\label{a_12}
\gamma&=&\frac{2}{3R^{8}}\biggl[\biggl(2R^{8} \biggl| \frac {r{a_2}^{4}}{{R}^{8} \left( a_0{R}^{2 }+2a_1{a_2}^{2}{R}^{2}-2a_1{a_2}^{4}{r}^{2 } \right) ^{2}} \biggl\{ 216{r}^{8}{a_1}^{2}{a_2}^{12}\epsilon-12{r}^{8}{a_1}^{2}{a_2}^{12}+56{r}^{6}{R}^{2}{a_1}^{2}{a_2}^{10}-880{r}^{6}{R}^{2}{a_1}^{2}{a_2}^{10}\epsilon\nonumber\\
&&+1352 {r}^{4}{R}^{4}{a_1}^{2}{a_2}^{8}\epsilon-100{r}^{4}{R}^{ 4}{a_1}^{2}{a_2}^{8}-168{r}^{6}{R}^{2}a_1\,{a_2} ^{8}a_0\epsilon-4\,{r}^{6}{R}^{2}a_1\,{a_2}^{8}a_0+80{r}^{2}{R}^{6}{a_1}^{2}{a_2}^{6}-928\,{r}^{2}{R}^{6} {a_1}^{2}{a_2}^{6}\epsilon+520{r}^{4}{R}^{4}a_1{a_2}^{6}a_0\epsilon\nonumber\\
&&+4\,{r}^{4}{R}^{4}a_1\,{a_2}^{ 6}a_0+240\,{R}^{8}{a_2}^{4}{a_1}^{2}\epsilon-24\,{R}^{8 }{a_2}^{4}{a_1}^{2}-544\,{R}^{6}{a_2}^{4}{r}^{2}a_0\,a_1\,\epsilon+8\,{R}^{6}{a_2}^{4}{r}^{2}a_0\,a_1+9\,{R}^{4}{a_2}^{4}{r}^{4}{a_0}^{2}+18\,{R}^{4}{a_2}^ {4}{r}^{4}{a_0}^{2}\epsilon\nonumber\\
&&-8\,a_0\,{R}^{8}{a_2}^{2}a_1+192\,a_0\,{R}^{8}{a_2}^{2}a_1\,\epsilon-16\,{a_0}^{2}{R}^{6}{a_2}^{2}{r}^{2}+24{a_0}^{2 }{R}^{8}\epsilon -40{a_0}^{2}{R}^{6}{a_2}^{2}{r}^{2}\epsilon+6{a_0}^{2}{R}^{8}\biggr\}  \biggr|+6{R}^{4}r{a_2}^{4}-8{R}^{2}{r}^{3}{ a_2}^{6}+3{a_2}^{8}{r}^{5}\biggr) \biggl| \nonumber\\
&&\biggl\{{R}^{8}  \biggl[ a_0{R}^{2}+2a_1\,{a_2}^{2}{R}^{ 2}-2a_1\,{a_2}^{4}{r}^{2} \biggr] ^{2}\biggr\}\biggl\{r{a_2}^{4} \biggl( 216\,{r}^{8}{a_1}^{2}{a_2}^{12}\epsilon-12\,{r}^{8}{a_1}^{2}{a_2}^{12}+56\,{r}^{6}{R}^{2}{a_1}^{2}{a_2}^{10}-880\,{r}^{6}{R}^{2}{a_1}^{2}{a_2}^{10}\epsilon\nonumber\\
&&+1352 {r}^{4}{R}^{4}{a_1}^{2}{a_2}^{8}\epsilon-100{r}^{4}{R}^{ 4}{a_1}^{2}{a_2}^{8}-168{r}^{6}{R}^{2}a_1{a_2} ^{8}a_0\epsilon-4\,{r}^{6}{R}^{2}a_1{a_2}^{8}a_0+80\,{r}^{2}{R}^{6}{a_1}^{2}{a_2}^{6}-928{r}^{2}{R}^{6} {a_1}^{2}{a_2}^{6}\epsilon+520{r}^{4}{R}^{4}a_1{a_2}^{6}a_0\epsilon\nonumber\\
&&+4\,{r}^{4}{R}^{4}a_1\,{a_2}^{ 6}a_0+240{R}^{8}{a_2}^{4}{a_1}^{2}\epsilon-24{R}^{8 }{a_2}^{4}{a_1}^{2}-544{R}^{6}{a_2}^{4}{r}^{2}a_0a_1\,\epsilon+8{R}^{6}{a_2}^{4}{r}^{2}a_0a_1+9{R}^{4}{a_2}^{4}{r}^{4}{a_0}^{2}+18{R}^{4}{a_2}^ {4}{r}^{4}{a_0}^{2}\epsilon\nonumber\\
&&-8a_0\,{R}^{8}{a_2}^{2}a_1+192\,a_0\,{R}^{8}{a_2}^{2}a_1\,\epsilon-16\,{a_0}^{2}{R}^{6}{a_2}^{2}{r}^{2}-40\,{a_0}^{2}{R}^{6}{a_2}^{2}{r}^{2}\epsilon+6\,{a_0}^{2}{R}^{8}+24\,{a_0}^{2 }{R}^{8}\epsilon \biggr) \biggr\}^{-1} \biggr|\biggr]\,.
\end{eqnarray}
\begin{figure}
\centering
\subfigure[~$\gamma$   in cases $\epsilon=0$ and  $\epsilon\neq 0$ ]{\label{fig:gamma}\includegraphics[scale=0.28]{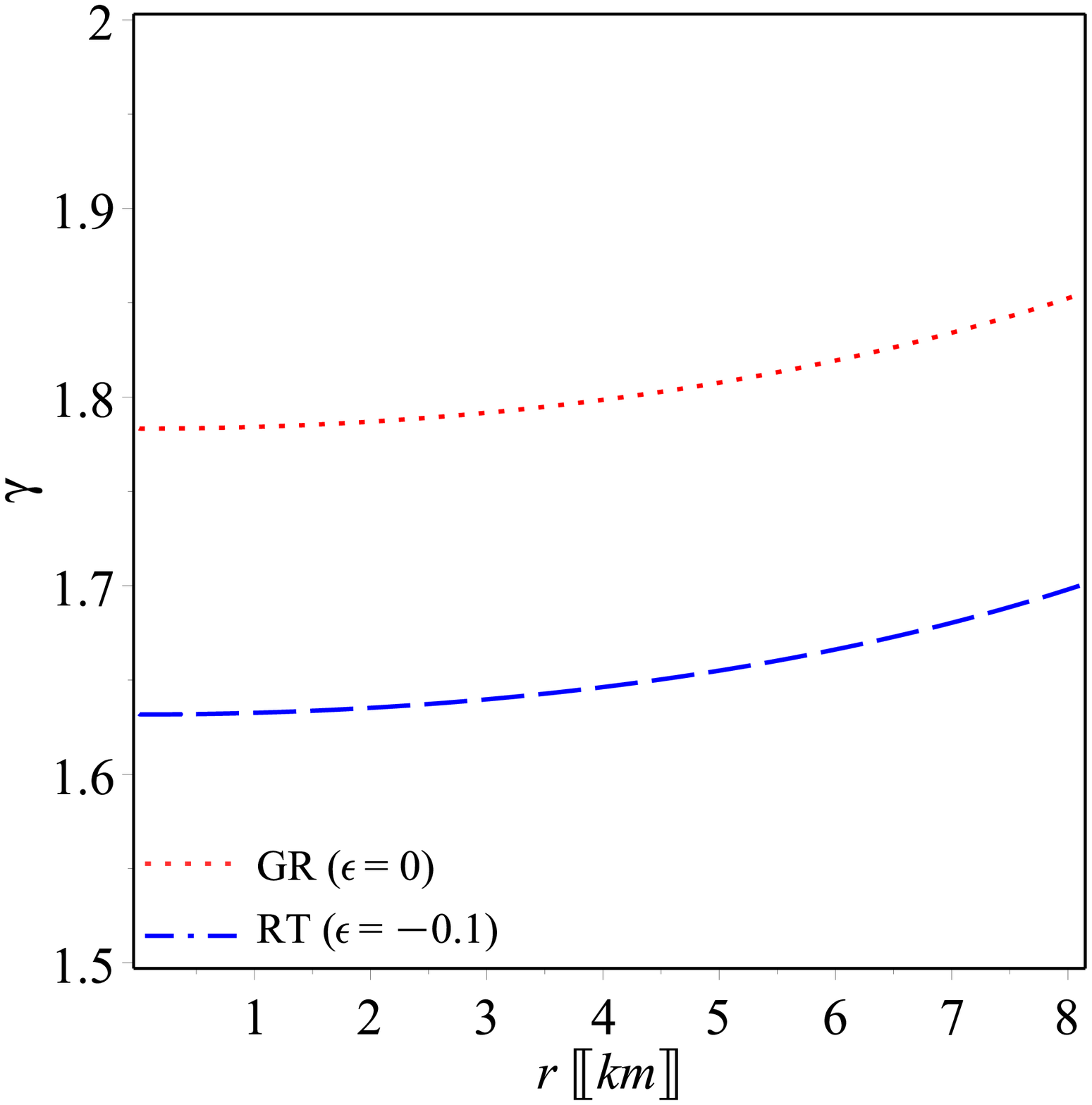}}
\subfigure[~$\Gamma_r$   in cases $\epsilon=0$ and  $\epsilon\neq 0$ ]{\label{fig:Gammar}\includegraphics[scale=0.28]{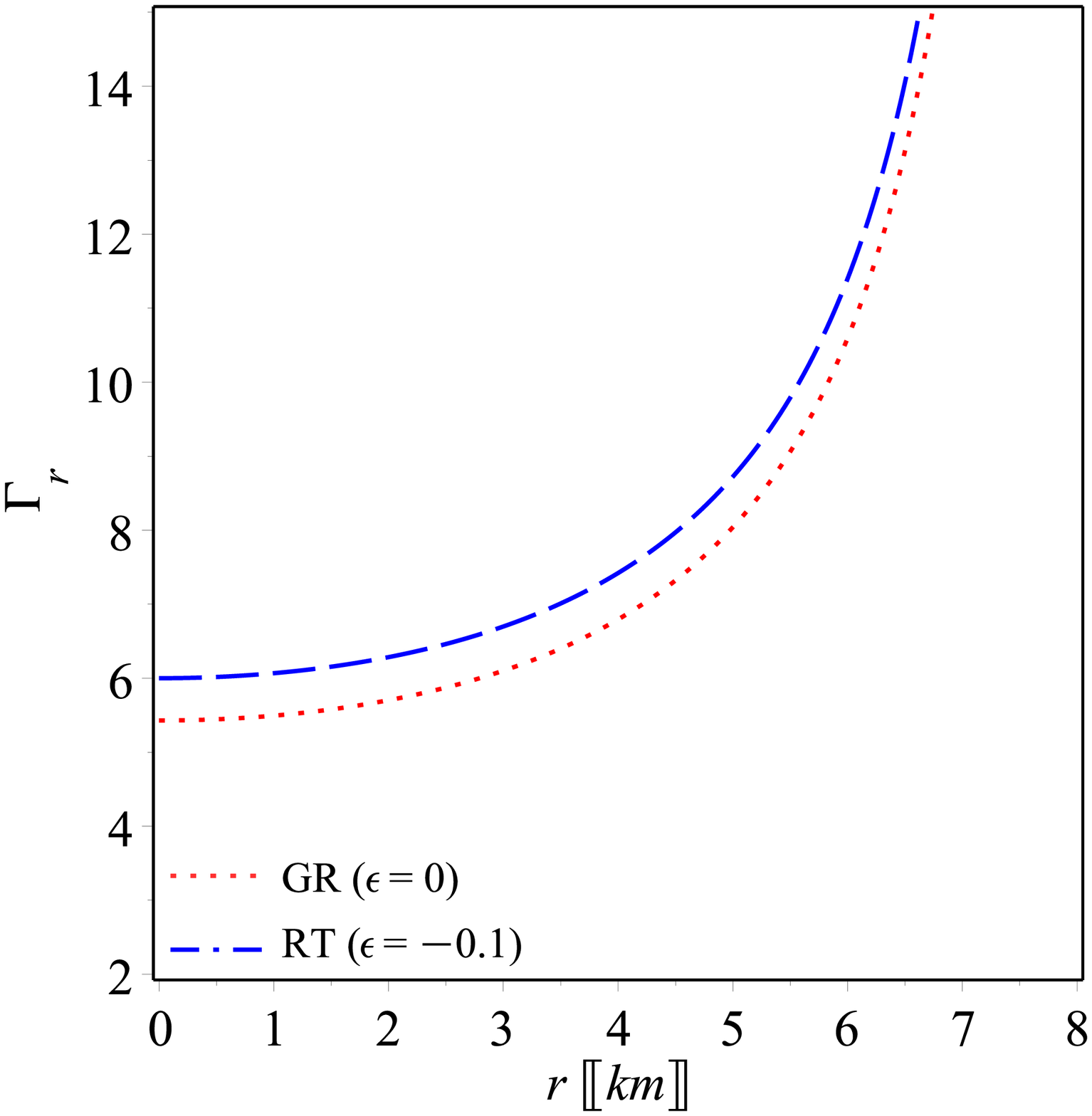}}
\subfigure[~$\Gamma_t$   in cases $\epsilon=0$ and  $\epsilon\neq 0$ ]{\label{fig:Gammat}\includegraphics[scale=0.28]{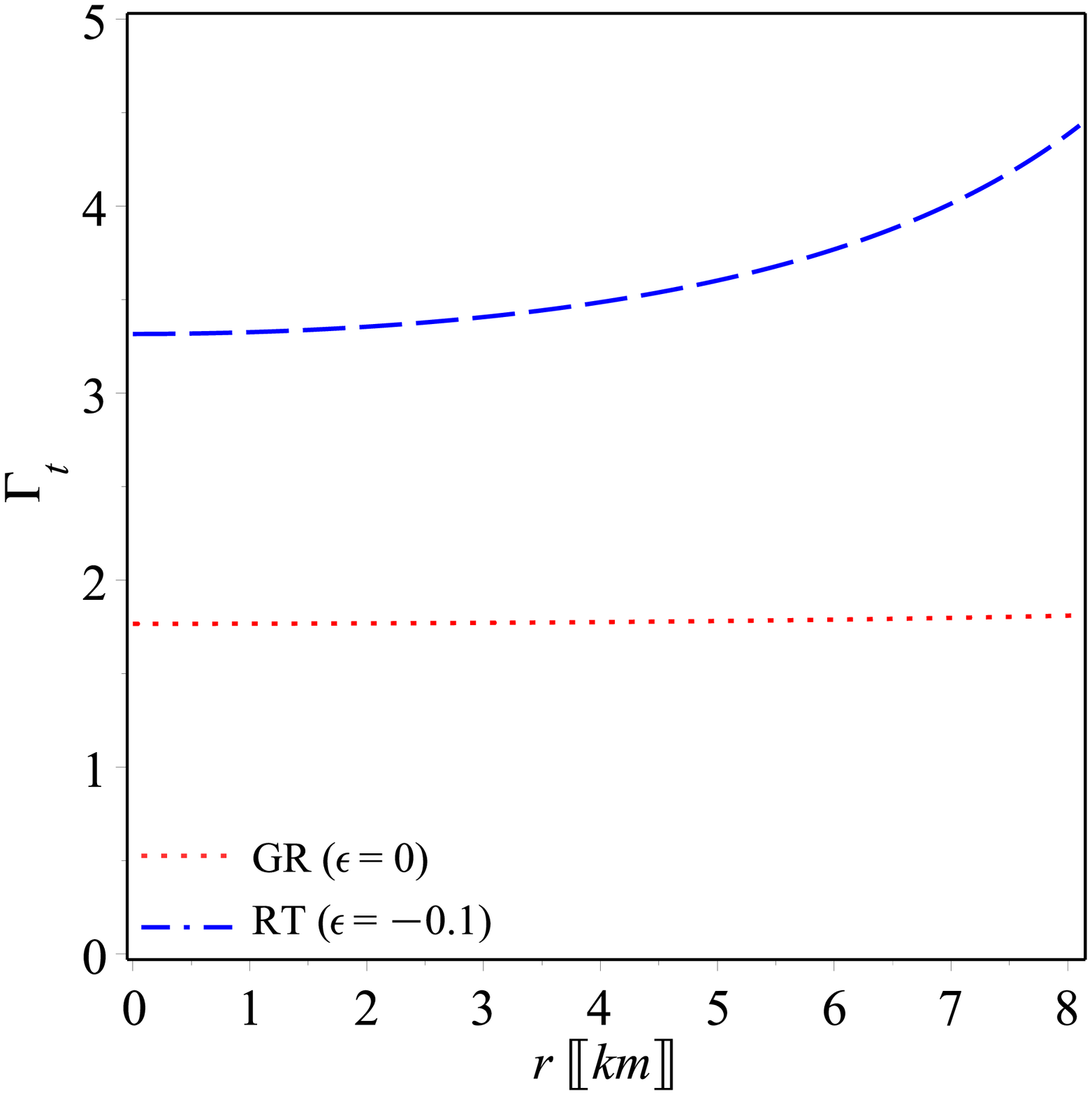}}
\caption[figtopcap]{\small{Plots of the Adiabatic indices $\gamma$, $\Gamma_r$ and $\Gamma_t$, namely \eqref{a_12}--\eqref{aic2}, versus the  radius  $r$ using the constants constrained from $\textit{Her X-1}$. For RT, the adiabatic index $\gamma$ less than the GR case but still greater than the  neutral equilibrium value $\gamma=\frac{4}{3}$. The radial and tangential adiabatic indices have higher values whereas the stability constraints $\Gamma_r>\gamma$ and $\Gamma_t>\gamma$ are fulfilled everywhere within the pulsar.}}
\label{Fig:8}
\end{figure}
From Eq. (\ref{a_11}),  we obtain  the adiabatic index of solution (\ref{sol})   in the form
\begin{eqnarray}\label{aic1}
\Gamma_r&=&-4\biggl\{ \left( 4a_1{a_2}^{2}{R}^{2}+3a_0{R }^{2}-4a_1{a_2}^{4}{r}^{2} \right)  \left( R{}^2-a_2{}^2r^2 \right) ^{3} \biggl[216{r}^{8}{a_1}^{2}{a_2}^{12} \epsilon -12{r}^{8}{a_1}^{2}{a_2}^{12}+56{r}^{6}{R}^{2}{a_1}^{2}{a_2}^{10}-880{r}^{6}{ R}^{2}{a_1}^{2}{a_2}^{10}\epsilon\nonumber\\
&&+1352\,{r}^{4}{R}^{4}{a_1}^{2}{a_2}^{8}\epsilon-100\,{r}^{4}{R}^{4}{a_1}^{2}{a_2}^{8}-168\,{r}^{6}{R}^{2}a_1\,{a_2}^{8}a_0\,\epsilon -4\,{r}^{6}{R}^{2}a_1\,{a_2}^{8}a_0+80\,{r}^{2}{R}^{6}{ a_1}^{2}{a_2}^{6}-928\,{r}^{2}{R}^{6}{a_1}^{2}{a_2 }^{6}\epsilon\nonumber\\
&&+520\,{r}^{4}{R}^{4}a_1\,{a_2}^{6}a_0\, \epsilon+4\,{r}^{4}{R}^{4}a_1\,{a_2}^{6}a_0+240\,{R}^{8 }{a_2}^{4}{a_1}^{2}\epsilon-24\,{R}^{8}{a_2}^{4}{a_1}^{2}-544\,{R}^{6}{a_2}^{4}{r}^{2}a_0\,a_1\,\epsilon+8\,{R}^{6}{a_2}^{4}{r}^{2}a_0\,a_1\nonumber\\
&& +9\,{R}^{4}{a_2 }^{4}{r}^{4}{a_0}^{2}+18\,{R}^{4}{a_2}^{4}{r}^{4}{a_0}^ {2}\epsilon-8\,a_0\,{R}^{8}{a_2}^{2}a_1+192\,a_0\, {R}^{8}{a_2}^{2}a_1\,\epsilon-16\,{a_0}^{2}{R}^{6}{a_2}^{2}{r}^{2}-40\,{a_0}^{2}{R}^{6}{a_2}^{2}{r}^{2} \epsilon+6\,{a_0}^{2}{R}^{8}\nonumber\\
&&+24\,{a_0}^{2}{R}^{8}\epsilon \biggr] \biggr\}\biggl\{ \biggl[36\,a_1\, {r}^{8}{a_2}^{10}\epsilon -2\,a_1\,{r}^{8}{a_2}^{10}+10\,{r}^{6}{R}^{2}a_1\,{a_2}^ {8}-148\,{r}^{6}{R}^{2}a_1\,{a_2}^{8}\epsilon-20\,{r}^{4}{R} ^{4}a_1\,{a_2}^{6}+232\,{R}^{4}{r}^{4}{a_2}^{6}a_1\,\epsilon\nonumber\\
&&-3\,{R}^{2}{r}^{6}{a_2}^{6}a_0-6\,{R}^{2}{r}^{6}{a_2}^{6}a_0\,\epsilon+
20\,{R}^{6}{r}^{2}{a_2}^{4}a_1-168\,{R}^{6}{r}^{2}{a_2}^{4}a_1\,\epsilon
+8\,{R}^{4}{r}^{4 }{a_2}^{4}a_0+20\,{R}^{4}{r}^{4}{a_2}^{4}a_0\, \epsilon-8\,{R}^{8}{a_2}^{2}a_1\nonumber\\
&&+48\,{R}^{8}{a_2}^{2}a_1\,\epsilon-6\,{R}^{6}{a_2}^{2}{r}^{2}a_0-24\,{R}^{6}{ a_2}^{2}{r}^{2}a_0\,\epsilon+12\,\epsilon\,{R}^{8}a_0 \biggr]  \biggl[216\,{r}^{ 8}{a_1}^{2}{a_2}^{12}\epsilon -108\,{r}^{8}{a_1}^{2}{a_2}^{12}+440\,{r}^{6}{R}^{2}{a_1}^ {2}{a_2}^{10}\nonumber\\
&&-880{r}^{6}{R}^{2}{a_1}^{2}{a_2}^{10} \epsilon-676{r}^{4}{R}^{4}{a_1}^{2}{a_2}^{8}+1352{r}^{4} {R}^{4}{a_1}^{2}{a_2}^{8}\epsilon-168{r}^{6}{R}^{2}a_1{a_2}^{8}a_0\epsilon+108\,{r}^{6}{R}^{2}a_1{a_2}^{8}a_0+464\,{r}^{2}{R}^{6}{a_1}^{2}{a_2}^{6}\nonumber\\
&&- 928\,{r}^{2}{R}^{6}{a_1}^{2}{a_2}^{6}\epsilon+520\,{r}^{4}{R }^{4}a_1\,{a_2}^{6}a_0\,\epsilon-332\,{r}^{4}{R}^{4}a_1\,{a_2}^{6}a_0-120\,{R}^{8}{a_2}^{4}{a_1}^{ 2}+240\,{R}^{8}{a_2}^{4}{a_1}^{2}\epsilon-544\,{R}^{6}{a_2}^{4}{r}^{2}a_0\,a_1\,\epsilon\nonumber\\
&&+344\,{R}^{6}{a_2}^{4} {r}^{2}a_0\,a_1-27\,{R}^{4}{a_2}^{4}{r}^{4}{a_0}^{ 2}+18\,{R}^{4}{a_2}^{4}{r}^{4}{a_0}^{2}\epsilon-120\,a_0\,{R}^{8}{a_2}^{2}a_1+192\,a_0\,{R}^{8}{a_2}^{2}a_1\,\epsilon+56\,{a_0}^{2}{R}^{6}{a_2}^{2}{r}^{2}
\nonumber\\
&&-40\,{ a_0}^{2}{R}^{6}{a_2}^{2}{r}^{2}\epsilon-30\,{a_0}^{2}{R }^{8}+24\,{a_0}^{2}{R}^{8}\epsilon \biggr] \biggr\}^{-1}\,,
\end{eqnarray}
and
\begin{eqnarray}\label{aic2}
\Gamma_t&=&-\biggl\{ \biggl[108\,{r}^{8 }{a_1}^{2}{a_2}^{12}\epsilon-24\,{r}^{8}{a_1}^{2}{a_2}^{12}+96\,{r}^{6}{R}^{2}{a_1}^{2 }{a_2}^{10}-440\,{r}^{6}{R}^{2}{a_1}^{2}{a_2}^{10} \epsilon+676\,{r}^{4}{R}^{4}{a_1}^{2}{a_2}^{8}\epsilon-144\, {r}^{4}{R}^{4}{a_1}^{2}{a_2}^{8}\nonumber\\
&&-84\,{r}^{6}{R}^{2}a_1 \,{a_2}^{8}a_0\,\epsilon+16\,{r}^{6}{R}^{2}a_1\,{a_2}^{8}a_0-464\,{r}^{2}{R}^{6}{a_1}^{2}{a_2}^{6} \epsilon+96\,{r}^{2}{R}^{6}{a_1}^{2}{a_2}^{6}+260\,{r}^{4}{R }^{4}a_1\,{a_2}^{6}a_0\,\epsilon-48\,{r}^{4}{R}^{4}a_1\,{a_2}^{6}a_0\nonumber\\
&&-24{R}^{8}{a_2}^{4}{a_1}^{2 }+120{R}^{8}{a_2}^{4}{a_1}^{2}\epsilon-272{R}^{6}{a_2}^{4}{r}^{2}a_0a_1\,\epsilon+48{R}^{6}{a_2}^{4}{ r}^{2}a_0a_1+9\,{R}^{4}{a_2}^{4}{r}^{4}{a_0}^{2} \epsilon-16a_0\,{R}^{8}{a_2}^{2}a_1+12\,{a_0}^{2}{R}^{8}\epsilon \nonumber\\
&&+96\,a_0\,{R} ^{8}{a_2}^{2}a_1\,\epsilon-20\,{a_0}^{2}{R}^{6}{a_2}^{2}{r}^{2}\epsilon\biggr] \biggl[ 10\,a_1\,{r}^{8}{a_2}^{10}-42\,{r}^{6}{R}^{2}a_1\,{a_2}^{8}-9\,{R}^{2}{r}^{6}{a_2}^{6}a_0+68\,{r}^{4}{ R}^{4}a_1\,{a_2}^{6}+28\,{R}^{4}{r}^{4}{a_2}^{4}a_0\nonumber\\
&&-52{R}^{6}{r}^{2}{a_2}^{4}a_1-30{R}^{6}{a_2}^{2}{r }^{2}a_0+16{R}^{8}{a_2}^{2}a_1+12a_0\,{R}^{8} \biggr] \biggr\}\biggl\{ \biggl[ 18a_1 {r}^{8}{a_2}^{10}\epsilon-4a_1\,{r}^{8}{a_2}^{10}-74{r}^{6}{R}^{2}a_1{a_2}^ {8}\epsilon+16\,{r}^{6}{R}^{2}a_1\,{a_2}^{8}\nonumber\\
&&+116\,{R}^{4}{r} ^{4}{a_2}^{6}a_1\,\epsilon-24\,{r}^{4}{R}^{4}a_1\,{a_2}^{6}-3\,{R}^{2}{r}^{6}{a_2}^{6}a_0\,\epsilon-84\,{R} ^{6}{r}^{2}{a_2}^{4}a_1\,\epsilon+16\,{R}^{6}{r}^{2}{a_2}^{4}a_1+10\,{R}^{4}{r}^{4}{a_2}^{4}a_0\,\epsilon-4\,{ R}^{8}{a_2}^{2}a_1\nonumber\\
&&+24\,{R}^{8}{a_2}^{2}a_1\, \epsilon-12\,{R}^{6}{a_2}^{2}{r}^{2}a_0\,\epsilon+6\, \epsilon\,{R}^{8}a_0 \biggr]  \biggl[ 216\,{r}^{8}{a_1}^{2}{a_2}^{12}\epsilon-108\,{r}^{8}{a_1}^{2}{ a_2}^{12}+440\, {r}^{6}{R}^{2}{a_1}^{2}{a_2}^{10}-880\,{r}^{6}{R}^{2}{a_1}^{2}{a_2}^{10}\epsilon\nonumber\\
&&-676\,{r}^{4}{R}^{4}{a_1}^{2}{a_2}^{8}+1352\,{r}^{4}{R}^{4}{a_1}^{2}{a_2}^{8}\epsilon- 168\,{r}^{6}{R}^{2}a_1\,{a_2}^{8}a_0\,\epsilon+108\,{r} ^{6}{R}^{2}a_1\,{a_2}^{8}a_0+464\,{r}^{2}{R}^{6}{a_1}^{2}{a_2}^{6}-928\,{r}^{2}{R}^{6}{a_1}^{2}{a_2}^{6} \epsilon\nonumber\\
&&+520\,{r}^{4}{R}^{4}a_1\,{a_2}^{6}a_0\,\epsilon -332\,{r}^{4}{R}^{4}a_1\,{a_2}^{6}a_0-120\,{R}^{8}{a_2}^{4}{a_1}^{2}+240\,{R}^{8}{a_2}^{4}{a_1}^{2} \epsilon-544\,{R}^{6}{a_2}^{4}{r}^{2}a_0\,a_1\,\epsilon +344\,{R}^{6}{a_2}^{4}{r}^{2}a_0\,a_1\nonumber\\
&&-27\,{R}^{4}{a_2}^{4}{r}^{4}{a_0}^{2}+18\,{R}^{4}{a_2}^{4}{r}^{4}{a_0}^{2}\epsilon-120\,a_0\,{R}^{8}{a_2}^{2}a_1+192\,a_0\,{R}^{8}{a_2}^{2}a_1\,\epsilon+56\,{a_0}^{2}{R}^{6} {a_2}^{2}{r}^{2}-40\,{a_0}^{2}{R}^{6}{a_2}^{2}{r}^{2} \epsilon\nonumber\\
&&-30\,{a_0}^{2}{R}^{8}+24\,{a_0}^{2}{R}^{8}\epsilon \biggr] \biggr\}\,.
\end{eqnarray}

In Figs. \ref{Fig:8}\subref{fig:gamma}--\subref{fig:Gammat}, we plot the profiles of the adiabatic indices $\gamma$, $\Gamma_r$ and  $\Gamma_t$ of the pulsar \textit{Her X-1}, respectively. It is clear that the value of the adiabatic index $\gamma$ as predicted by RT is less than the corresponding GR value everywhere within the pulsar. On the other hand, the other adiabatic indices $\Gamma_r$ and $\Gamma_t$ are higher in RT than GR, whereas the stability constraints, namely $\Gamma_r>\gamma$ and $\Gamma_t>\gamma$, as shown in the figures are clearly fulfilled.
%%%%%%%%%%%%%%%%%%%%%%% Section 8 %%%%%%%%%%%%%%%%%%%%%%%%%%%%%%%%%
\section{More Observational Constraints}\label{S8}
%%%%%%%%%%%%%%%%%%%%%%%%%%%%%%%%%%%%%%%%%%%%%%%%%%%%%%%%%%%%%%%%%%%
In this section we confront the present model with other pulsars' data to examine its validity with wide range of astrophysical observations. Also, we plot the mass-radius profile corresponding to different choices of the surface density compatible with the nuclear saturation density as boundary conditions showing the capability of the model to predict masses in the lower mass gap $2.5-5 M_\odot$. On the other hand, we confront the compactness parameter as predicted by the model with Buchdahl compactness bound.
%%%%%%%%%%%%%%%%%%%%%%%%%%%%%%%%%%%%%%
\subsection{Pulsars' data}
%%%%%%%%%%%%%%%%%%%%%%%%%%%%%%%%%%%%%%
In addition to $\textit{Her X-1}$, a similar analysis is developed for other twenty one pulsars cover a wide range of masses from $0.8 M_\odot$ to heavy pulsars of mass $2.01 M_\odot$. In Table \ref{Table1}, we give the observed masses and radii for each associated with the corresponding model parameters \{$a_0$, $a_1$, $a_2$\} assuming that the Rastall parameter $\epsilon=-0.1$. We confirm that the model at hand can predict masses of those pulsars compatible with their observed values. In Table \ref{Table2}, we give the calculated values of physical quantities of most interest. As seen the values of the density are consistent with the nuclear density. It is to be mentioned that there are no EoSs are assumed, the obtained results however fit well with the linear behaviour, whereas the slopes $dp_r/d\rho$ and $dp_t/d\rho$ of the best fit lines are consistent with the values as obtained in Table \ref{Table2}. On the other hand these values confirm that the model is stable and causal for each case and satisfies all the physical constraints in Sec. \ref{S4}. We additionally give the surface redshift values as predicted by the present model for the listed pulsars. All values are clearly in agreement with the upper bound constraint $Z_{R}\leq 2$ as given by Buchdahl \cite{PhysRev.116.1027}, also for anisotropic spheres see \cite{Ivanov:2002xf,PhysRevD.67.064003}.
\begin{table*}
\caption{Observed mass-radius of twenty pulsars and the corresponding model parameters ($\epsilon=-0.1$).}
\label{Table1}
\begin{tabular*}{\textwidth}{@{\extracolsep{\fill}}llcccccc@{}}
\hline
%Equil. & \multicolumn{1}{c}{$x$} & \multicolumn{1}{c}{$y$} & \multicolumn{1}{c}{$z$} & \multicolumn{1}{c}{$C$} & S \\
%Points \\
{{Pulsar}}   &Ref.& observed mass ($M_{\odot}$) &  observed radius [{km}]& estimated mass ($M_{\odot}$) &  {$a_0$}    & {$a_1$}     & {$a_2$}   \\
\hline
Her X-1 &\cite{Abubekerov_2008}         &  $0.85\pm 0.15$    &  $8.1\pm 0.41$   &$0.905$&  $0.369$    & $-0.622$    & $0.298$     \\
RX J185635-3754 &\cite{2002ApJ...564..981P}     &  $0.9\pm 0.2$      &  $6$             &$0.949$&  $0.517$    & $-0.706$    & $0.369$     \\
LMC X-4 &\cite{Rawls:2011jw}         &  $1.04\pm 0.09$    &  $8.301\pm 0.2$  &$1.103$&  $0.375$    & $-0.658$    & $0.330$     \\
GW170817-2  &\cite{LIGOScientific:2018cki}     &  $1.27\pm 0.09$    &  $11.9\pm 1.4$   &$1.351$&  $0.437$    & $-0.625$    & $0.301$     \\
EXO 1785-248 &\cite{Ozel:2008kb}    &  $1.3\pm 0.2$      &  $8.849\pm 0.4$  &$1.372$&  $0.507$    & $-0.699$    & $0.364$     \\
PSR J0740+6620 &\cite{Raaijmakers:2019qny}  &  $1.34\pm 0.16$    &  $12.71\pm 1.19$ &$1.426$&  $0.370$    & $-0.623$    & $0.298$     \\
M13 &\cite{Webb:2007tc}            &  $1.38\pm 0.2$     &  $9.95\pm 0.27$  &$1.459$&  $0.481$    & $-0.683$    & $0.351$     \\
LIGO    &\cite{LIGOScientific:2020zkf}         &  $1.4$             &  $12.9\pm 0.8$   &$1.489$&  $0.381$    & $-0.628$    & $0.303$     \\
X7  &\cite{Rybicki:2005id}             &  $1.4$             &  $14.5\pm 1.8$   &$1.492$&  $0.340$    & $-0.607$    & $0.284$     \\
PSR J0037-4715 &\cite{Reardon:2015kba}   &  $1.44\pm 0.07$    &  $13.6\pm 0.9$   &$1.532$&  $0.372$    & $-0.623$    & $0.299$     \\
PSR J0740+6620 &\cite{Miller:2019cac}  &  $1.44\pm 0.16$    &  $13.02\pm 1.24$ &$1.531$&  $0.388$    & $-0.632$    & $0.307$     \\
GW170817-1  &\cite{LIGOScientific:2018cki}     &  $1.45\pm 0.09$    &  $11.9\pm 1.4$   &$1.539$&  $0.425$    & $-0.652$    & $0.325$     \\
4U 1820-30 &\cite{G_ver_2010}      &  $1.46\pm 0.2$     &  $11.1\pm 1.8$   &$1.546$&  $0.457$    & $-0.670$    & $0.440$     \\
Cen X-3 &\cite{Naik:2011qc}         &  $1.49\pm 0.49$    &  $9.178\pm 0.13$ &$1.566$&  $0.556$    & $-0.731$    & $0.388$     \\
4U 1608-52  &\cite{1996IAUC.6331....1M}     &  $1.57\pm 0.3$     &  $9.8\pm 1.8$    &$1.651$&  $0.550$    & $-0.727$    & $0.385$     \\
KS 1731-260 &\cite{Ozel:2008kb}     &  $1.61\pm 0.37$    &  $10\pm 2.2$     &$1.692$&  $0.552$    & $-0.728$    & $0.386$     \\
EXO 1745-268  &\cite{Ozel:2008kb}   &  $1.65\pm 0.25$    &  $10.5\pm 1.8$   &$1.736$&  $0.540$    & $-0.720$    & $0.380$     \\
Vela X-1 &\cite{Rawls:2011jw}        &  $1.77\pm 0.08$    &  $9.56\pm 0.08$  &$1.845$&  $0.627$    & $-0.781$    & $0.424$     \\
4U 1724-207 &\cite{Ozel:2008kb}     &  $1.81\pm 0.27$    &  $12.2\pm 1.4$   &$1.909$&  $0.512$    & $-0.702$    & $0.366$     \\
SAX J1748.9-2021 &\cite{Ozel:2008kb} &  $1.81\pm 0.3$     &  $11.7\pm 1.7$   &$1.906$&  $0.532$    & $-0.715$    & $0.376$     \\
PSR J1614-2230\footnote{We note that the estimated mass for massive pulsars slightly exceeds the observational value which would impose more strict constraints on Rastall parameter to be $\epsilon=0.06$.} &\cite{Demorest:2010bx}  &  $1.97\pm 0.04$    &  $13\pm 2$       &$2.076$&  $0.522$    & $-0.709$    & $0.371$     \\
PSR J0348+0432 &\cite{Antoniadis:2013pzd}   &  $2.01\pm 0.04$    &  $13\pm 2$       &$2.117$&  $0.532$    & $-0.715$    & $0.376$     \\
\hline
\end{tabular*}
\end{table*}
\begin{table*}
\caption{Calculated physical quantities of the most interest.}
\label{Table2}
\begin{tabular*}{\textwidth}{@{\extracolsep{\fill}}lcccccccccc@{}}
\hline
%Equil. & \multicolumn{1}{c}{$x$} & \multicolumn{1}{c}{$y$} & \multicolumn{1}{c}{$z$} & \multicolumn{1}{c}{$C$} & S \\
%Points \\
{{Pulsar}}                              &{$\rho(0)$} &      {$\rho_R$} &   \multicolumn{1}{c}{$v_r^2(0)/c^2$}  &    \multicolumn{1}{c}{$v_r^2(R)/c^2$}  & \multicolumn{1}{c}{$v_t^2(0)/c^2$} & \multicolumn{1}{c}{$v_t^2(R)/c^2$}  &  {$\rho c^2-p_r-2p_t|_0$}&{$\rho-p_r-2p_t|_R$}& \multicolumn{1}{c}{$Z_R$}\\
                                        &{[$g/cm^3$]}&     {[$g/cm^3$]} &   {}                &    {}                & {}               & {}                &  {[$Pa$]}                &{[$Pa$]}            & {}\\
\hline
Her X-1            &9.18$\times10^{14}$     &7.39$\times10^{14}$  &  0.445   &   0.376     &  0.246 & 0.195 & 6.27$\times10^{34}$ & 5.73$\times10^{34}$ & 0.204  \\
RX J185635-3754       &2.53$\times10^{15}$     &1.81$\times10^{15}$  &  0.608   &   0.451     &  0.402 & 0.276 & 1.25$\times10^{35}$ & 1.26$\times10^{35}$ & 0.340  \\
LMC X-4            &1.07$\times10^{15}$     &8.19$\times10^{14}$  &  0.506   &   0.406     &  0.304 & 0.228 & 6.53$\times10^{34}$ & 6.08$\times10^{34}$ & 0.260  \\
GW170817-2            &4.33$\times10^{14}$     &3.48$\times10^{14}$  &  0.450   &   0.379     &  0.250 & 0.198 & 2.93$\times10^{34}$ & 2.68$\times10^{34}$ & 0.208  \\
EXO 1785-248       &1.13$\times10^{15}$     &8.21$\times10^{14}$  &  0.593   &   0.444     &  0.388 & 0.269 & 5.80$\times10^{34}$ & 5.76$\times10^{34}$ & 0.329  \\
PSR J0740+6620       &3.75$\times10^{14}$     &3.01$\times10^{14}$  &  0.446   &   0.377     &  0.247 & 0.196 & 2.55$\times10^{34}$ & 2.33$\times10^{34}$ & 0.205  \\
M13       &8.39$\times10^{14}$     &6.20$\times10^{14}$  &  0.556   &   0.429     &  0.353 & 0.253 & 4.63$\times10^{34}$ & 4.45$\times10^{34}$ & 0.302  \\
LIGO       &3.76$\times10^{14}$     &3.00$\times10^{14}$  &  0.454   &   0.381     &  0.255 & 0.201 & 2.52$\times10^{34}$ & 2.31$\times10^{34}$ & 0.213  \\
X7       &2.61$\times10^{14}$     &2.14$\times10^{14}$  &  0.424   &   0.366     &  0.226 & 0.183 & 1.85$\times10^{34}$ & 1.69$\times10^{34}$ & 0.183  \\
PSR J0037-4715       &3.29$\times10^{14}$     &2.64$\times10^{14}$  &  0.447   &   0.378     &  0.248 & 0.197 & 2.24$\times10^{34}$ & 2.04$\times10^{34}$ & 0.206  \\
PSR J0740+6620       &3.77$\times10^{14}$     &2.99$\times10^{14}$  &  0.460   &   0.384     &  0.260 & 0.204 & 2.50$\times10^{34}$ & 2.29$\times10^{34}$ & 0.219  \\
GW170817-1       &5.04$\times10^{14}$     &3.89$\times10^{14}$  &  0.494   &   0.401     &  0.293 & 0.222 & 3.14$\times10^{34}$ & 2.91$\times10^{34}$ & 0.250  \\
4U 1820-30         &6.33$\times10^{14}$     &4.77$\times10^{14}$  &  0.528   &   0.416     &  0.325 & 0.239 & 3.70$\times10^{34}$ & 3.49$\times10^{34}$ & 0.279  \\
Cen X-3            &1.19$\times10^{15}$     &8.24$\times10^{14}$  &  0.677   &   0.477     &  0.469 & 0.305 & 4.98$\times10^{34}$ & 5.51$\times10^{34}$ & 0.386  \\
4U 1608-52         &1.03$\times10^{15}$     &7.15$\times10^{14}$  &  0.664   &   0.472     &  0.456 & 0.300 & 4.45$\times10^{34}$ & 4.82$\times10^{34}$ & 0.378  \\
KS 1731-260         &9.92$\times10^{14}$     &6.89$\times10^{14}$  &  0.669   &   0.474     &  0.461 & 0.302 & 4.24$\times10^{34}$ & 4.63$\times10^{34}$ & 0.381  \\
EXO 1745-268         &8.74$\times10^{14}$     &6.14$\times10^{14}$  &  0.646   &   0.466     &  0.439 & 0.293 & 3.95$\times10^{34}$ & 4.18$\times10^{34}$ & 0.366  \\
Vela X-1           &1.29$\times10^{15}$     &8.33$\times10^{14}$  &  0.861   &   0.538     &  0.645 & 0.371 & 3.01$\times10^{34}$ & 5.11$\times10^{34}$ & 0.486  \\
4U 1724-207      &6.04$\times10^{14}$     &4.35$\times10^{14}$  &  0.600   &   0.447     &  0.394 & 0.273 & 3.04$\times10^{34}$ & 3.04$\times10^{34}$ & 0.334  \\
SAX J1748.9-2021  &6.91$\times10^{14}$     &4.89$\times10^{14}$  &  0.633   &   0.460     &  0.426 & 0.287 & 3.22$\times10^{34}$ & 3.35$\times10^{34}$ & 0.357  \\
PSR J1614-2230     &5.46$\times10^{14}$     &3.90$\times10^{14}$  &  0.616   &   0.454     &  0.410 & 0.280 & 2.65$\times10^{34}$ & 2.70$\times10^{34}$ & 0.346  \\
PSR J0348+0432     &5.59$\times10^{14}$     &3.96$\times10^{14}$  &  0.632   &   0.460     &  0.425 & 0.287 & 2.61$\times10^{34}$ & 2.71$\times10^{34}$ & 0.357 \\
\hline
\end{tabular*}
\end{table*}
%
%%%%%%%%%%%%%%%%%%%%%%%%%%%%%%%%%%%%%%%%%%%%%%%%%%%%%%%
\subsection{Mass-Radius Profile}\label{Sec:MR-relation}
\begin{figure}
\centering
%\subfigure[~$\gamma$   in cases $\epsilon=0$ and  $\epsilon\neq 0$ ]{\label{fig:gamma}\includegraphics[scale=0.25]{Fig8a.eps}}
\subfigure[~Compactness-Radius ]{\label{fig:MR1}\includegraphics[scale=0.26]{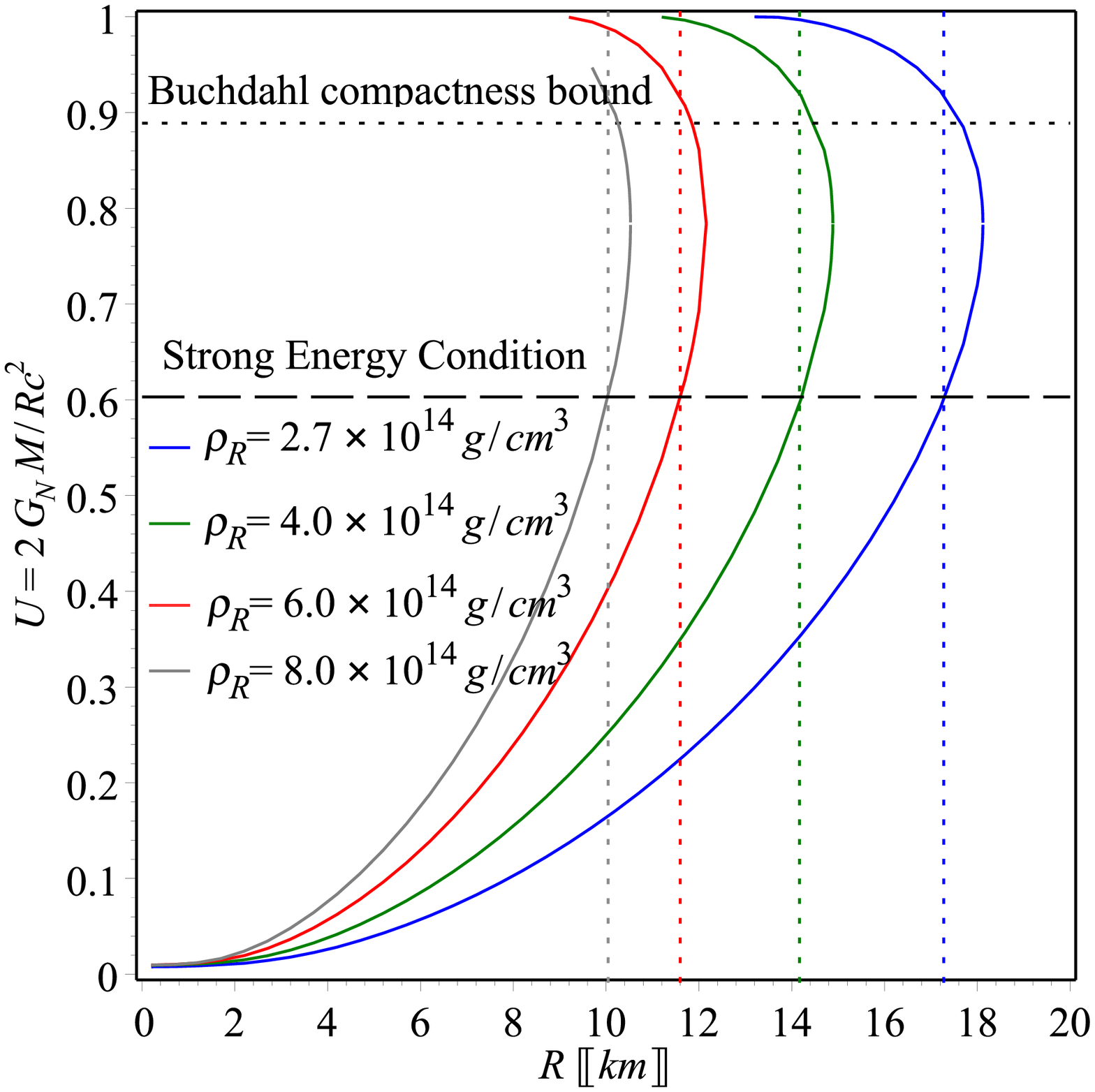}}\hspace{0.5cm}
\subfigure[~Mass-Radius ]{\label{fig:MR2}\includegraphics[scale=0.26]{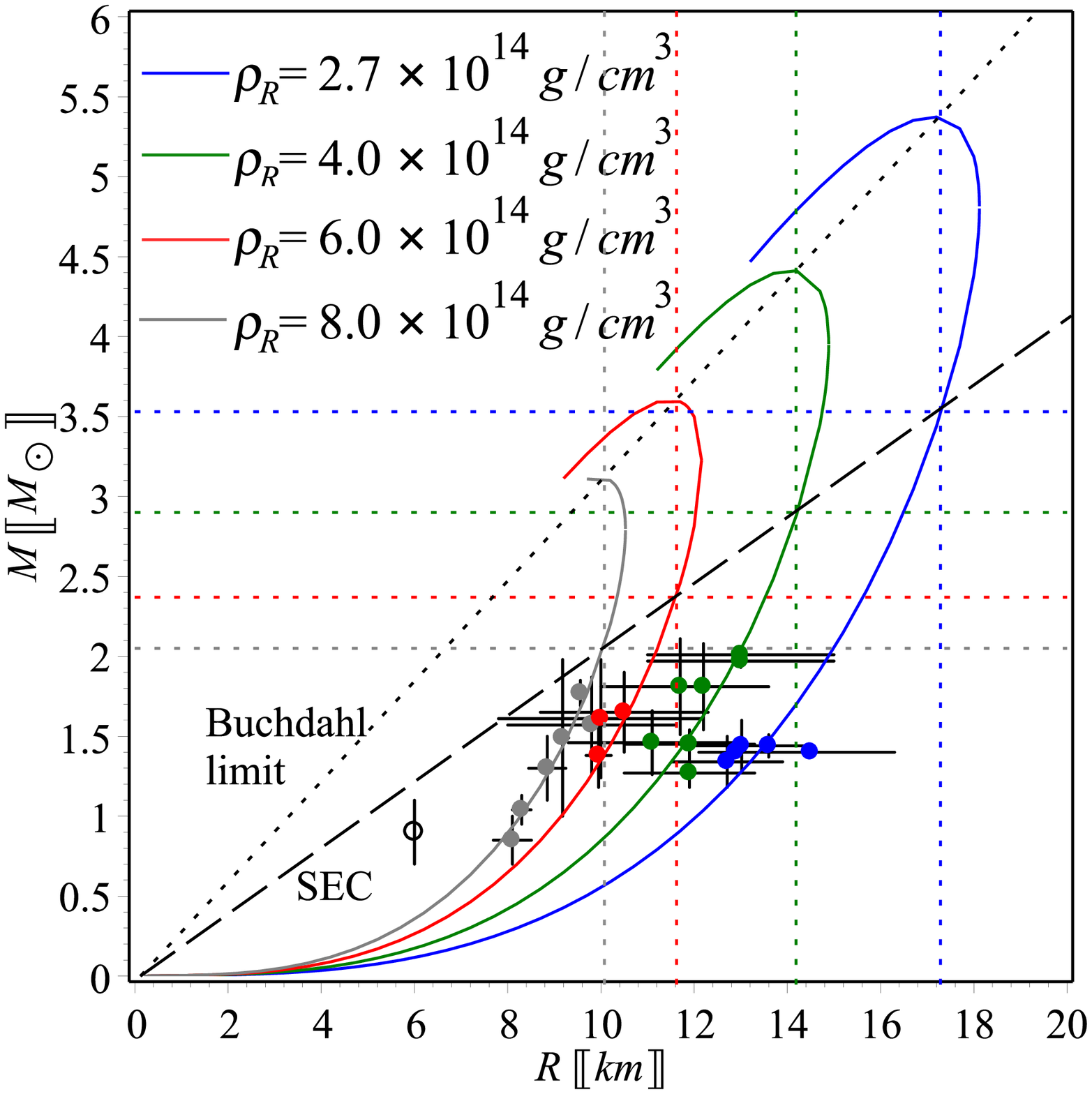}}\hspace{0.5cm}
\subfigure[~Neutron star core ]{\label{fig:MR3}\includegraphics[scale=0.26]{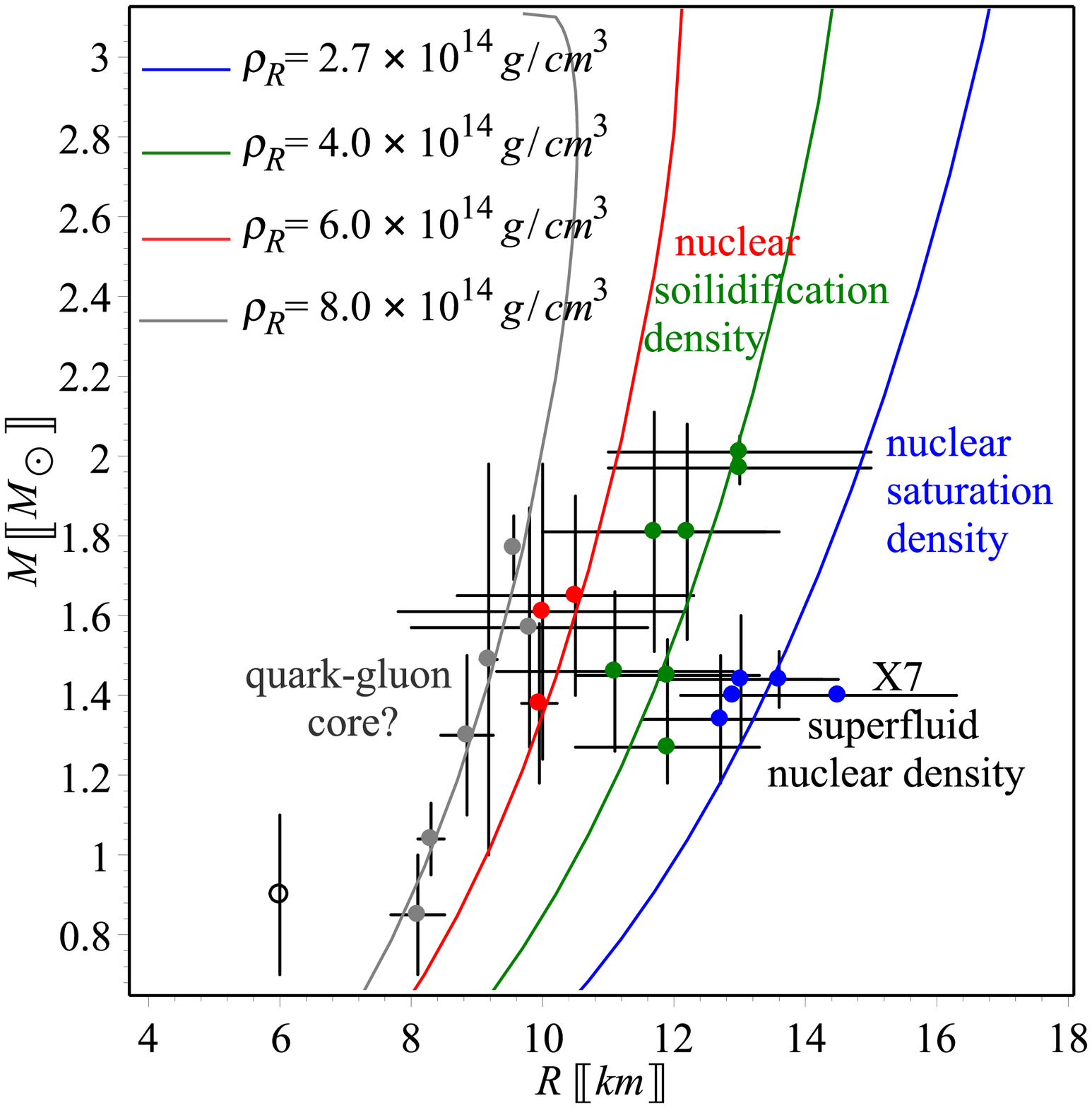}}
\caption[figtopcap]{\small{\subref{fig:MR1} Compactness-radius profiles for four surface densities, the horizontal dot and dash lines visualize Buchdahl ($U=8/9$) and the SEC bound ($U(\epsilon=-0.1)=0.603$) on the compactness parameter. Clearly both constraints give almost the same maximum radii. \subref{fig:MR2} Mass-radius profiles for four surface densities combined with observed mass-radius values of the pulsars in Table \ref{Table1}. The diagonal dot and dash lines set Buchdahl and SEC physical regions. Clearly all pulsars are below the SEC exclusion limit. The horizontal dot lines give the maximum possible mass as obtained by the SEC. \subref{fig:MR3} Pulsars on the red, green and blue mass-radius profiles are suggested to have neutron cores whereas the surface densities match superfluid, saturated, solidified nuclear densities. The pulsars on the gray mass-radius profile match perfectly a surface density boundary condition $\rho_R=8\times 10^{14}$ g/cm$^3$ which may suggest quark-gluon cores for those pulsars.}}
\label{Fig:9}
\end{figure}

%%%%%%%%%%%%%%%%%%%%%%%%%%%%%%%%%%%%%%%%%%%%%%%%%%%%%%%
As is shown in Table \ref{Table2} the surface densities of the listed pulsars, $2.14\times 10^{14}\lesssim \rho_R \lesssim 1.81\times 10^{15}$ g/cm$^{3}$, are mostly compatible with a neutron core. For four different values of the surface density of the pulsars $\rho_R=2.7\times 10^{14}$ g/cm$^{3}$, $4\times 10^{14}$ g/cm$^{3}$, $6\times 10^{14}$ g/cm$^{3}$ and $8\times 10^{14}$ g/cm$^{3}$ we plot the corresponding compactness-radius curve as in Fig. \ref{Fig:9}\subref{fig:MR1}. In all cases the maximum compactness values do not exceed unity. However for a compact object to be stable it should satisfy Buchdahl compactness bound $U=\frac{2G_N M}{c^2 R}\leq 8/9$ (for isotropic sphere \cite{PhysRev.116.1027}). We visualize Buchdahl upper bound on the compactness parameter with the corresponding maximum radii as obtained for the four surface densities. It is convenient to give the model parameters \{$a_0$, $a_1$, {\color{red} $a_2$}\} in terms of the total compactness parameter $U$. Recalling the matching conditions \eqref{Eq2} and \eqref{Eq3} we write
\begin{eqnarray}
a_0&=&\small{3\left\{\left[(U-\frac{8}{9})(1-U)^{\frac{1}{4}}+\frac{8}{9}(1-U)\right]\epsilon-\frac{1}{18}(1-U)^{\frac{1}{4}}U\right\}\sqrt{-(1-U)^2 (U-2+4(1-U)^{\frac{3}{4}}-6\sqrt{1-U}+4(1-U)^{\frac{1}{4}})} \over \left[-\frac{1}{2}(1-U)^{\frac{1}{4}}+\sqrt{1-U}-\frac{1}{2}(1-U)^{\frac{3}{4}}\right](\epsilon-\frac{1}{3})(U-1)},\nonumber\\[5pt]
a_1&=&-\frac{1}{2}{\left(\sqrt{1-U}-(1-U)^{\frac{3}{4}}\right)a_0 +4\sqrt{-(1-U)^2 (U-2+4(1-U)^{\frac{3}{4}}-6\sqrt{1-U}+4(1-U)^{\frac{1}{4}})} \over U-\sqrt{1-U}+2(1-U)^{\frac{3}{4}}-1},\nonumber\\[5pt]
a_2&=&\sqrt{1-(1-U)^{1/4}}.
\end{eqnarray}
Notably for the pulsar \textit{Her X-1} with $\epsilon=-0.1$, the compactness parameter $U=0.31$, which reproduces the set of constants $a_0 \approx 0.369$, $a_1\approx -0.622$ and $a_2\approx 0.298$ as previously obtained in Sec. \ref{S6}. Using Buchdahl limit $U=8/9$, we obtain the maximum radii $R=17.66$ km for $\rho_R=2.7\times 10^{14}$ g/cm$^{3}$, $R=14.51$ km for $\rho_R=4\times 10^{14}$ g/cm$^{3}$, $R=11.85$ km for $\rho_R=6\times 10^{14}$ g/cm$^{3}$ and $R=10.26$ km for $\rho_R=8\times 10^{14}$ g/cm$^{3}$. However, more restricted values on the maximum compactness can be imposed by utilizing the SEC as obtained in the present model. Thus we write the density, radial and tangential pressures, namely \eqref{sol}, in terms of the compactness parameter $U$, after straight forward calculations we find that the SEC is satisfied when the compactness $U\leq 0.603$ (using $\epsilon=-0.1$). Higher $U$ values can be obtained for higher values of $\epsilon$, e.g. $U(\epsilon=-0.05) \lesssim 0.639$. However, we restrict our discussion to $\epsilon=-0.1$. We visualize the SEC constraint on the upper compactness value as in Fig. \ref{Fig:9}\subref{fig:MR1}. It can be seen that the lower $U$ value as obtained by the SEC gives almost the same maximum radii as Buchdahl compactness bound or slightly lesser. However, it puts more restricted values on the maximum masses as will be discussed shortly. For a compact object to be physically stable we use $U\leq 0.603$ as obtained by the SEC. This determines the maximum radii as allowed by the present model as $R=17.30$ km for $\rho_R=2.7\times 10^{14}$ g/cm$^{3}$, $R=14.22$ km for $\rho_R=4\times 10^{14}$ g/cm$^{3}$, $R=11.61$ km for $\rho_R=6\times 10^{14}$ g/cm$^{3}$ and $R=10.05$ km for $\rho_R=8\times 10^{14}$ g/cm$^{3}$.

In Fig. \ref{Fig:9}\subref{fig:MR2}, we plot the mass-radius relation as obtained by the present model for the four surface densities along with the scatter plot of the observed mass-radius of the pulsars listed in Table \ref{Table1}. Consequently, we determine the maximum masses allowed by the present model according to Buchdahl limit for the surface densities $\rho_R=2.7\times 10^{14}$ g/cm$^{3}$, $4\times 10^{14}$ g/cm$^{3}$, $6\times 10^{14}$ g/cm$^{3}$ and $8\times 10^{14}$ g/cm$^{3}$, which have been found to be $5.31 M_\odot$, $4.37 M_\odot$, $3.56 M_\odot$ and $3.09 M_\odot$ respectively. In comparison with the GR case as obtained by Das et al. \cite{Das:2019dkn}, RT allows more masses. Indeed these values provide unstable models, since they do not satisfy the SEC. We use the maximum compactness as obtained by the SEC, $U=0.603$, which determines the physically stable pulsars with maximum masses $3.53 M_\odot$, $2.90 M_\odot$, $2.37 M_\odot$ and $2.05 M_\odot$. Obviously from Fig. \ref{Fig:9}\subref{fig:MR2} all pulsars are below the SEC exclusion limit.

In Fig. \ref{Fig:9}\subref{fig:MR3}, we split the pulsars into four groups according their matching with the mass-radius profiles of the surface densities. Notably, the nuclear saturation density $\rho_\text{sat} = 2.7 \times 10^{14}$ g/cm$^3$ and solidification density $\rho_\text{solid}$ is expected to be from $2.8 \times 10^{14}$ g/cm$^3$ to $5 \times 10^{14}$ g/cm$^3$ whereas the maximum density of cold matter $\rho_\text{max} \lesssim 4\times 10^{15}$ g/cm$^3$ based on an EoS-independent analytic solutions of Einstein field equations. In the present work, we find that the pulsar X7 has a neutron core with density less than $\rho_\text{sat}$ which matches anisotropic superfluid nuclear density $\sim 1.5 \times 10^{14}$ g/cm$^3$. The pulsars PSR J0030+0451, PSR J0037-4715 and the LIGO constraints have a neutron core with density matches $\rho_\text{sat}$. The pulsars M13, KS 1731-260, EXO 1745-268, GW170817-2, GW170817-1, 4U 1820-30, 4U 1724-207 and SAX J1748.9-2021 have a neutron core with density $\sim \rho_\text{solid}$. Notably, the most heaviest pulsars PSR J1614-2230 and PSR J0348+0432 fit well with the mass-radius curve corresponding to surface density $4 \times 10^{14}$ g/cm$^3$ which suggests that these pulsars are having solidified neutron cores. Similarly, the pulsars Her X-1, LMC X-4, EXO 1785-248, Cen X-3, 4U 1608-52 and Vela X-1 lie beyond neutron core density and match perfectly with mass-radius curve with a surface density of $\rho_{R} = 8 \times 10^{14}$ g/cm$^3$, this suggests that those pulsars may have quark-gluon cores. This claim should be verified by gravitational wave signals and by experimental tests. It is to be mentioned that the four categories presented here fit well with linear EoS as is shown earlier in Fig. \ref{Fig:5} whereas their slopes are given in Table \ref{Table2}.

Interestingly, the obtained curves allow for a heavy neutron star to lie in the lower mass gap $2.5-5 M_\odot$ which keeps an open window for the binary secondary companion GW190814 \cite{LIGOScientific:2020zkf} to be a neutron star with an anisotropic core whereas the EoS $p_r(\rho)$ and $p_t(\rho)$ are obtained by simple linear relations. Clearly RT (with $\epsilon < 0$), using Buchdahl limit, allows for higher maximum masses than the corresponding results as obtained by Das et al. within the GR framework \cite{Das:2019dkn}. This result is consistent with our earlier conclusion after Fig. \ref{Fig:6}\subref{fig:mass} and also in agreement with the previous study of isotropic neutron stars in RT by Oliveira et al. \cite{Oliveira:2015lka}. However, in the present work we investigate anisotropic compact objects without assuming any EoS. In the following section we summarize and conclude the present work.
%%%%%%%%%%%%%%%%%%%%%%%%%%%%%%%%%%%%%%%%%
\section{Summary and conclusions}\label{S9}
%%%%%%%%%%%%%%%%%%%%%%%%%%%%%%%%%%%%%%%%%
One way to modify GR is to drop one of its main assumptions. Rastall questioned the local conservation in presence of curved spacetime, which led him to a set of field equations consistent with a different assumption, that is a non-minimal coupling between matter and geometry whereas the energy-momentum tensor $\mathcal{T}{^\alpha}{_{\beta;\alpha}}\propto \mathcal{R}_{;\beta}$. Vissar, recently, claimed that RT is completely equivalent to GR \cite{Visser:2017gpz}. On the contrary, Darabi et al. investigated Visser's claim but they concluded that Visser misinterpreted the matter-geometry coupling term, and consequently led him to a wrong conclusion \cite{Darabi:2017coc}. In addition, they showed that by applying Visser's approach to $f(R)$ theory one may conclude that it is equivalent to GR as well, which is not true. One of the good examples which may reveal the contribution of the matter-geometry coupling in RT in contrast to GR is the stellar models when the presence of matter play a crucial role. In this study, we applied RT to anisotropic static star with spherical symmetry seeking for a non-trivial interior solution. We showed that, for static spherically symmetric stellar model, the Rastall parameter in general does not contribute to the anisotropy parameter while it contributes non-trivially to the density and the radial/tangential pressures. Applying appropriate ansatz we derived the corresponding density and pressure forms.

We matched the interior solution to the exterior Schwarzschild one (which is a solution in RT as well) in the case of the pulsar $\textit {Her X-1 }$ ($M= 0.85 \pm 0.15\, M_{\odot},\, R= 8.1 \pm 0.41$ km) and by recalling Zeldovich constraint, we determine the numerical values of the set of constants \{$a_0$, $a_1$ and $a_2$\} for a particular choice of Rastall parameter which has been found as $\epsilon=-0.1$. We verified that the obtained model satisfies the physical constraints as listed in Sec. \ref{S4} and their predictions are slightly different from GR ones. We give detailed patterns of all physical quantities in Figs. \ref{Fig:1}--\ref{Fig:8}. In addition, we confronted the model at hand with other pulsars' observations and we confirm that all physical conditions are fulfilled in all cases. In Table \ref{Table1}, we give the numerical values of the set of constants for the same Rastall parameter value $\epsilon=-0.1$ of twenty additional pulsars as is done in the case of the pulsar $\textit {Her X-1 }$. Also, in Table \ref{Table2}, we give the corresponding physical quantities of the most interest for all pulsar at the present study. All results confirm that the model satisfies the physical conditions as listed in Sec. \ref{S4}.

For four surface densities, we utilized the SEC to set an upper limit of the compactness $U\sim 0.603$ (where $\epsilon=-0.1$) less than Buchdahl compactness $U=8/9$, see Fig. \ref{Fig:9}. Consequently we determine the maximum possible masses with a stable configuration. For a boundary density compatible with neutron star core at saturation nuclear density the maximum mass $\sim 3.53 M_\odot$. This results have been obtained without assuming a specific EoS, however, the pressure-density relation fits well with linear behaviour whereas the slopes keep a stable and causal structures. We split the twenty pulsars into four groups, three groups fit well with neutron cores with anisotropic superfluid, saturated and solidified cores. We found that the most massive pulsars PSR J1614-2230 and PSR J0348+0432 fit will with the mass-radius curve corresponding to surface density $4 \times 10^{14}$ g/cm$^3$ which suggests that those pulsars are having solidified neutron cores. Furthermore, we suggest that the following group of pulsars to have a quark-gluon cores, Her X-1, LMC X-4, EXO 1785-248, Cen X-3, 4U 1608-52 and Vela X-1. This group fit perfectly with a boundary density $8\times 10^{14}$ g/cm$^{3}$ lies beyond the neutron core and suggested to have quark-gluon core. Even for those of relatively small masses we find that corresponding radii are much smaller than what is expected for a typical size of neutron star $10$ km, which characterizes low-mass quark star \cite{Alaverdyan:2001num,Xu2005}. In our case as we showed that the additional Rastall force in TOV equation contributes to resize the compact object to be smaller than GR prediction for the same mass. This in return results in higher core density $\sim 10^{15}$ g/cm$^3$. Noting that the nuclear saturation density $\rho_\text{nuc} \approx 2.7-2.8 \times 10^{14}$ g/cm$^3$, while the typical solidification density of neutron matter $\rho_\text{solid}\gtrsim 4\times 10^{14}$ g/cm$^3$. 

To conclude: We investigated Rastall gravity, for an anisotropic star with a static spherical symmetry, whereas the matter-geometry coupling as assumed in RT is expected to play a crucial role differentiating RT predictions from GR ones. Indeed, all results confirm that RT is not equivalent to GR. However, Rastall parameter does not contribute to the anisotropy parameter, in general, i.e. RT predicts same amount of anisotropy as GR in the static spherically stellar models. Interestingly RT predicts more mass relative GR within same size which reflects the role of matter-geometry coupling to allow masses in the lower mass gap. This keeps the opportunity that the binary secondary companion GW190814 to be a neutron star with an anisotropic core whereas the EoS are obtained by simple linear relations.

%\begin{acknowledgements}
%If you'd like to thank anyone, place your comments here and remove the percent signs.
%\end{acknowledgements}

%\bibliography{Ref}   % name your BibTeX data base

\begin{thebibliography}{100}
\providecommand{\url}[1]{{#1}}
\providecommand{\urlprefix}{URL }
\expandafter\ifx\csname urlstyle\endcsname\relax
  \providecommand{\doi}[1]{DOI \discretionary{}{}{}#1}\else
  \providecommand{\doi}{DOI \discretionary{}{}{}\begingroup
  \urlstyle{rm}\Url}\fi

\bibitem{Schwarzschild:1916uq}
K.~Schwarzschild, Sitzungsber. Preuss. Akad. Wiss. Berlin (Math. Phys. )
  \textbf{1916}, 189 (1916)

\bibitem{Schwarzschild:1916ae}
K.~Schwarzschild, Sitzungsber. Preuss. Akad. Wiss. Berlin (Math. Phys. )
  \textbf{1916}, 424 (1916)

\bibitem{ruderman1972pulsars}
M.~Ruderman, Annual Review of Astronomy and Astrophysics \textbf{10}(1), 427
  (1972)

\bibitem{PhysRevD.65.104011}
B.V. Ivanov, Phys. Rev. D \textbf{65}, 104011 (2002).
\newblock \doi{10.1103/PhysRevD.65.104011}.
\newblock \urlprefix\url{https://link.aps.org/doi/10.1103/PhysRevD.65.104011}

\bibitem{Schunck:2003kk}
F.E. Schunck, E.W. Mielke, Class. Quant. Grav. \textbf{20}, R301 (2003).
\newblock \doi{10.1088/0264-9381/20/20/201}

\bibitem{Mak:2001eb}
M.K. Mak, T.~Harko, Proc. Roy. Soc. Lond. A \textbf{459}, 393 (2003).
\newblock \doi{10.1098/rspa.2002.1014}

\bibitem{Usov:2004iz}
V.V. Usov, Phys. Rev. D \textbf{70}, 067301 (2004).
\newblock \doi{10.1103/PhysRevD.70.067301}

\bibitem{Rahaman:2010mr}
F.~Rahaman, S.~Ray, A.K. Jafry, K.~Chakraborty, Phys. Rev. D \textbf{82},
  104055 (2010).
\newblock \doi{10.1103/PhysRevD.82.104055}

\bibitem{Rahaman:2010bt}
F.~Rahaman, P.K.F. Kuhfittig, M.~Kalam, A.A. Usmani, S.~Ray, Class. Quant.
  Grav. \textbf{28}, 155021 (2011).
\newblock \doi{10.1088/0264-9381/28/15/155021}

\bibitem{Varela:2010mf}
V.~Varela, F.~Rahaman, S.~Ray, K.~Chakraborty, M.~Kalam, Phys. Rev. D
  \textbf{82}, 044052 (2010).
\newblock \doi{10.1103/PhysRevD.82.044052}

\bibitem{Rahaman:2011hd}
F.~Rahaman, R.~Maulick, A.K. Yadav, S.~Ray, R.~Sharma, Gen. Rel. Grav.
  \textbf{44}, 107 (2012).
\newblock \doi{10.1007/s10714-011-1262-y}

\bibitem{Kalam:2012sh}
M.~Kalam, F.~Rahaman, S.~Ray, S.M. Hossein, I.~Karar, J.~Naskar, Eur. Phys. J.
  C \textbf{72}, 2248 (2012).
\newblock \doi{10.1140/epjc/s10052-012-2248-y}

\bibitem{Deb:2015vda}
D.~Deb, S.~Roy~Chowdhury, S.~Ray, F.~Rahaman, Gen. Rel. Grav. \textbf{50}(9),
  112 (2018).
\newblock \doi{10.1007/s10714-018-2434-9}

\bibitem{Shee:2015kqa}
D.~Shee, F.~Rahaman, B.K. Guha, S.~Ray, Astrophys. Space Sci. \textbf{361}(5),
  167 (2016).
\newblock \doi{10.1007/s10509-016-2753-9}

\bibitem{Maurya:2016oml}
S.K. Maurya, Y.K. Gupta, S.~Ray, D.~Deb, Eur. Phys. J. C \textbf{76}(12), 693
  (2016).
\newblock \doi{10.1140/epjc/s10052-016-4527-5}

\bibitem{Abbas:2014rja}
G.~Abbas, S.~Nazeer, M.A. Meraj, Astrophys. Space Sci. \textbf{354}, 449
  (2014).
\newblock \doi{10.1007/s10509-014-2110-9}

\bibitem{Abbas:2015yma}
G.~Abbas, A.~Kanwal, M.~Zubair, Astrophys. Space Sci. \textbf{357}(2), 109
  (2015).
\newblock \doi{10.1007/s10509-015-2337-0}

\bibitem{Abbas:2015wea}
G.~Abbas, S.~Qaisar, A.~Jawad, S.~Qaisar, A.~Jawad, Astrophys. Space Sci.
  \textbf{359}(2), 57 (2015).
\newblock \doi{10.1007/s10509-015-2509-y}

\bibitem{Zubair:2015cpa}
M.~Zubair, G.~Abbas, Astrophys. Space Sci. \textbf{361}(1), 27 (2016).
\newblock \doi{10.1007/s10509-015-2610-2}

\bibitem{zubair2016some}
M.~Zubair, G.~Abbas, Astrophysics and Space Science \textbf{361}(10), 1 (2016)

\bibitem{zubair2016possible}
M.~Zubair, G.~Abbas, I.~Noureen, Astrophysics and Space Science
  \textbf{361}(1), 1 (2016)

\bibitem{Ilyas:2018tht}
M.~Ilyas, Eur. Phys. J. C \textbf{78}(9), 757 (2018).
\newblock \doi{10.1140/epjc/s10052-018-6232-z}

\bibitem{Yousaf:2017lto}
Z.~Yousaf, M.~Sharif, M.~Ilyas, M.Z. Bhatti, Eur. Phys. J. C \textbf{77}(10),
  691 (2017).
\newblock \doi{10.1140/epjc/s10052-017-5280-0}

\bibitem{Shamir:2017rjz}
M.F. Shamir, M.~Ahmad, Eur. Phys. J. C \textbf{77}(10), 674 (2017).
\newblock \doi{10.1140/epjc/s10052-017-5239-1}

\bibitem{Shamir:2017yza}
M.F. Shamir, S.~Zia, Eur. Phys. J. C \textbf{77}(7), 448 (2017).
\newblock \doi{10.1140/epjc/s10052-017-5010-7}

\bibitem{Das:2016mxq}
A.~Das, F.~Rahaman, B.K. Guha, S.~Ray, Eur. Phys. J. C \textbf{76}(12), 654
  (2016).
\newblock \doi{10.1140/epjc/s10052-016-4503-0}

\bibitem{Palmer:1974hb}
R.G. Palmer, P.W. Anderson, Phys. Rev. D \textbf{9}, 3281 (1974).
\newblock \doi{10.1103/PhysRevD.9.3281}

\bibitem{Ramanan:2019kwf}
S.~Ramanan, DAE Symp. Nucl. Phys. \textbf{64}, 45 (2019)

\bibitem{Weber:2006ep}
F.~Weber, R.~Negreiros, P.~Rosenfield, M.~Stejner, Prog. Part. Nucl. Phys.
  \textbf{59}, 94 (2007).
\newblock \doi{10.1016/j.ppnp.2006.12.008}

\bibitem{Rahmansyah:2020gar}
A.~Rahmansyah, A.~Sulaksono, A.B. Wahidin, A.M. Setiawan, Eur. Phys. J. C
  \textbf{80}(8), 769 (2020).
\newblock \doi{10.1140/epjc/s10052-020-8361-4}

\bibitem{Sawyer:1972cq}
R.F. Sawyer, Phys. Rev. Lett. \textbf{29}, 382 (1972).
\newblock \doi{10.1103/PhysRevLett.29.382}

\bibitem{bowers1974anisotropic}
R.L. Bowers, E.~Liang, The Astrophysical Journal \textbf{188}, 657 (1974)

\bibitem{herrera1997local}
L.~Herrera, N.O. Santos, Physics Reports \textbf{286}(2), 53 (1997)

\bibitem{herrera1992cracking}
L.~Herrera, Physics Letters A \textbf{165}(3), 206 (1992)

\bibitem{Herrera:2007kz}
L.~Herrera, J.~Ospino, A.~Di~Prisco, Phys. Rev. D \textbf{77}, 027502 (2008).
\newblock \doi{10.1103/PhysRevD.77.027502}

\bibitem{Herrera:2008bt}
L.~Herrera, N.O. Santos, A.~Wang, Phys. Rev. D \textbf{78}, 084026 (2008).
\newblock \doi{10.1103/PhysRevD.78.084026}

\bibitem{Herrera:2011cr}
L.~Herrera, A.~Di~Prisco, J.~Ibanez, Phys. Rev. D \textbf{84}, 107501 (2011).
\newblock \doi{10.1103/PhysRevD.84.107501}

\bibitem{misner1973gravitation}
C.W. Misner, K.S. Thorne, J.A. Wheeler, \emph{Gravitation} (Macmillan, 1973)

\bibitem{EventHorizonTelescope:2019dse}
K.~Akiyama, et~al., Astrophys. J. Lett. \textbf{875} (2019).
\newblock \doi{10.3847/2041-8213/ab0ec7}

\bibitem{Abbott:2016blz}
B.P. Abbott, et~al., Phys. Rev. Lett. \textbf{116}(6), 061102 (2016).
\newblock \doi{10.1103/PhysRevLett.116.061102}

\bibitem{Abbott:2017oio}
B.P. Abbott, et~al., Phys. Rev. Lett. \textbf{119}(14), 141101 (2017).
\newblock \doi{10.1103/PhysRevLett.119.141101}

\bibitem{TheLIGOScientific:2017qsa}
B.P. Abbott, et~al., Phys. Rev. Lett. \textbf{119}(16), 161101 (2017).
\newblock \doi{10.1103/PhysRevLett.119.161101}

\bibitem{Riess:1998cb}
A.G. Riess, et~al., Astron. J. \textbf{116}, 1009 (1998).
\newblock \doi{10.1086/300499}

\bibitem{Perlmutter:1998np}
S.~Perlmutter, et~al., Astrophys. J. \textbf{517}, 565 (1999).
\newblock \doi{10.1086/307221}

\bibitem{deBernardis:2000sbo}
P.~de~Bernardis, et~al., Nature \textbf{404}, 955 (2000).
\newblock \doi{10.1038/35010035}

\bibitem{Knop:2003iy}
R.A. Knop, et~al., Astrophys. J. \textbf{598}, 102 (2003).
\newblock \doi{10.1086/378560}

\bibitem{Verde:2019ivm}
L.~Verde, T.~Treu, A.G. Riess, Nature Astron. \textbf{3}, 891 (2019).
\newblock \doi{10.1038/s41550-019-0902-0}

\bibitem{Hashim:2020sez}
M.~Hashim, W.~El~Hanafy, A.~Golovnev, A.A. El-Zant, JCAP \textbf{07}, 052
  (2021).
\newblock \doi{10.1088/1475-7516/2021/07/052}

\bibitem{Hashim:2021pkq}
M.~Hashim, A.A. El-Zant, W.~El~Hanafy, A.~Golovnev, JCAP \textbf{07}, 053
  (2021).
\newblock \doi{10.1088/1475-7516/2021/07/053}

\bibitem{DiValentino:2021izs}
E.~Di~Valentino, O.~Mena, S.~Pan, L.~Visinelli, W.~Yang, A.~Melchiorri, D.F.
  Mota, A.G. Riess, J.~Silk, Class. Quant. Grav. \textbf{38}(15), 153001
  (2021).
\newblock \doi{10.1088/1361-6382/ac086d}

\bibitem{Clifton:2011jh}
T.~Clifton, P.G. Ferreira, A.~Padilla, C.~Skordis, Phys. Rept. \textbf{513}, 1
  (2012).
\newblock \doi{10.1016/j.physrep.2012.01.001}

\bibitem{Nojiri:2006ri}
S.~Nojiri, S.D. Odintsov, eConf p.~06 (2006).
\newblock \doi{10.1142/S0219887807001928}

\bibitem{DeFelice:2010aj}
A.~De~Felice, S.~Tsujikawa, Living Rev. Rel. \textbf{13}, 3 (2010).
\newblock \doi{10.12942/lrr-2010-3}

\bibitem{Nashed:2019tuk}
G.G.L. Nashed, S.~Capozziello, Phys. Rev. (10), 104018 (2019).
\newblock \doi{10.1103/PhysRevD.99.104018}

\bibitem{Nashed:2011fg}
G.G.L. Nashed, Annalen Phys. \textbf{523}, 450 (2011).
\newblock \doi{10.1002/andp.201100030}

\bibitem{2014IJTP...53.3901W}
M.I. {Wanas}, H.A. {Hassan}, International Journal of Theoretical Physics
  \textbf{53}(11), 3901 (2014).
\newblock \doi{10.1007/s10773-014-2141-6}

\bibitem{Nashed:2018efg}
G.G.L. Nashed, Int. J. Mod. Phys. D \textbf{27}(7), 1850074 (2018).
\newblock \doi{10.1142/S0218271818500748}

\bibitem{Nashed:2020kdb}
G.G.L. Nashed, E.N. Saridakis, Phys. Rev. D \textbf{102}(12), 124072 (2020).
\newblock \doi{10.1103/PhysRevD.102.124072}

\bibitem{Cognola:2006eg}
G.~Cognola, E.~Elizalde, S.~Nojiri, S.D. Odintsov, S.~Zerbini, Phys. Rev. D
  \textbf{73}, 084007 (2006).
\newblock \doi{10.1103/PhysRevD.73.084007}

\bibitem{Li:2007jm}
B.~Li, J.D. Barrow, D.F. Mota, Phys. Rev. D \textbf{76}, 044027 (2007).
\newblock \doi{10.1103/PhysRevD.76.044027}

\bibitem{DeFelice:2008wz}
A.~De~Felice, S.~Tsujikawa, Phys. Lett. B \textbf{675}, 1 (2009).
\newblock \doi{10.1016/j.physletb.2009.03.060}

\bibitem{Astashenok:2014nua}
A.V. Astashenok, S.~Capozziello, S.D. Odintsov, JCAP \textbf{01}, 001 (2015).
\newblock \doi{10.1088/1475-7516/2015/01/001}

\bibitem{Linder:2010py}
E.V. Linder, Phys. Rev. p. 127301 (2010).
\newblock \doi{10.1103/PhysRevD.81.127301, 10.1103/PhysRevD.82.109902}.
\newblock [Erratum: Phys. Rev.D82,109902(2010)]

\bibitem{Cai:2015emx}
Y.F. Cai, S.~Capozziello, M.~De~Laurentis, E.N. Saridakis, Rept. Prog. Phys.
  \textbf{79}(10), 106901 (2016).
\newblock \doi{10.1088/0034-4885/79/10/106901}

\bibitem{Awad:2017tyz}
A.M. Awad, S.~Capozziello, G.G.L. Nashed, JHEP \textbf{07}, 136 (2017).
\newblock \doi{10.1007/JHEP07(2017)136}

\bibitem{Nashed:uja}
G.G.L. Nashed, Gen. Rel. Grav. \textbf{45}, 1887 (2013).
\newblock \doi{10.1007/s10714-013-1566-1}

\bibitem{Nashed:2013bfa}
G.G.L. Nashed, Phys. Rev. p. 104034 (2013).
\newblock \doi{10.1103/PhysRevD.88.104034}

\bibitem{Nashed:2018qag}
G.G.L. Nashed, W.~El~Hanafy, K.~Bamba, JCAP \textbf{01}, 058 (2019).
\newblock \doi{10.1088/1475-7516/2019/01/058}

\bibitem{Nashed:2019yto}
G.G.L. Nashed, W.~El~Hanafy, S.D. Odintsov, V.K. Oikonomou, Int. J. Mod. Phys.
  D \textbf{29}(13), 2050090 (2020).
\newblock \doi{10.1142/S021827182050090X}

\bibitem{ElHanafy:2017sih}
W.~El~Hanafy, G.G.L. Nashed, Int. J. Mod. Phys. D \textbf{26}(14), 1750154
  (2017).
\newblock \doi{10.1142/S0218271817501541}

\bibitem{ElHanafy:2020pek}
W.~El~Hanafy, E.N. Saridakis, JCAP \textbf{09}, 019 (2021).
\newblock \doi{10.1088/1475-7516/2021/09/019}

\bibitem{Rastall:1972swe}
P.~Rastall, Phys. Rev. D \textbf{6}, 3357 (1972).
\newblock \doi{10.1103/PhysRevD.6.3357}

\bibitem{Rastall:1976uh}
P.~Rastall, Can. J. Phys. \textbf{54}, 66 (1976).
\newblock \doi{10.1139/p76-008}

\bibitem{PhysRevD.85.084008}
C.E.M. Batista, M.H. Daouda, J.C. Fabris, O.F. Piattella, D.C. Rodrigues, Phys.
  Rev. D \textbf{85}, 084008 (2012).
\newblock \doi{10.1103/PhysRevD.85.084008}.
\newblock \urlprefix\url{https://link.aps.org/doi/10.1103/PhysRevD.85.084008}

\bibitem{fabris2012rastall}
J.C. Fabris, O.F. Piattella, D.C. Rodrigues, C.E. Batista, M.H. Daouda, in
  \emph{International Journal of Modern Physics: Conference Series}, vol.~18
  (World Scientific, 2012), vol.~18, pp. 67--76

\bibitem{moradpour2016thermodynamics}
H.~Moradpour, Physics Letters B \textbf{757}, 187 (2016)

\bibitem{heydarzade2017black}
Y.~Heydarzade, F.~Darabi, Physics Letters B \textbf{771}, 365 (2017)

\bibitem{ma2017noncommutative}
M.S. Ma, R.~Zhao, The European Physical Journal C \textbf{77}(9), 1 (2017)

\bibitem{lobo2018thermodynamics}
I.P. Lobo, H.~Moradpour, J.~Morais~Graca, I.~Salako, International Journal of
  Modern Physics D \textbf{27}(07), 1850069 (2018)

\bibitem{bamba2018thermodynamics}
K.~Bamba, A.~Jawad, S.~Rafique, H.~Moradpour, The European Physical Journal C
  \textbf{78}(12), 1 (2018)

\bibitem{xu2018kerr}
Z.~Xu, X.~Hou, X.~Gong, J.~Wang, The European Physical Journal C
  \textbf{78}(6), 1 (2018)

\bibitem{kumar2018rotating}
R.~Kumar, S.G. Ghosh, The European Physical Journal C \textbf{78}(9), 1 (2018)

\bibitem{moradpour2016thermodynamic}
H.~Moradpour, I.G. Salako, Advances in High Energy Physics \textbf{2016}

\bibitem{Soroushfar2019}
S.~Soroushfar, R.~Saffari, S.~Upadhyay, General Relativity and Gravitation
  \textbf{51}(10) (2019).
\newblock \doi{10.1007/s10714-019-2614-2}.
\newblock Cited By 9

\bibitem{Cruz:2019jiq}
M.~Cruz, S.~Lepe, G.~Morales-Navarrete, Class. Quant. Grav. \textbf{36}(22),
  225007 (2019).
\newblock \doi{10.1088/1361-6382/ab45ab}

\bibitem{Visser:2017gpz}
M.~Visser, Phys. Lett. B \textbf{782}, 83 (2018).
\newblock \doi{10.1016/j.physletb.2018.05.028}

\bibitem{Darabi:2017coc}
F.~Darabi, H.~Moradpour, I.~Licata, Y.~Heydarzade, C.~Corda, Eur. Phys. J. C
  \textbf{78}, 25 (2018).
\newblock \doi{10.1140/epjc/s10052-017-5502-5}

\bibitem{Hansraj:2018zwl}
S.~Hansraj, A.~Banerjee, P.~Channuie, Annals Phys. \textbf{400}, 320 (2019).
\newblock \doi{10.1016/j.aop.2018.12.003}

\bibitem{Hansraj:2020clg}
S.~Hansraj, A.~Banerjee, Mod. Phys. Lett. A \textbf{35}(13), 2050105 (2020).
\newblock \doi{10.1142/S0217732320501059}

\bibitem{abbas2018new}
G.~Abbas, M.~Shahzad, The European Physical Journal A \textbf{54}(12), 1 (2018)

\bibitem{abbas2018isotropic}
G.~Abbas, M.~Shahzad, Astrophysics and Space Science \textbf{363}(12), 1 (2018)

\bibitem{Abubekerov_2008}
M.K. Abubekerov, E.A. Antokhina, A.M. Cherepashchuk, V.V. Shimanskii, Astronomy
  Reports \textbf{52}(5), 379 (2008).
\newblock \doi{10.1134/s1063772908050041}.
\newblock \urlprefix\url{https://doi.org/10.1134%2Fs1063772908050041}

\bibitem{Oliveira:2015lka}
A.M. Oliveira, H.E.S. Velten, J.C. Fabris, L.~Casarini, Phys. Rev. D
  \textbf{92}(4), 044020 (2015).
\newblock \doi{10.1103/PhysRevD.92.044020}

\bibitem{Nashed:2020buf}
G.G.L. Nashed, A.~Abebe, K.~Bamba, Eur. Phys. J. C \textbf{80}(12), 1109
  (2020).
\newblock \doi{10.1140/epjc/s10052-020-08671-8}

\bibitem{Nashed:2020kjh}
G.G.L. Nashed, S.~Capozziello, Eur. Phys. J. C \textbf{80}(10), 969 (2020).
\newblock \doi{10.1140/epjc/s10052-020-08551-1}

\bibitem{Das:2019dkn}
S.~Das, F.~Rahaman, L.~Baskey, Eur. Phys. J. (10), 853 (2019).
\newblock \doi{10.1140/epjc/s10052-019-7367-2}

\bibitem{Singh:2019ykp}
K.~Newton~Singh, F.~Rahaman, A.~Banerjee, Phys. Rev. D \textbf{100}(8), 084023
  (2019).
\newblock \doi{10.1103/PhysRevD.100.084023}

\bibitem{Roupas:2020mvs}
Z.~Roupas, G.G.L. Nashed, Eur. Phys. J. C \textbf{80}(10), 905 (2020).
\newblock \doi{10.1140/epjc/s10052-020-08462-1}

\bibitem{HERRERa1992206}
L.~Herrera, Physics Letters A \textbf{165}(3), 206  (1992).
\newblock \doi{https://doi.org/10.1016/0375-9601(92)90036-L}.
\newblock
  \urlprefix\url{http://www.sciencedirect.com/science/article/pii/037596019290036L}

\bibitem{1971reas.book.....Z}
Y.B. {Zeldovich}, I.D. {Novikov}, \emph{{Relativistic astrophysics. Vol.1:
  Stars and relativity}} (1971)

\bibitem{Gangopadhyay:2013gha}
T.~Gangopadhyay, S.~Ray, X.D. Li, J.~Dey, M.~Dey, Mon. Not. Roy. Astron. Soc.
  \textbf{431}, 3216 (2013).
\newblock \doi{10.1093/mnras/stt401}

\bibitem{PhysRev.116.1027}
H.A. Buchdahl, Phys. Rev. \textbf{116}, 1027 (1959).
\newblock \doi{10.1103/PhysRev.116.1027}.
\newblock \urlprefix\url{https://link.aps.org/doi/10.1103/PhysRev.116.1027}

\bibitem{Ivanov:2002xf}
B.V. Ivanov, Phys. Rev. D \textbf{65}, 104011 (2002).
\newblock \doi{10.1103/PhysRevD.65.104011}

\bibitem{PhysRevD.67.064003}
D.E. Barraco, V.H. Hamity, R.J. Gleiser, Phys. Rev. D \textbf{67}, 064003
  (2003).
\newblock \doi{10.1103/PhysRevD.67.064003}.
\newblock \urlprefix\url{https://link.aps.org/doi/10.1103/PhysRevD.67.064003}

\bibitem{Bohmer2006}
C.G. B\"{o}hmer, T.~Harko, Classical and Quantum Gravity \textbf{23}(22), 6479
  (2006).
\newblock \doi{10.1088/0264-9381/23/22/023}.
\newblock \urlprefix\url{https://doi.org/10.1088%2F0264-9381%2F23%2F22%2F023}

\bibitem{PhysRev.55.364}
R.C. Tolman, Phys. Rev. \textbf{55}, 364 (1939).
\newblock \doi{10.1103/PhysRev.55.364}.
\newblock \urlprefix\url{https://link.aps.org/doi/10.1103/PhysRev.55.364}

\bibitem{PhysRev.55.374}
J.R. Oppenheimer, G.M. Volkoff, Phys. Rev. \textbf{55}, 374 (1939).
\newblock \doi{10.1103/PhysRev.55.374}.
\newblock \urlprefix\url{https://link.aps.org/doi/10.1103/PhysRev.55.374}

\bibitem{PoncedeLeon1993}
J.~Ponce~de Leon, General Relativity and Gravitation \textbf{25}(11), 1123
  (1993).
\newblock \doi{10.1007/BF00763756}.
\newblock \urlprefix\url{https://doi.org/10.1007/BF00763756}

\bibitem{1964ApJ...140..417C}
S.~{Chandrasekhar}, Astrophys. J. \textbf{140}, 417 (1964).
\newblock \doi{10.1086/147938}

\bibitem{1989A&A...221....4M}
M.~{Merafina}, R.~{Ruffini}, Astronomy and Astrophysics \textbf{221}(1), 4
  (1989)

\bibitem{10.1093/mnras/265.3.533}
R.~Chan, L.~Herrera, N.O. Santos, Monthly Notices of the Royal Astronomical
  Society \textbf{265}(3), 533 (1993).
\newblock \doi{10.1093/mnras/265.3.533}.
\newblock \urlprefix\url{https://doi.org/10.1093/mnras/265.3.533}

\bibitem{1975A&A....38...51H}
H.~{Heintzmann}, W.~{Hillebrandt}, aap \textbf{38}(1), 51 (1975)

\bibitem{2002ApJ...564..981P}
J.A. {Pons}, F.M. {Walter}, J.M. {Lattimer}, M.~{Prakash}, R.~{Neuh{\"a}user},
  P.~{An}, Astrophys. J. \textbf{564}(2), 981 (2002).
\newblock \doi{10.1086/324296}

\bibitem{Rawls:2011jw}
M.L. Rawls, J.A. Orosz, J.E. McClintock, M.A.P. Torres, C.D. Bailyn, M.M.
  Buxton, Astrophys. J. \textbf{730}, 25 (2011).
\newblock \doi{10.1088/0004-637X/730/1/25}

\bibitem{LIGOScientific:2018cki}
B.P. Abbott, et~al., Phys. Rev. Lett. \textbf{121}(16), 161101 (2018).
\newblock \doi{10.1103/PhysRevLett.121.161101}

\bibitem{Ozel:2008kb}
F.~Ozel, T.~Guver, D.~Psaltis, Astrophys. J. \textbf{693}, 1775 (2009).
\newblock \doi{10.1088/0004-637X/693/2/1775}

\bibitem{Raaijmakers:2019qny}
G.~Raaijmakers, et~al., Astrophys. J. Lett. \textbf{887}(1) (2019).
\newblock \doi{10.3847/2041-8213/ab451a}

\bibitem{Webb:2007tc}
N.A. Webb, D.~Barret, Astrophys. J. \textbf{671}, 727 (2007).
\newblock \doi{10.1086/522877}

\bibitem{LIGOScientific:2020zkf}
R.~Abbott, et~al., Astrophys. J. Lett. \textbf{896}(2) (2020).
\newblock \doi{10.3847/2041-8213/ab960f}

\bibitem{Rybicki:2005id}
G.B. Rybicki, C.O. Heinke, R.~Narayan, J.E. Grindlay, Astrophys. J.
  \textbf{644}, 1090 (2006).
\newblock \doi{10.1086/503701}

\bibitem{Reardon:2015kba}
D.J. Reardon, et~al., Mon. Not. Roy. Astron. Soc. \textbf{455}(2), 1751 (2016).
\newblock \doi{10.1093/mnras/stv2395}

\bibitem{Miller:2019cac}
M.C. Miller, et~al., Astrophys. J. Lett. \textbf{887}(1) (2019).
\newblock \doi{10.3847/2041-8213/ab50c5}

\bibitem{G_ver_2010}
T.~G\"{u}ver, P.~Wroblewski, L.~Camarota, F.~\"{O}zel, The Astrophysical
  Journal \textbf{719}(2), 1807 (2010).
\newblock \doi{10.1088/0004-637x/719/2/1807}.
\newblock \urlprefix\url{https://doi.org/10.1088%2F0004-637x%2F719%2F2%2F1807}

\bibitem{Naik:2011qc}
S.~Naik, B.~Paul, Z.~Ali, Astrophys. J. \textbf{737}, 79 (2011).
\newblock \doi{10.1088/0004-637X/737/2/79}

\bibitem{1996IAUC.6331....1M}
F.E. {Marshall}, L.~{Angelini}, IAU Circ. \textbf{6331}, 1 (1996)

\bibitem{Demorest:2010bx}
P.~Demorest, T.~Pennucci, S.~Ransom, M.~Roberts, J.~Hessels, Nature
  \textbf{467}, 1081 (2010).
\newblock \doi{10.1038/nature09466}

\bibitem{Antoniadis:2013pzd}
J.~Antoniadis, et~al., Science \textbf{340}, 6131 (2013).
\newblock \doi{10.1126/science.1233232}

\bibitem{Alaverdyan:2001num}
G.~Alaverdyan, A.R. Harutyunyan, Y.L. Vartanyan, Astrofiz. \textbf{44}, 265
  (2001).
\newblock \doi{10.1023/A:1010957111993}

\bibitem{Xu2005}
R.~Xu, \emph{Low-Mass Quark Stars} (Springer Netherlands, Dordrecht, 2005), pp.
  179--190.
\newblock \doi{{10.1007/1-4020-3881-X\_18}}.
\newblock \urlprefix\url{https://doi.org/10.1007/1-4020-3881-X\_18}

\end{thebibliography}

\end{document}